\documentclass[a4paper,11pt]{article}
\usepackage{amssymb}
\usepackage{graphicx}
\usepackage{amsmath}

\DeclareFontFamily{U}{rsf}{}
\DeclareFontShape{U}{rsf}{m}{n}{
  <5> <6> rsfs5 <7> <8> <9> rsfs7 <10-> rsfs10}{}
\DeclareMathAlphabet\Scr{U}{rsf}{m}{n} \makeatletter
\@addtoreset{equation}{section} \makeatother

\def\be{\begin{equation}}
\def\ee{\end{equation}}
\def\ba{\begin{array}}
\def\ea{\end{array}}

\newcommand{\bea}{\begin{eqnarray}}
\newcommand{\eea}{\end{eqnarray}}

\begin{document}

\begin{titlepage}

\begin{flushright}
CERN-PH-TH/2012-026\\ SU-ITP-2012-06
\end{flushright}

\vskip 2.0 cm
\begin{center}  {\huge{\bf      Degeneration of Groups of Type $E_{7}$ \\
     \vspace{5pt}and Minimal Coupling in Supergravity
    }}

\vskip 1.5 cm

{\Large{\bf Sergio Ferrara$^{1,2}$}, {\bf Renata Kallosh$^3$}, {\bf Alessio Marrani$^1$}}

\vskip 1.0 cm

$^1${\sl Physics Department, Theory Unit, CERN,  CH 1211, Geneva 23, Switzerland }\\

\vskip 0.5 cm

$^2${\sl INFN - Laboratori Nazionali di Frascati, Via Enrico Fermi
40, 00044 Frascati, Italy}\\

\vskip 0.5 cm


$^3${\sl Department of Physics, Stanford University, Stanford,
California 94305 USA}

 \end{center}

 \vskip 3.5 cm

\begin{abstract}
We study properties of $D=4$ $\mathcal{N}\geqslant 2$ extended supergravities  (and related compactifications of superstring theory) and their consistent truncation to the phenomenologically interesting models of $\mathcal{N}=1$ supergravity.  This involves a detailed classification of the \textquotedblleft degenerations" of the duality groups of type $E_{7}$, when the
corresponding \textit{%
quartic} invariant polynomial built from the symplectic irreducible representation
of $G_{4}$ \textquotedblleft degenerates" into a \textit{perfect square}. With regard to cosmological applications, \textit{minimal coupling} of vectors in consistent
truncation to $\mathcal{N}=1$ from higher-dimensional or higher-$\mathcal{N}$
theory is non-generic. On the other hand,  \textit{non-minimal coupling} involving vectors coupled to scalars and axions is generic. These features of supergravity, following from the electric-magnetic duality,  may be useful in other applications, like stabilization of moduli, and in studies of non-perturbative black-hole solutions of supergravity/string theory.

 \end{abstract}
\vspace{24pt} \end{titlepage}


\newpage \tableofcontents \newpage

\section{\label{Intro}Introduction}

In the present investigation, we relate a physical property of supergravity
couplings to a mathematical property of the underlying electric-magnetic
duality symmetries\footnote{%
Further below, we use the term $U$-duality, meaning the \textquotedblleft
continuous\textquotedblright\ symmetries of \cite{CJ-1}. Their discrete
versions are the $U$-duality non-perturbative string theory symmetries \cite%
{HT-1}.} of $\mathcal{N}\geqslant 2$ extended supergravity in $D=4$
space-time dimensions.

In the textbook \cite{Bagger-Witten}, the coupling of $\mathcal{N}=1$ vector
and chiral multiplets to supergravity is presented in its \textit{minimal}
form, \textit{i.e.} it is assumed that the vector kinetic term
\begin{equation}
-{\frac{1}{4}}\delta _{\alpha \beta }F_{\mu \nu }^{\alpha }F^{\beta \mid \mu
\nu }  \label{min}
\end{equation}%
is scalar independent. However, supersymmetry allows for the replacement of
the constant kinetic vector matrix $\delta _{\alpha \beta }$ by an
holomorphic function of the scalar fields $z$, $\delta _{\alpha \beta
}\rightarrow f_{\alpha \beta }\left( z\right) $, such that kinetic vector
term reads
\begin{equation}
-\frac{1}{4}\Big (\text{Re}\,f_{\alpha \beta }(z)\Big )F_{\mu \nu }^{\alpha
}F^{\beta \mu \nu }+\frac{i}{4}\Big (\text{Im}\,f_{\alpha \beta }(z)\Big )%
F_{\mu \nu }^{\alpha }\tilde{F}^{\beta \mu \nu }\ .  \label{nonminimal}
\end{equation}%
Here the function $f_{\alpha \beta }(z)$ is holomorphic, so that a \textit{%
non-minimal} coupling is introduced. For example, for one vector, in the
simplest case, $f(z)=\phi +ia$ and we have a vector-vector-scalar, $\phi
F^{2}$, and a vector-vector-axion, $aF\tilde{F}$, couplings.

In theories with global supersymmetry the choice of the minimal coupling is
often preferred since only for constant, scalar independent $f_{\alpha \beta
}$ the theory is renormalizable. It is the same consideration which
suggested that a preferred K\"{a}hler potential is canonical. In the context
of supergravity, however, the requirement of renormalizability is less
relevant, the issue we address here is: what kind of vector coupling is
preferred in the models originating from higher supersymmetries/higher
dimensions.

\textit{Non-minimal} vector scalar couplings may play an important rule in
inflationary cosmology, because a direct coupling of the inflaton scalar
field to matter vector fields (as heavy vector bosons, or photons) may
provide the only way to complete the creation of matter in the early
Universe. This problem was recently addressed in \cite{FK-creation}, where
it was pointed out that in $\mathcal{N}=1$ supergravity obtained by
reduction from higher-dimensional and/or higher-supersymmetric theories the
\textit{non-minimal} vector scalar couplings (\ref{nonminimal}) are generic.

The present paper is intended to generalize the results of \cite{FK-creation}%
, because we believe that the issue of \textit{minimal coupling} in $%
\mathcal{N}\geqslant 2$ extended supergravities deserves some attention.
Indeed, such theories never \footnote{%
With exception of \textquotedblleft pure" $\mathcal{N}=2$ and $\mathcal{N}=3$
supergravity theories, which have no scalars, with $U(1)$ and $U\left(
3\right) $ $U$-duality group, respectively, consistent with the analysis of
\cite{GZreview}.} exhibit a constant $f_{\alpha \beta }$, and in \cite%
{FK-creation} this fact was pointed out to be \textit{a consequence of
electric-magnetic-duality, which requires a special coupling of the
non-linear sigma model of scalars to the vector sector} \cite{GZreview}. The
kinetic vector matrix $\mathcal{N}_{\Lambda \Sigma }$ which occurs in $%
\mathcal{N}\geqslant 2$, $D=4$ extended supergravities is not holomorphic,
\begin{equation}
\mathrm{Im}\mathcal{N}_{\Lambda \Sigma }F_{\mu \nu }^{\Lambda }F^{\mu \nu
\Sigma }+i\mathrm{Re}\mathcal{N}_{\Lambda \Sigma }F_{\mu \nu }^{\Lambda }%
\tilde{F}^{\Sigma \mu \nu }\,.  \label{0}
\end{equation}%
Here the kinetic term for vectors $\mathcal{N}_{\Lambda \Sigma }$ in general
depends on scalars. The matrix $\mathrm{Im}\mathcal{N}_{\Lambda \Sigma }$ is
a metric in the vector moduli space. Comparing the Maxwell term, $\mathcal{N}%
_{\Lambda \Sigma }$ should reduce to $-\frac{i}{4}\overline{f}_{\alpha \beta
}\left( \overline{z}\right) $ in the $\mathcal{N}=1$ theory \cite{CFGVP-1}.
Consistent truncations of $\mathcal{N}\geqslant 2$ extended supergravities
to $\mathcal{N}=1$ have been studied in \cite{ADF-1,ADF-2}, where it was
shown how the non-holomorphic $\mathcal{N}_{\Lambda \Sigma }$ reduces to an
anti-holomorphic $f_{\alpha \beta }$ in the corresponding truncated theories.

Let us remind that in $\mathcal{N}=2$ special K\"{a}hler geometry, in a
symplectic frame in which an holomorphic prepotential function $F\left(
X\right) $ exists (such that $X^{\Lambda }\partial _{\Lambda }F=2F$), the
kinetic vector matrix is given by (see \textit{e.g. }\cite{a1}, and Refs.
therein):%
\begin{equation}
\mathcal{N}_{\Lambda \Sigma }=\overline{F}_{\Lambda \Sigma }-2i\overline{T}%
_{\Lambda }\overline{T}_{\Sigma }\left( L^{\Xi }\text{Im}F_{\Xi \Omega
}L^{\Omega }\right) ,  \label{1.4}
\end{equation}%
where $F_{\Lambda \Sigma }=\partial _{\Lambda }\partial _{\Sigma }F$, $%
L^{\Lambda }=e^{K/2}X^{\Lambda }$ is the covariantly holomorphic
contravariant symplectic section, and%
\begin{equation}
T_{\Lambda }=2i\text{Im}\mathcal{N}_{\Lambda \Sigma }L^{\Sigma }
\end{equation}%
is the projector on the graviphoton ($T_{\mu \nu }^{-}=T_{\Lambda }F_{\mu
\nu }^{\Lambda \mid -}$), whose \textquotedblleft flux" define the $\mathcal{%
N}=2$ central charge $Z$ (see \textit{e.g.} \cite{CDFVP,CDF-rev} and Refs.
therein). Note that $\mathcal{N}_{\Lambda \Sigma }$ is not anti-holomorphic
because of the presence of the second term in the r.h.s. of (\ref{1.4}). In
order to have a consistent $\mathcal{N}=1$ reduction, one needs to impose $%
T_{\Lambda }=0$, \textit{i.e.} that the graviphoton projection vanishes
(when $\Lambda $ is restricted to the index running on $\mathcal{N}=1$
vector multiplets). One then obtains that \textit{minimal coupling} demands $%
F(X)$ to be \textit{quadratic} in the truncated scalars of the corresponding
would-be $\mathcal{N}=1$ vector multiplets.

It is here worth observing that, while \textit{minimal coupling} seems
natural in $\mathcal{N}=1$ supergravity \cite{Bagger-Witten}, its relaxation
is actually natural if one considers $\mathcal{N}=1$ theories coming from
supergravity theory \cite{W} or from higher dimensions \cite{CCF}. In the
systematic approach of the present paper, we will provide a detailed list of
examples in which \textit{minimal coupling} is impossible in the
higher-dimensional or higher-$\mathcal{N}$ theory, but it can be achieved by
a further suitable consistent truncation to $\mathcal{N}=1$.

This is related to the mathematical property of the $U$-duality group $G_{4}$
of type $E_{7}$ \cite{brown}. Simple, \textit{non-degenerate} groups $G_{4}$
are related to Freudenthal triple systems $\mathfrak{M}\left( J_{3}\right) $
on simple rank-$3$ Jordan algebras $J_{3}$. In general, $G_{4}\equiv $Conf$%
\left( J_{3}\right) =$Aut$\left( \mathfrak{M}\left( J_{3}\right) \right) $
(see \textit{e.g.} \cite{G-Lects,Small-Orbits-Phys,Small-Orbits-Maths} for a
recent introduction, and a list of Refs.). When considering a consistent
reduction to a subgroup, $G_{4}$ groups of type $E_{7}$ may admit a \textit{%
\textquotedblleft degeneration\textquotedblright } in which the rank-$4$
invariant symmetric structure $\mathbf{q}$ is \textit{reducible}, namely it
is the product of two symmetric invariant tensors. As a consequence, the
corresponding \textit{quartic} invariant polynomial built from the
symplectic irrep. $\mathbf{R}$ of $G_{4}$ \textquotedblleft degenerates"
into a \textit{perfect square}\footnote{%
An analysis at the level of quartic invariant polynomial, and dependent on
charge configurations, has been considered in \cite{Duff-FD-1}.}. Here $%
\mathbf{R}$ denotes the symplectic representation of the $U$-duality group $%
G_{4}$ formed by a the chiral (or anti-chiral) vector field strengths $%
F^{\Lambda \mid \pm }$ and their duals $G_{\Lambda }^{\mp }\equiv \mp \frac{i%
}{2}\delta \mathcal{L}/\delta F^{\Lambda \mid \mp }$:%
\begin{equation}
\mathbf{R=}\left( F^{\Lambda \mid \pm },G_{\Lambda }^{\pm }\right) ,
\end{equation}%
such that \textquotedblleft fluxes" of suitably defined projections defines
the central charge (matrix) and matter charges (\textit{if any}; see (see
\textit{e.g.} \cite{A-central} and Refs. therein). Sometimes, in order to
simplify the analysis, in the treatment below we will switch to the basis of
the fluxes of the corresponding field strengths, defining the dyonic vector
of magnetic and electric charges (\cite{GZreview}; see \textit{e.g.} the
treatment of \cite{a1}):%
\begin{equation}
\mathbf{R=}\left( p^{\Lambda },q_{\Lambda }\right) \equiv \mathcal{Q},
\label{Q-call}
\end{equation}%
even if our analysis does not only restrict to charged states, such as black
holes. By truncation of the charged fluxes $\mathcal{Q}$ we here mean the
reduction of the group $G_{4}$ and its irrep. $\mathbf{R}\left( G_{4}\right)
$ to some proper subgroup $G_{4}^{\prime }$ and its irrep. $\mathbf{R}\left(
G_{4}^{\prime }\right) \equiv \mathbf{R}^{\prime }$.

Since $\mathcal{N}>2$ theories are related to scalar manifolds which are
symmetric spaces, we will consider $\mathcal{N}=2$ theories with symmetric
cosets. Therefore, $\mathcal{N}=1$ truncations are simpler to investigate,
because the $\mathcal{N}=2$ theory leading to $\mathcal{N}=1$ \textit{%
minimal coupling} are the so-called $\mathcal{N}=2$ \textit{minimally coupled%
} Maxwell-Einstein supergravities \cite{Luciani}, whose scalar manifold is a
(non-compact) $\mathbb{CP}^{n}$ space. In a scalar-dressed symplectic frame
of $\mathcal{N}=2$ special K\"{a}hler geometry, the \textit{%
\textquotedblleft degeneration"} of the quartic polynomial invariant to a
quadratic one corresponds to setting the $C$-tensor to zero ($C_{ijk}=0$).
Also for $\mathcal{N}>2$, we will then consider those cases in which the
reduction to $\mathcal{N}=2$ gives rise to a $\mathbb{CP}^{n}$ special K\"{a}%
hler geometry ($C_{ijk}=0$), in which the $U$-duality group $G_{4}=U\left(
1,n\right) $ is a \textit{degenerate}\footnote{%
In \cite{FK-creation} these groups were called \textit{\textquotedblleft not
of type }$E_{7}$\textit{"}.} group of type $E_{7}$ \cite{Garibaldi-deg},
with the rank-$4$ completely symmetric invariant $\mathbf{q}$-structure r%
\textit{educible}, as pointed out above.

As recalled in Example 1.2 of \cite{Garibaldi-deg} and proved in \cite%
{brown,mey}, all degenerate Freudenthal triple systems are isomorphic to the
degenerate triple system in which the resulting quartic invariant polynomial
$\mathcal{I}_{4}$ is the square of a \textit{quadratic} invariant polynomial
$\mathcal{I}_{2}$ which, as pointed out above, also corresponds to the case
relevant for $D=4$ supergravity with symmetric scalar manifold (see the
treatment of Sec. \ref{Identities}, as well). The \textit{degeneration} of a
$U$-duality group $G_{4}$ of type $E_{7}$ is also confirmed by the fact that
the fundamental identity characterizing simple, \textit{non-degenerate}
groups of type $E_{7}$ (proved in Sec. 2 of \cite{Garibaldi-deg} for $E_{7}$%
, and generalized in formula (\ref{Id-2}) further below \textit{at least}
for all groups listed in Table 1) does \textit{not} hold in these cases; see
Sec. \ref{Identities}. The cases of $U$-duality groups as \textit{semi-simple%
}, \textit{non-degenerate} groups of type $E_{7}$ relevant to $D=4$
supergravity theories with symmetric (vector multiplets') scalar manifolds
are also analyzed in Subsec. \ref{Semi-Simple-Non-Degenerate}.

Simple, \textit{degenerate} groups of type $E_{7}$ relevant to $D=4$
supergravity (namely, $U\left( 1,n\right) $ or $U\left( 3,n\right) $) share
the property that the dyonic charge vector $\mathcal{Q}$ (\ref{Q-call})
(element of the Freudenthal triple system) fits into the sum of the
fundamental and anti-fundamental irrep.
\begin{equation}
\mathcal{Q}\in \mathbf{R}\equiv \mathbf{Fund}+\overline{\mathbf{Fund}},
\label{Fund+AntiFund}
\end{equation}%
thus naturally admitting a \textit{complex} representation, endowed with an
invariant Hermitian quadratic structure (see \textit{e.g.} \cite%
{Gilmore,Helgason}), whose real part gives rise to the aforementioned
quadratic invariant polynomial $\mathcal{I}_{2}$; see the discussion in Sec. %
\ref{Identities}.\medskip

It should be stressed that the conditions on truncations of fluxes and
embeddings of scalar manifolds, under consideration in the treatment below,
are generally only \textit{necessary}, but \textit{not sufficient} for
minimal coupling. An analysis of the consistency of the truncations at the
level of supersymmetry transformations, along the lines exploited in \cite%
{ADF-1} and \cite{ADF-2} (this latter on the further truncation $\mathcal{N}%
=2\rightarrow 1$) is required to determine also a sufficient condition; we
give a brief general account of this analysis at the start of Sec. \ref{N>4}%
, and we provide an explicit example in Subsec. \ref{N=8--->N=6}.\smallskip \medskip

The plan of the paper is as follows.

After axiomatically introducing groups of type $E_{7}$ in Sec. \ref%
{Identities}, we analyze various truncations to \textit{minimal coupling}
models in subsequent Sections. It is here worth pointing out that by
truncation of a theory we here mean a sub-theory obtained from the original
one by reducing the amount of supersymmetry. For \textit{\textquotedblleft
pure"} ($\mathcal{N}\geqslant 5$) supergravities, this means to consistently
truncate away the extra gravitino multiplet(s); these cases are considered
in Secs. \ref{N>4} and \ref{N=6}. On the other hand, for matter-coupled ($%
2\leqslant \mathcal{N}\leqslant 4$) theories the truncation also requires to
consistently truncate the matter multiplets' sector; such cases are analyzed
in Secs. \ref{4--->3}, \ref{4--->2+hypers} and \ref{3--->2CPn+hypers}. In
presence of matter coupling, there is another way of obtaining sub-theories,
namely to consistently reduce the matter sector but not the gravitino
multiplet(s); Sec. \ref{From-N=2} deals with such cases. The list of
examples produced by the systematic approach of the present investigation is
much larger than the ones given in \cite{ADF-1,ADF-2,FK-creation}, and it is
of some interest also because some truncations correspond to orbifolds and
orientifolds of string theories with larger supersymmetry, as discussed in
Sec. \ref{N=2--->N=1}, in which the further truncation $\mathcal{N}%
=2\rightarrow \mathcal{N}=1$ is considered. Comments on the \textit{%
\textquotedblleft degeneration"} of the so-called Freudenthal duality are
then given in Sec. \ref{Comment-FD}. Sec. \ref{Fermions} contain some
remarks on fermions and minimal coupling. Conclusive remarks and an outlook
are given in Sec. \ref{Conclusion}. Appendix \ref{Pauli}, containing some
details on the structure of Pauli terms, concludes the paper.

\section{\label{Identities}On Groups of Type $E_{7}$}

\subsection{Axiomatic Characterization}

The first axiomatic characterization of groups \textit{\textquotedblleft of
type }$E_{7}$\textit{"} through a module (irreducible representation) was
given in 1967 by Brown \cite{brown}.

A group $G$ of type $E_{7}$ is a Lie group endowed with a representation $%
\mathbf{R}$ such that:

\begin{enumerate}
\item $\mathbf{R}$ is \textit{symplectic}, \textit{i.e.} (the subscripts
\textquotedblleft $s$\textquotedblright\ and \textquotedblleft $a$%
\textquotedblright\ stand for symmetric and skew-symmetric throughout):
\begin{equation}
\exists !\mathbb{C}_{\left[ MN\right] }\equiv \mathbf{1\in R\times }_{a}%
\mathbf{R;}  \label{sympl-metric}
\end{equation}%
$\mathbb{C}_{\left[ MN\right] }$ defines a non-degenerate skew-symmetric
bilinear form (\textit{symplectic product}); given two different charge
vectors $\mathcal{Q}_{x}$ and $\mathcal{Q}_{y}$ in $\mathbf{R}$, such a
bilinear form is defined as
\begin{equation}
\left\langle \mathcal{Q}_{x},\mathcal{Q}_{y}\right\rangle \equiv \mathcal{Q}%
_{x}^{M}\mathcal{Q}_{y}^{N}\mathbb{C}_{MN}=-\left\langle \mathcal{Q}_{y},%
\mathcal{Q}_{x}\right\rangle .  \label{WW}
\end{equation}

\item $\mathbf{R}$ admits a unique rank-$4$ completely symmetric primitive $%
G $-invariant structure, usually named $K$-tensor
\begin{equation}
\exists !\mathbb{K}_{\left( MNPQ\right) }\equiv \mathbf{1\in }\left[ \mathbf{%
R\times R\times R\times R}\right] _{s}\mathbf{;}
\end{equation}%
thus, by contracting the $K$-tensor with the same charge vector $\mathcal{Q}$
in $\mathbf{R}$, one can construct a rank-$4$ homogeneous $G$-invariant
polynomial (whose $\varsigma $ is the normalization constant):
\begin{equation}
\mathbf{q}\left( \mathcal{Q}\right) \equiv \varsigma \mathbb{K}_{MNPQ}%
\mathcal{Q}^{M}\mathcal{Q}^{N}\mathcal{Q}^{P}\mathcal{Q}^{Q},  \label{I4}
\end{equation}%
which corresponds to the evaluation of the rank-$4$ symmetric invariant $%
\mathbf{q}$-structure induced by the $K$-tensor on four identical modules $%
\mathbf{R}$:
\begin{equation}
\mathbf{q}\left( Q\right) \equiv \left. \mathbf{q}\left( \mathcal{Q}_{x},%
\mathcal{Q}_{y},\mathcal{Q}_{z},\mathcal{Q}_{w}\right) \right\vert _{%
\mathcal{Q}_{x}=\mathcal{Q}_{y}=\mathcal{Q}_{z}=\mathcal{Q}_{w}\equiv
\mathcal{Q}}\equiv \varsigma \left[ \mathbb{K}_{MNPQ}\mathcal{Q}_{x}^{M}%
\mathcal{Q}_{y}^{N}\mathcal{Q}_{z}^{P}\mathcal{Q}_{w}^{Q}\right] _{\mathcal{Q%
}_{x}=\mathcal{Q}_{y}=\mathcal{Q}_{z}=\mathcal{Q}_{w}\equiv \mathcal{Q}}.
\end{equation}%
A famous example of \textit{quartic} invariant in $G=E_{7}$ is the \textit{%
Cartan-Cremmer-Julia} invariant (\cite{Cartan}, p. 274), constructed out of
the fundamental representation $\mathbf{R}=\mathbf{56}$.

\item if a trilinear map $T\mathbf{:R\times R\times R}\rightarrow \mathbf{R}$
is defined such that
\begin{equation}
\left\langle T\left( \mathcal{Q}_{x},\mathcal{Q}_{y},\mathcal{Q}_{z}\right) ,%
\mathcal{Q}_{w}\right\rangle =\mathbf{q}\left( \mathcal{Q}_{x},\mathcal{Q}%
_{y},\mathcal{Q}_{z},\mathcal{Q}_{w}\right) ,
\end{equation}%
then it holds that
\begin{equation}
\left\langle T\left( \mathcal{Q}_{x},\mathcal{Q}_{x},\mathcal{Q}_{y}\right)
,T\left( \mathcal{Q}_{y},\mathcal{Q}_{y},\mathcal{Q}_{y}\right)
\right\rangle =-2\left\langle \mathcal{Q}_{x},\mathcal{Q}_{y}\right\rangle
\mathbf{q}\left( \mathcal{Q}_{x},\mathcal{Q}_{y},\mathcal{Q}_{y},\mathcal{Q}%
_{y}\right) .  \label{FTS3}
\end{equation}%
This last property makes the group of type $E_{7}$ amenable to a treatment
in terms of (rank-$3$) Jordan algebras and related Freudenthal triple
systems.
\end{enumerate}

Remarkably, groups of type $E_{7}$, appearing in $D=4$ supergravity as $U$%
-duality groups, admit a $D=5$ uplift to groups of type $E_{6}$, as well as
a $D=3$ downlift to groups of type $E_{8}$. It should also be recalled that
split form of exceptional $E$ - Lie groups appear in the exceptional
Cremmer-Julia \cite{CJ-1} sequence $E_{11-D\left( 11-D\right) }$ of $U$%
-duality groups of $M$-theory compactified on a $D$-dimensional torus, in $%
D=3,4,5$. Other sequences, composed by non-split, non-compact real forms of
exceptional groups, are also relevant to non-maximal supergravity in various
dimensions (see \textit{e.g.} the treatment in \cite{Exc-Reds}, also for a
list of related Refs.).

The connection of groups of type $E_{7}$ to supergravity can be summarized
by stating that all $2\leqslant N\leqslant 8$-extended supergravities in $%
D=4 $ with symmetric scalar manifolds ${\frac{G_{4}}{H_{4}}}$ have $G_{4}$
of type $E_{7}$ \cite{Duff-FD-1,FMY-FD-1}. It is intriguing to notice that
the first paper on groups of type $E_{7}$ was written about a decade before
the discovery of of extended ($\mathcal{N}=2$) supergravity \cite{FVN}, in
which electromagnetic duality symmetry was observed \cite{FSZ}.

An example of Lie group which is not of type $E_{7}$ is the exceptional Lie
group $E_{6}$ in its fundamental representation\footnote{%
Strictly speaking, the pair $\left( G,\mathbf{R}\right) =\left( E_{6},%
\mathbf{27}\right) $ is the prototype of the so-called groups \textit{%
\textquotedblleft of type }$E_{6}$\textit{"}.} $\mathbf{27}$; this is
relevant to both maximal ($\mathcal{N}=8$) and exceptional ($\mathcal{N}=2$)
supergravity theories in $D=5$. The representation $\mathbf{27}$ is \textit{%
not} symplectic, but rather it is conjugated to its contra-gradient
counterpart ($a=1,...,27$):%
\begin{equation}
\exists !\delta _{b}^{a}\equiv \mathbf{1\in 27\times }\overline{\mathbf{27}}%
\mathbf{.}
\end{equation}%
Furthermore, $\mathbf{27}$ admits a unique rank-$3$ completely symmetric
primitive $E_{6}$-invariant structure, usually named $d$-tensor
\begin{equation}
\exists !d_{abc}\equiv \mathbf{1\in }\left[ \mathbf{27\times 27\times 27}%
\right] _{s}\mathbf{;}
\end{equation}%
thus, by contracting the $d$-tensor with the same charge vector $Q$ in $%
\mathbf{27}$, one can construct a rank-$3$ homogeneous $E_{6}$-invariant
polynomial (whose $\vartheta $ is the normalization constant):
\begin{equation}
\mathbf{d}\left( Q\right) \equiv \vartheta d_{abc}Q^{a}Q^{b}Q^{c},
\end{equation}%
which corresponds to the evaluation of the rank-$3$ symmetric invariant $%
\mathbf{d}$-structure induced by the $d$-tensor on four identical modules $%
\mathbf{27}$:
\begin{equation}
\mathbf{d}\left( Q\right) \equiv \left. \mathbf{d}\left(
Q_{x},Q_{y},Q_{z}\right) \right\vert _{Q_{x}=Q_{y}=Q_{z}\equiv \mathcal{Q}%
}\equiv \varsigma \left[ \vartheta d_{abc}Q_{x}^{a}Q_{y}^{b}Q_{z}^{c}\right]
_{\mathcal{Q}_{x}=\mathcal{Q}_{y}=\mathcal{Q}_{z}\equiv \mathcal{Q}}.
\end{equation}%
\medskip

Focussing on the relevance to supergravity theories in $D=4$, in the
remaining part of this Section we will characterize various classes of
groups of type $E_{7}$ in terms of (tensor and) scalar identities, along the
lines of \cite{Garibaldi-deg} and exploiting results of previous
investigations, such as \cite{Exc-Reds} and \cite{FMY-T-CV}.

\subsection{\label{Simple-Non-Deg}Simple, Non-Degenerate}

In \textit{simple}, \textit{non-degenerate} groups $G_{4}$ of type $E_{7}$
\cite{brown} relevant to $D=4$ (super)gravity with symmetric scalar
manifolds (listed in Table 1\footnote{%
We only consider rank-$3$ Jordan algebras related to locally supersymmetric
theories of gravity.}), the following identity holds (\textit{cfr.} (5.18)
of \cite{Exc-Reds}):
\begin{equation}
\mathbb{K}_{MNPQ}\mathbb{K}_{RSTU}\mathbb{C}^{PT}\mathbb{C}^{QU}=\xi \left[
\left( 2\tau -1\right) \mathbb{K}_{MNRS}+\xi \tau \left( \tau -1\right)
\mathbb{C}_{M(R}\mathbb{C}_{S)N}\right] .  \label{Id-1}
\end{equation}%
$\mathbb{C}_{MN}$ is the symplectic metric, and $\mathbb{K}_{MNPQ}$ denotes
the completely symmetric, rank-$4$ invariant \textquotedblleft $K$%
-tensor\textquotedblright\ in the relevant symplectic irrep. $\mathbf{R}%
\left( G_{4}\right) $ ($M$ is an index in $\mathbf{R}$):
\begin{eqnarray}
\mathbb{C} &\equiv &\exists !\mathbf{1}\in \left[ \mathbf{R}\times \mathbf{R}%
\right] _{a}; \\
\mathbb{K} &\equiv &\exists !\mathbf{1}\in \left[ \mathbf{R}\times \mathbf{%
R\times R}\times \mathbf{R}\right] _{s},
\end{eqnarray}%
where the subscript \textquotedblleft $s$\textquotedblright\
(\textquotedblleft $a$\textquotedblright ) denotes the (anti)symmetric part
of the tensor product. Moreover, the $G_{4}$-dependent parameters are
defined as \cite{Exc-Reds,ADFMT-1}
\begin{eqnarray}
\tau &\equiv &\frac{2d}{f\left( f+1\right) };  \label{csi} \\
\xi &\equiv &-\frac{1}{3\tau },  \label{tau}
\end{eqnarray}%
where
\begin{eqnarray}
f &\equiv &\text{dim}_{\mathbb{R}}\left( \mathbf{R}\left( G_{4}\right)
\right) ;  \label{f} \\
d &\equiv &\text{dim}_{\mathbb{R}}\left( \mathbf{Adj}\left( G_{4}\right)
\right) .  \label{d}
\end{eqnarray}

By using (\ref{Id-1}), one can show that the following identity holds:
\begin{equation}
tr\left( p\left( x\otimes x\right) p\left( y\otimes y\right) \right) =\beta
\left[ \mathbf{q}\left( x,x,y,y\right) -2b\left( y,x\right) ^{2}\right] ,
\label{Id-2}
\end{equation}%
where (recall definition (\ref{WW}))
\begin{eqnarray}
b\left( x,y\right) &\equiv &-\mathbb{C}_{MN}\mathcal{Q}_{x}^{M}\mathcal{Q}%
_{y}^{N}=-\left\langle \mathcal{Q}_{x},\mathcal{Q}_{y}\right\rangle . \\
\mathbf{q}\left( x,y,z,w\right) &\equiv &-6\mathbb{K}_{MNPQ}\mathcal{Q}%
_{x}^{M}\mathcal{Q}_{y}^{N}\mathcal{Q}_{z}^{P}\mathcal{Q}_{w}^{Q};  \label{q}
\\
\beta &\equiv &\frac{2}{\tau },  \label{beta}
\end{eqnarray}%
and $p$ denotes the following vector space map (\textit{cfr.} Sec. 2 of \cite%
{Garibaldi-deg} for further detail)
\begin{equation}
p\left( x\otimes y\right) z\equiv t\left( x,y,z\right) -b\left( z,x\right)
y-b\left( z,y\right) x,
\end{equation}%
where $t\left( x,y,z\right) $ is the trilinear product related to $\mathbf{q}%
\left( x,y,z,w\right) $ as
\begin{equation}
\mathbf{q}\left( x,y,z,w\right) \equiv b\left( x,t\left( y,z,w\right)
\right) .
\end{equation}%
The scalar identity (\ref{Id-2}) holds \textit{at least} for all simple,
\textit{non-degenerate} groups $G_{4}$ of type $E_{7}$ listed in Table 1
(and for all their other non-compact forms, as well as for the corresponding
compact Lie group $G_{4,c}$), and it is a consequence of the tensor identity
(\ref{Id-1}), which in turn follows from the identity for the $K$-tensor
given by (5.17) of \cite{Exc-Reds}. In the particular case of $E_{7}$ (see
Tables 1 and 2), it holds $\tau =1/12\Rightarrow \beta =24$, and the
identity proved in Theorem 2.3 of \cite{Garibaldi-deg} is retrieved.

\begin{table}[p]
\begin{center}
\begin{tabular}{|c||c|c|c|}
\hline
$%
\begin{array}{c}
\\
J_{3}%
\end{array}%
$ & $%
\begin{array}{c}
\\
G_{4} \\
~~%
\end{array}%
$ & $%
\begin{array}{c}
\\
\mathbf{R} \\
~~%
\end{array}%
$ & $%
\begin{array}{c}
\\
\mathcal{N} \\
~~%
\end{array}%
$ \\ \hline\hline
$%
\begin{array}{c}
\\
J_{3}^{\mathbb{O}} \\
~%
\end{array}%
$ & $E_{7\left( -25\right) }~$ & $\mathbf{56}$ & $2~$ \\ \hline
$%
\begin{array}{c}
\\
J_{3}^{\mathbb{O}_{s}} \\
~%
\end{array}%
$ & $E_{7\left( 7\right) }$ & $\mathbf{56}$ & $8$ \\ \hline
$%
\begin{array}{c}
\\
J_{3}^{\mathbb{H}} \\
~%
\end{array}%
$ & $SO^{\ast }\left( 12\right) $ & $\mathbf{32}$ & $2,~6$ \\ \hline
$%
\begin{array}{c}
\\
J_{3}^{\mathbb{C}} \\
~%
\end{array}%
$ & $SU\left( 3,3\right) $ & $\mathbf{20}$ & $2~$ \\ \hline
$%
\begin{array}{c}
\\
M_{1,2}\left( \mathbb{O}\right) \\
~%
\end{array}%
$ & $SU\left( 1,5\right) $ & $\mathbf{20}$ & $5$ \\ \hline
$%
\begin{array}{c}
\\
J_{3}^{\mathbb{R}} \\
~%
\end{array}%
$ & $Sp\left( 6,\mathbb{R}\right) $ & $\mathbf{14}^{\prime }$ & $2$ \\ \hline
$%
\begin{array}{c}
\\
\mathbb{R} \\
(T^{3}\text{~model})~~%
\end{array}%
$ & $SL\left( 2,\mathbb{R}\right) $ & $\mathbf{4}$ & $2$ \\ \hline
\end{tabular}%
\end{center}
\caption{Simple, \textit{non-degenerate} groups $G_{4}$ related to
Freudenthal triple systems $\mathfrak{M}\left( J_{3}\right) $ on simple rank-%
$3$ Jordan algebras $J_{3}$. The relevant symplectic irrep. $\mathbf{R}$ of $%
G_{4}$ is also reported. $\mathbb{O}$, $\mathbb{H}$, $\mathbb{C}$ and $%
\mathbb{R}$ respectively denote the four division algebras of octonions,
quaternions, complex and real numbers, and $\mathbb{O}_{s}$, $\mathbb{H}_{s}$%
, $\mathbb{C}_{s}$ are the corresponding split forms. Note that the $G_{4}$
related to split forms $\mathbb{O}_{s}$, $\mathbb{H}_{s}$, $\mathbb{C}_{s}$
is the \textit{maximally non-compact} (\textit{split}) real form of the
corresponding compact Lie group. The corresponding scalar manifolds are the
\textit{symmetric} cosets $\frac{G_{4}}{H_{4}}$, where $H_{4}$ is the
maximal compact subgroup (with symmetric embedding) of $G_{4}$. The number
of supercharges of the resulting supergravity theory in $D=4$ is also
listed. $M_{1,2}\left( \mathbb{O}\right) $ is the Jordan triple system
generated by $2\times 1$ vectors over $\mathbb{O}$ \protect\cite{GST}. The $%
D=5$ uplift of the $T^{3}$ model based on $J_{3}=\mathbb{R}$ is the \textit{%
pure} $\mathcal{N}=2$, $D=5$ supergravity. $J_{3}^{\mathbb{H}}$ is related
to both $8$ and $24$ supersymmetries, because the corresponding supergravity
theories are \textit{\textquotedblleft twin"}, namely they share the very
same bosonic sector \protect\cite{GST,ADF-fixed,Gnecchi-1,Samtleben-Twin}. }
\end{table}

\begin{table}[tbp]
\begin{center}
\begin{tabular}{|c||c|c|c|c|c|c|}
\hline
\rule[-1mm]{0mm}{6mm} $G_{4,c}$ & $q$ & $f$ & $d$ & $\tau $ & $\xi $ & $%
\beta $ \\ \hline\hline
\rule[-1mm]{0mm}{6mm} $E_{7}$ & $%
\begin{array}{ccc}
~ & 8 & ~%
\end{array}%
$ & $%
\begin{array}{ccc}
~ & 56 & ~%
\end{array}%
$ & $%
\begin{array}{ccc}
~ & 133 & ~%
\end{array}%
$ & $%
\begin{array}{ccc}
~ & 1/12 & ~%
\end{array}%
$ & $%
\begin{array}{ccc}
~ & -4 & ~%
\end{array}%
$ & $%
\begin{array}{ccc}
~ & 24 & ~%
\end{array}%
$ \\ \hline
\rule[-1mm]{0mm}{6mm} $SO\left( 12\right) $ & $4$ & $32$ & $66$ & $1/8$ & $%
-8/3$ & ${16}$ \\ \hline
\rule[-1mm]{0mm}{6mm} $SU\left( 6\right) $ & $2$ & $20$ & $35$ & $1/6$ & $-2$
& ${12}$ \\ \hline
\rule[-1mm]{0mm}{6mm} $USp\left( 6\right) $ & $1$ & $14$ & $21$ & $1/5$ & $%
-5/3$ & ${10}$ \\ \hline
$SU\left( 2\right) $ & $-2/3$ & $4$ & $3$ & $3/10$ & $-10/9$ & ${20/3}$ \\
\hline
\end{tabular}%
\end{center}
\caption{The parameter $q$ and the related $q$-parametrized quantites $f$ (%
\protect\ref{f-2}), $d$ (\protect\ref{d-2}), $\protect\tau $(\protect\ref%
{tau-2}), $\protect\xi $ (\protect\ref{csi-2}) and $\protect\beta $ (\protect
\ref{beta-2}). The corresponding compact form $G_{4,c}$ of $G_{4}$ is
listed. }
\label{grouptheorytable}
\end{table}

It is worth remarking that, by defining the parameter $q$ as specified in
Table 2, the values of $f$ (\ref{f}), $d$ (\ref{d}), $\tau $ (\ref{tau}), $%
\xi $ (\ref{csi}) and $\beta $ (\ref{beta}) can be easily $q$-parametrized
as follows ((\ref{d-2}) was noticed in \cite{Exc-Reds}):
\begin{eqnarray}
f &=&2\left( 3q+4\right) ;  \label{f-2} \\
d &=&\frac{3\left( 3q+4\right) \left( 2q+3\right) }{q+4};  \label{d-2} \\
\tau &=&\frac{1}{q+4};  \label{tau-2} \\
\xi &=&-\frac{\left( q+4\right) }{3};  \label{csi-2} \\
\beta &=&2\left( q+4\right) .  \label{beta-2}
\end{eqnarray}%
The specific values for the groups listed in Table 1 are reported in Table
2. Note that, speaking in terms of compact form $G_{4,c}$ of $G_{4}$, for $%
G_{4,c}=E_{7}$, $SO\left( 12\right) $, $SU\left( 6\right) $ and $USp\left(
6\right) $, $q$ can be defined as
\begin{equation}
q\equiv \text{dim}_{\mathbb{R}}\mathbb{A},  \label{def-q}
\end{equation}%
where $\mathbb{A}$ denotes the division algebra on which the corresponding
rank-$3$ simple Jordan algebra $J_{3}^{\mathbb{A}}$ is constructed ($q=8$, $%
4 $, $2$, $1$ for $\mathbb{A}=\mathbb{O}$, $\mathbb{H}$, $\mathbb{C}$, $%
\mathbb{R}$, respectively). Note that the \textit{triality symmetric}
so-called $\mathcal{N}=2$ $STU$ model \cite{stu}, based on $J_{3}=\mathbb{R}%
\oplus \mathbb{R}\oplus \mathbb{R}$, can be obtained by setting $q=0$;
however, since the corresponding $G_{4}$ is \textit{semi-simple}, it will be
considered further below.

Also, note that the dimensions $f$ and $d$ of $G_{4}$'s listed in Table 1
satisfy the relation \cite{Exc-Reds}
\begin{equation}
d=\frac{3f\left( f+1\right) }{f+16}.
\end{equation}

\subsection{\label{Simple-Deg}Simple, Degenerate}

As pointed out in Sec. 2 of \cite{Garibaldi-deg}, the story changes for
\textit{degenerate} groups of type $E_{7}$.

Confining ourselves to the ones relevant in $D=4$ supergravity with
symmetric scalar manifold, they are nothing but $G_{4}=U\left( r,s\right) $
with $r=1$ ($\mathcal{N}=2$ \textit{minimally coupled} to $s$ vector
multiplets \cite{Luciani}) or $r=3$ ($\mathcal{N}=3$ coupled to $s$ vector
multiplets \cite{N=3}), and the relevant (complex) symplectic representation
is $\mathbf{R}\left( G_{4}\right) =\mathbf{r}+\mathbf{s}$. In these cases,
it can be computed that
\begin{equation}
\mathbb{K}_{MNPQ}=\frac{\zeta ^{2}}{3}\mathbb{S}_{M(N}\mathbb{S}_{PQ)},
\label{K-deg}
\end{equation}%
where $\zeta $ is a real constant, and the rank-$2$ symmetric invariant
symplectic tensor $\mathbb{S}$ ($\mathbb{S}^{T}=\mathbb{S}$, $\mathbb{SCS}=%
\mathbb{C}$) is defined by the following formula:
\begin{equation}
\mathbf{Q}_{x}^{i}\overline{\mathbf{Q}}_{y}^{\overline{j}}\eta _{i\overline{j%
}}=\mathbb{S}_{MN}\mathcal{Q}_{x}^{M}\mathcal{Q}_{y}^{N}+i\mathbb{C}_{MN}%
\mathcal{Q}_{x}^{M}\mathcal{Q}_{y}^{N},
\end{equation}%
where $\eta _{i\overline{j}}$ is the invariant metric of the fundamental
irrep. $\mathbf{r}+\mathbf{s}$ of $U\left( r,s\right) $, and $\mathbf{Q}%
_{x}^{i}$ and $\mathbf{Q}_{x}^{i}$ are the charge vectors in the \textit{%
complex} (manifestly $U\left( r,s\right) $-covariant) symplectic frame. By
introducing
\begin{equation}
\mathcal{I}_{2}\left( x,y\right) \equiv \zeta \mathbb{S}_{MN}\mathcal{Q}%
_{x}^{M}\mathcal{Q}_{y}^{N},
\end{equation}%
it is immediate to check the \textit{degenerate} nature of the quartic
invariant $\mathbf{q}$-structure (\ref{q}):
\begin{gather}
\mathbf{q}\left( x,y,z,w\right) \equiv -6\mathbb{K}_{MNPQ}\mathcal{Q}_{x}^{M}%
\mathcal{Q}_{y}^{N}\mathcal{Q}_{z}^{P}\mathcal{Q}_{w}^{Q}  \notag \\
=-2\left[ \mathcal{I}_{2}\left( x,y\right) \mathcal{I}_{2}\left( z,w\right) +%
\mathcal{I}_{2}\left( x,z\right) \mathcal{I}_{2}\left( y,w\right) +\mathcal{I%
}_{2}\left( x,w\right) \mathcal{I}_{2}\left( y,z\right) \right] ; \\
\Downarrow  \notag \\
\mathbf{q}\left( x,x,y,y\right) =-2\left[ 2\mathcal{I}_{2}\left( x,y\right)
^{2}+\mathcal{I}_{2}\left( x,x\right) \mathcal{I}_{2}\left( y,y\right) %
\right] ;  \label{q-x-x-y-y} \\
\Downarrow  \notag \\
-\frac{1}{6}\mathbf{q}\left( x,x,x,x\right) =\mathcal{I}_{2}\left(
x,x\right) ^{2}.  \label{q-x-x-x-x}
\end{gather}

The analogue of identity (\ref{Id-2}) for such \textit{degenerate} groups of
type $E_{7}$ enjoys a very simple form ($\mathbb{C}_{MN}\mathbb{C}%
^{MN}=2\left( r+s\right) $):
\begin{gather}
\mathbb{K}_{QPNR}\mathbb{K}_{SMTU}\mathbb{C}^{RS}=\zeta ^{4}\mathbb{S}_{(QP}%
\mathbb{C}_{N)(M}\mathbb{S}_{TU)};  \label{Id-3} \\
\Downarrow  \notag \\
\mathbb{K}_{QPNR}\mathbb{K}_{SMTU}\mathbb{C}^{NM}\mathbb{C}^{RS}=\frac{\zeta
^{4}}{9}\left[ \left( 2\left( r+s\right) +4\right) \mathbb{S}_{PQ}\mathbb{S}%
_{TU}+2\mathbb{C}_{PT}\mathbb{C}_{QU}+2\mathbb{C}_{PU}\mathbb{C}_{QT}\right]
.  \label{Id-4}
\end{gather}

By exploiting (\ref{Id-4}), one can thus compute:
\begin{eqnarray}
tr\left( p\left( x\otimes x\right) p\left( y\otimes y\right) \right) &=&4
\left[ \mathbf{q}\left( x,x,y,y\right) -\left( 4\zeta ^{4}+1\right) b\left(
y,x\right) ^{2}\right]  \notag \\
&&-4\zeta ^{2}\left[ 2\left( r+s\right) +4\right] \mathcal{I}_{2}\left(
x,x\right) \mathcal{I}_{2}\left( y,y\right) ,  \label{Id-5}
\end{eqnarray}%
which can be considered the analogue of (\ref{Id-2}) for the \textit{%
degenerate} groups of type $E_{7}$ under consideration. The validity of the
postulate (\ref{FTS3}) implies $\zeta ^{2}=1/2$.

It should be remarked that, according to the discussion in Example 1.2 of
\cite{Garibaldi-deg} (and to the whole treatment therein), the invariant $%
\mathbf{q}$-structure of \textit{any} degenerate Freudenthal triple system
enjoys the form (\ref{q-x-x-x-x}), up to isomorphisms. Therefore, the
simple, \textit{degenerate} groups of type $E_{7}$ mentioned above (relevant
to $\mathcal{N}=2$ \textit{minimally coupled} and $\mathcal{N}=3$
supergravity in $D=4$; see also the treatment below) can be regarded as
\textquotedblleft prototypes\textquotedblright\ (up to isomorphisms) of
(simple) \textit{degenerate} groups of type $E_{7}$.

\subsection{\label{Semi-Simple-Non-Degenerate}Semi-Simple, Non-Degenerate}

Let us now consider \textit{semi-simple,} \textit{non-degenerate} groups of
type $E_{7}$.

Confining ourselves to the ones relevant in $D=4$ supergravity with
symmetric scalar manifold, they are nothing but $G_{4}=SL\left( 2,\mathbb{R}%
\right) \times SO\left( m,n\right) $ with $m=2$ ($\mathcal{N}=2$ coupled to $%
n+1$ vector multiplets) or $m=6$ ($\mathcal{N}=4$ coupled to $n$ vector
multiplets), and the relevant symplectic representation is the
bi-fundamental $\mathbf{R}\left( G_{4}\right) =\left( \mathbf{2},\mathbf{m}+%
\mathbf{n}\right) $. They are respectively related to \textit{semi-simple}
rank-$3$ Jordan algebras $\mathbb{R}\oplus \mathbf{\Gamma }_{m-1,n-1}$,
where $\mathbf{\Gamma }_{m-1,n-1}$ is a Jordan algebra with a quadratic form
of pseudo-Euclidean $\left( m-1,n-1\right) $ signature, \textit{i.e.} the
Clifford algebra of $O(m-1,n-1)$ \cite{Jordan}.The aforementioned $\mathcal{N%
}=2$ $STU$ model \cite{stu}, based on $J_{3}=\mathbb{R}\oplus \mathbf{\Gamma
}_{1,1}\sim \mathbb{R}\oplus \mathbb{R}\oplus \mathbb{R}$, is recovered by
setting $m=n=2$.

In these cases, electro-magnetic splitting of the symplectic representation $%
\mathbf{R}$ can be implemented in a \textit{manifestly }$G_{4}$-\textit{%
covariant} fashion. Namely, $\mathcal{Q}$ is an electro-magnetic doublet $%
\mathbf{2}$ of the $SL\left( 2,\mathbb{R}\right) $ factor of $G_{4}$ itself.
The symplectic index $M$ thus splits as follows (\textit{cfr.} Eq. (3.7) of
\cite{FMOSY-1})
\begin{equation}
\left.
\begin{array}{l}
M=\alpha \Lambda , \\
\alpha =1,2,~\Lambda =1,...,m+n-2.%
\end{array}%
\right\} \Rightarrow \mathcal{Q}^{M}\equiv \mathcal{Q}_{\alpha }^{\Lambda },
\label{CV-split}
\end{equation}%
and it should be pointed out that in the $\mathcal{N}=2$ case usually $%
\Lambda =0,1,...,n-1$, with \textquotedblleft $0$\textquotedblright\
pertaining to the $D=4$ graviphoton vector. The manifestly $G_{4}$-covariant
symplectic frame (\ref{CV-split}) is usually dubbed \textit{Calabi-Vesentini
}frame \cite{CV}, and it was firstly introduced in supergravity in \cite%
{CDFVP}.

The symplectic metric $\mathbb{C}_{MN}=\mathbb{C}_{\Lambda \Sigma }^{\alpha
\beta }$ and rank-$4$ completely symmetric $\mathbb{K}$-tensor $\mathbb{K}%
_{MNPQ}=\mathbb{K}_{\Lambda \Sigma \Xi \Omega }^{\alpha \beta \gamma \delta
} $ enjoy the following expression in term of the invariant structures $%
\epsilon ^{\alpha \beta }$ and $\eta _{\Lambda \Xi }$ of $SL_{v}\left( 2,%
\mathbb{R}\right) $ and of $SO\left( m,n-2\right) $, respectively \cite%
{FMY-T-CV}:
\begin{eqnarray}
\mathbb{C}_{\Lambda \Sigma }^{\alpha \beta } &=&\eta _{\Lambda \Sigma
}\epsilon ^{\alpha \beta };  \label{C-T} \\
\mathbb{K}_{\Lambda \Sigma \Xi \Omega }^{\alpha \beta \gamma \delta } &=&%
\frac{1}{12}\left[ \left( \epsilon ^{\alpha \beta }\epsilon ^{\gamma \delta
}+\epsilon ^{\alpha \delta }\epsilon ^{\beta \gamma }\right) \eta _{\Lambda
\Xi }\eta _{\Sigma \Omega }+\left( \epsilon ^{\alpha \beta }\epsilon
^{\delta \gamma }+\epsilon ^{\alpha \gamma }\epsilon ^{\delta \beta }\right)
\eta _{\Lambda \Omega }\eta _{\Sigma \Xi }+\left( \epsilon ^{\alpha \gamma
}\epsilon ^{\beta \delta }+\epsilon ^{\alpha \delta }\epsilon ^{\beta \gamma
}\right) \eta _{\Lambda \Sigma }\eta _{\Xi \Omega }\right] .  \notag \\
&&  \label{K-T}
\end{eqnarray}%
From this, one can compute the analogue of identities (\ref{Id-1}) and (\ref%
{Id-4}) for the \textit{semi-simple}, non-degenerate groups of type $E_{7}$
under consideration ($\epsilon _{\alpha \beta }\epsilon ^{\alpha \beta }=2$,
$\eta _{\Lambda \Sigma }\eta ^{\Lambda \Sigma }=m+n$):%
\begin{eqnarray}
\mathbb{K}_{MNPQ}\mathbb{K}_{RSTU}\mathbb{C}^{PT}\mathbb{C}^{QU} &=&\mathbb{K%
}_{\Lambda \Sigma \Xi \Omega }^{\alpha \beta \gamma \delta }\mathbb{K}%
_{\Delta \Theta \Phi \Psi }^{\eta \xi \lambda \rho }\mathbb{C}_{\delta \eta
}^{\Omega \Delta }\mathbb{C}_{\gamma \xi }^{\Xi \Theta }  \notag \\
&=&\frac{1}{6}\mathbb{K}_{\Lambda \Sigma \Phi \Psi }^{\alpha \beta \lambda
\rho }-\frac{1}{36}\mathbb{C}_{\Lambda (\Phi }^{\alpha \lambda }\mathbb{C}%
_{\Psi )\Sigma }^{\rho \beta }  \notag \\
&&-\frac{1}{72}\left[
\begin{array}{l}
\eta _{\Lambda \Psi }\eta _{\Sigma \Phi }\left( \epsilon ^{\alpha \beta
}\epsilon ^{\lambda \rho }+\epsilon ^{\alpha \lambda }\epsilon ^{\rho \beta
}\right) \\
+\eta _{\Lambda \Phi }\eta _{\Sigma \Psi }\left( \epsilon ^{\alpha \beta
}\epsilon ^{\rho \lambda }+2\epsilon ^{\alpha \lambda }\epsilon ^{\beta \rho
}+\epsilon ^{\alpha \rho }\epsilon ^{\beta \lambda }\right) \\
+\left( m+n-1\right) \eta _{\Lambda \Sigma }\eta _{\Phi \Psi }\left(
\epsilon ^{\alpha \rho }\epsilon ^{\lambda \beta }+\epsilon ^{\alpha \lambda
}\epsilon ^{\rho \beta }\right)%
\end{array}%
\right] .  \label{Id-CV-1}
\end{eqnarray}%
By exploiting (\ref{Id-CV-1}), one can thus compute:
\begin{eqnarray}
tr\left( p\left( x\otimes x\right) p\left( y\otimes y\right) \right) &=&5%
\mathbf{q}\left( x,x,y,y\right) -4b\left( y,x\right) ^{2}  \notag \\
&&-\left[ \epsilon ^{\alpha \beta }\epsilon ^{\rho \lambda }\eta _{\Lambda
\Psi }\eta _{\Sigma \Phi }+\left( m+n-2\right) \epsilon ^{\alpha \rho
}\epsilon ^{\beta \lambda }\eta _{\Lambda \Sigma }\eta _{\Phi \Psi }\right]
\mathcal{Q}_{\alpha \mid x}^{\Lambda }\mathcal{Q}_{\beta \mid x}^{\Sigma }%
\mathcal{Q}_{\lambda \mid y}^{\Phi }\mathcal{Q}_{\rho \mid y}^{\Psi },
\notag \\
&&  \label{Id-CV-2}
\end{eqnarray}%
where we recall that the quartic invariant form $\mathbf{q}$ is defined by (%
\ref{q}). The identity (\ref{Id-CV-2}) can be considered the analogue of (%
\ref{Id-2}) and (\ref{Id-5}) for the \textit{semi-simple}, \textit{%
non-degenerate} groups of type $E_{7}$ under consideration, and it is
different from them both.

\subsection{The Unified Limit}

The different structure exhibited by the scalar identities (\ref{Id-2})
(holding for simple, \textit{non-degenerate} groups of type $E_{7}$), (\ref%
{Id-5}) (holding for simple, \textit{degenerate} groups of type $E_{7}$) and
(\ref{Id-CV-2}) (holding for semi-simple, \textit{non-degenerate} groups of
type $E_{7}$) is manifest : the structure of (\ref{Id-2}) is the same as the
structure of the first line of (\ref{Id-5}) and of (\ref{Id-CV-2}), but the
second line of (\ref{Id-5}) and of(\ref{Id-CV-2}) is not compatible with
such a structure.

Therefore, along the lines of \cite{Garibaldi-deg}, the scalar identities (%
\ref{Id-2}), (\ref{Id-5}) and (\ref{Id-CV-2}) (or the corresponding tensor
identities) can be considered as defining identities for \textit{simple}
\textit{non-degenerate}, \textit{simple} \textit{degenerate}, and \textit{%
semi-simple} \textit{non-degenerate} groups of type $E_{7}$, respectively.

However, it should be also noted that (\ref{Id-2}), (\ref{Id-5}) and (\ref%
{Id-CV-2}) share the very same $x\equiv y$ limit:
\begin{equation}
tr\left( p\left( x\otimes x\right) p\left( x\otimes x\right) \right) =\beta
\mathbf{q}\left( x,x,x,x\right) ,  \label{x=y-limit}
\end{equation}%
\textit{modulo} the renamings
\begin{equation}
\beta \equiv 4\left[ 1+\frac{\zeta ^{2}}{3}\left( r+s+2\right) \right]
\equiv \left[ 5+\frac{1}{3}\left( m+n\right) \right] .
\end{equation}%
\bigskip

Before proceeding to the analysis of various truncation patterns to \textit{%
minimal coupling} models, it is worth stressing a peculiar feature of the $%
\mathcal{N}=2$ theory among $D=4$ extended supergravity theories.

$\mathcal{N}=2$ supergravity is the unique extended supergravity which
admits two different types of matter multiplets, namely vector and hyper
multiplets. Thus, out of the three classes (simple non-degenerate, simple
degenerate and semi-simple non-degenerate, respectively treated in Subsecs. %
\ref{Simple-Non-Deg}, \ref{Simple-Deg} and \ref{Semi-Simple-Non-Degenerate})
of groups $G_{4}$ of type $E_{7}$ treated above, one can always construct a
semi-simple group of type $E_{7}$ with the following structure:%
\begin{eqnarray}
&&G_{4}\times \mathcal{G}_{4};  \notag \\
&&\left( \mathbf{R}\left( G_{4}\right) ,~\mathcal{R}\left( \mathcal{G}%
_{4}\right) =\mathbf{1}\right) .  \label{semi-simple-trivial-type}
\end{eqnarray}%
As pointed out above, $G_{4}$ is the $U$-duality group of the $\mathcal{N}=2$
theory (which is also the global isometry group of the special K\"{a}hler
vector multiplets' scalar manifold), whereas $\mathcal{G}_{4}$ is the global
isometry group of the quaternonic K\"{a}hler hypermultiplets' scalar
manifold. The various truncations analyzed in subsequent Sections provide a
number of examples of truncations of simple non-degenerate groups of type $%
E_{7}$ down to $\mathcal{N}=2$ semi-simple degenerate (see \textit{e.g.}
Sec. \ref{4--->2+hypers}) or semi-simple non-degenerate (see \textit{e.g.}
the models with $n_{V},n_{H}\neq 0$ in Table \ 5 below) groups of type $%
E_{7} $ of type (\ref{semi-simple-trivial-type}).

\section{\label{N>4}Maximal Truncations from $\mathcal{N}=8$ ($J_{3}^{%
\mathbb{O}_{s}}$)}

One can perform the kinematical reduction of $\mathcal{N}$-extended
supergravity multiplets down to $\mathcal{N}^{\prime }<\mathcal{N}$
multiplets (massless multiplets in $\mathcal{N}$-extended $D=4$ supergravity
are reported in Tables 3 and 4). The reduction is subjected to the following
dynamical conditions : the inclusion of $U$-duality groups: $G_{4}\supset
G_{4}^{\prime }$, as well as of the stabilizers of the scalar manifold: $%
H_{4}\supset H_{4}^{\prime }$, such that the scalar manifold of the
truncated theory is a proper sub-manifold of the scalar manifold of the
starting theory: $G_{4}/H_{4}\supset G_{4}^{\prime }/H_{4}^{\prime }$. At
the level of electric and magnetic fluxes, the branching $\mathbf{R}(G_{4})=%
\mathbf{R}^{\prime }(G_{4}^{\prime })+\mathbf{...}$ has to hold, where $%
\mathbf{R}^{\prime }(G_{4}^{\prime })$ is the relevant symplectic
representation of $G_{4}^{\prime }$ itself.

If $\mathcal{N}$, $\mathcal{N}^{\prime }\geqslant 4$, the kinematical
multiplet truncation actually coincides with the dynamical truncation,
because there is a unique choice of matter multiplets in these cases. On the
other hand, already for $\mathcal{N}^{\prime }=3$ this is no longer true for
$\mathcal{N}\geqslant 6$, and for $\mathcal{N}=8\rightarrow \mathcal{N}%
^{\prime }=2$ many possibilities exist; the \textit{maximal} truncations (in
the sense of $G_{4}\supset G_{4}^{\prime }$ specified above) are listed in
Table 5. Kinematical truncations $\mathcal{N}=6\rightarrow 5$, $\mathcal{N}%
=6\rightarrow 4$ and $\mathcal{N}=5\rightarrow 4$ actually coincide with the
corresponding dynamical reduction. The two latter cases yield $2$ and no
matter multiplets, respectively. Further truncation of these theories down
to $\mathcal{N}=1$ reduces to some of the general examples we consider
further below.

\begin{table}[h]
\begin{center}
\begin{tabular}{|l|l|l|}
\hline
$\mathcal{N}$ & massless $\lambda _{MAX}=2$ multiplet & massless $\lambda
_{MAX}=3/2$ multiplet \\ \hline
&  &  \\
$8$ & $\bigl[(2),8(\frac{3}{2}),28(1),56(\frac{1}{2}),70(0)\bigr]$ & none \\
&  &  \\
$6$ & $\bigl[(2),6(\frac{3}{2}),16(1),26(\frac{1}{2}),30(0)\bigr]$ & $\bigl[(%
\frac{3}{2}),6(1),15(\frac{1}{2}),20(0)\bigr]$ \\
&  &  \\
$5$ & $\bigl[(2),5(\frac{3}{2}),10(1),11(\frac{1}{2}),10(0)\bigr]$ & $\bigl[(%
\frac{3}{2}),6(1),15(\frac{1}{2}),20(0)\bigr]$ \\
&  &  \\
$4$ & $\bigl[(2),4(\frac{3}{2}),6(1),4(\frac{1}{2}),2(0)\bigr]$ & $\bigl[(%
\frac{3}{2}),4(1),7(\frac{1}{2}),8(0)\bigr]$ \\
&  &  \\
$3$ & $\bigl[(2),3(\frac{3}{2}),3(1),(\frac{1}{2})\bigr]$ & $\bigl[(\frac{3}{%
2}),3(1),3(\frac{1}{2}),2(0)\bigr]$ \\
&  &  \\
$2$ & $\bigl[(2),2(\frac{3}{2}),(1)\bigr]$ & $\bigl[(\frac{3}{2}),2(1),(%
\frac{1}{2})\bigr]$ \\
&  &  \\
$1$ & $\bigl[(2),(\frac{3}{2})\bigr]$ & $\bigl[(\frac{3}{2}),(1)\bigr]$ \\
&  &  \\ \hline
\end{tabular}%
\end{center}
\caption{Massless multiplets with maximal helicity $\protect\lambda _{MAX}=2$%
, $3/2$ \protect\cite{ADFL-Higgs}.}
\label{hel23/2}
\end{table}

\begin{table}[h]
\begin{center}
\begin{tabular}{|l|l|l|}
\hline
$\mathcal{N}$ & massless $\lambda _{MAX}=1$ multiplet & massless $\lambda
_{MAX}=1/2$ multiplet \\ \hline
&  &  \\
$8$,$6$,$5$ & none & none \\
&  &  \\
$4$ & $\bigl[(1),4(\frac{1}{2}),6(0)\bigr]$ & none \\
&  &  \\
$3$ & $\bigl[(1),4(\frac{1}{2}),6(0)\bigr]$ & none \\
&  &  \\
$2$ & $\bigl[(1),2(\frac{1}{2}),2(0)\bigr]$ & $\bigl[2(\frac{1}{2}),4(0)%
\bigr]$ \\
&  &  \\
$1$ & $\bigl[(1),(\frac{1}{2})\bigr]$ & $\bigl[(\frac{1}{2}),2(0)\bigr]$ \\
&  &  \\ \hline
\end{tabular}%
\end{center}
\caption{Massless multiplets with maximal helicity $\protect\lambda _{MAX}=1$%
, $1/2$ \protect\cite{ADFL-Higgs}.}
\label{hel11/2}
\end{table}
\medskip

Before proceeding with the analysis of the various truncations $\mathcal{N}%
=8\rightarrow \mathcal{N}^{\prime }<8$, we would like here to add a brief
discussion of the general consistency conditions yielded by the
supersymmetry transformations of the $\mathcal{N}=8$ fermionic fields, along
the lines of the treatment in Sec. 5 of \cite{ADF-1} (a similar discussion
related to the Attractor Mechanism has been given also in Sec. 5 of \cite%
{FK-Univ}). Neglecting three fermion terms, the transformations of the
gravitinos $\psi _{A}$ and of spin $\frac{1}{2}$ fermions $\chi _{ABC}$ read
as follows ($A=1,...,8$; \textit{cfr.} \textit{e.g.} (5.2)-(5.3) of \cite%
{ADF-1}):%
\begin{eqnarray}
\delta \psi _{A\mu } &=&\nabla _{\mu }\epsilon _{A}+T_{AB\mid \nu \rho
}^{-}\gamma _{\mu }^{~\nu }\gamma ^{\rho }\epsilon ^{B};  \label{var-SUSY-1}
\\
\delta \chi _{ABC} &=&P_{ABCD,\alpha }\partial _{\mu }\phi ^{\alpha }\gamma
^{\mu }\epsilon ^{D}+T_{[AB\mid \mu \nu }^{-}\gamma ^{\mu \nu }\epsilon
_{\left\vert C\right] },  \label{var-SUSY-2}
\end{eqnarray}%
where $\nabla _{\mu }\epsilon _{A}\equiv \mathcal{D}_{\mu }\epsilon
_{A}+\omega _{A}^{~B}\epsilon _{B}$, $T_{AB}^{-}$ is the (dressed)
graviphotonic field strengths' 2-form, $P_{ABCD}$ is the \textit{Vielbein}
1-form, and $\phi ^{\alpha }$ are the $70$ real scalars of the rank-$7$
symmetric $\mathcal{N}=8$ scalar manifold $E_{7(7)}/SU(8)/\mathbb{Z}_{2}$.

When considering a truncation $\mathcal{N}=8\longrightarrow \mathcal{N}%
^{\prime }<8$, it holds that%
\begin{eqnarray}
SU(8) &\supset &SU(\mathcal{N}^{\prime })\times SU(8-\mathcal{N}^{\prime
})\times U(1); \\
\mathbf{8} &=&\left( \mathcal{N}^{\prime },\mathbf{1}\right) _{\mathcal{N}%
^{\prime }-8}+\left( \mathbf{1},\mathbf{8-}\mathcal{N}^{\prime }\right) _{%
\mathcal{N}^{\prime }}.
\end{eqnarray}%
Correspondingly, the supersymmetry parameters, the gravitinos and the spin $%
1/2$ fermions branch as ($a=1,...,\mathcal{N}^{\prime }$, $i=1,...,8-%
\mathcal{N}^{\prime }$):%
\begin{eqnarray}
\epsilon _{A} &=&\epsilon _{a},~~\epsilon _{i}; \\
\psi _{A} &=&\psi _{a},~~\psi _{i}; \\
\chi _{ABC} &=&\chi _{abc},~\chi _{abi},~\chi _{aij},~\chi _{ijk}.
\end{eqnarray}%
The conditions of consistent truncation read%
\begin{equation}
\left\{
\begin{array}{l}
\epsilon _{i}=0; \\
\psi _{i}=0; \\
\chi _{abi}=\chi _{aij}=\chi _{ijk}=0,%
\end{array}%
\right. \text{such~that~}\left\{
\begin{array}{l}
\delta \psi _{i}=0; \\
\\
\delta \chi _{abi}=\delta \chi _{aij}=\delta \chi _{ijk}=0,%
\end{array}%
\right.  \label{conds-SUSY}
\end{equation}%
with the exception of the case $\mathcal{N}^{\prime }=6$ (discussed in
Subsec. \ref{N=8--->N=6}), for which $\chi _{aij}$, as well as its
corresponding supersymmetry variation, does not vanish.

By exploiting (\ref{conds-SUSY}) and (\ref{var-SUSY-1})-(\ref{var-SUSY-2}),
one obtains the following general consistency conditions:%
\begin{equation}
\left\{
\begin{array}{l}
i):\omega _{i}^{~a}=0; \\
\\
ii):T_{ai}^{-}=0; \\
\\
iii):P_{abci}=P_{aijk}=0; \\
\\
iv):P_{abij}=0; \\
\\
v):T_{ij}^{-}=0,%
\end{array}%
\right.   \label{cconds-SUSY}
\end{equation}%
where conditions $iv)$ and $v)$ do not hold for $\mathcal{N}\prime =6$.
Condition $i)$ (on the spin connection $\omega $) confirms the consystency
condition on the reduction of the holonomy ($\mathcal{R}$-symmetry) group,
as discussed in Secs. 3 and 4 of \cite{ADF-1}. Conditions $iii)$ and $iv)$
(on the \textit{Vielbein }$P$) confirm the consistency conditions from the
embedding of the scalar manifold of the $\mathcal{N}^{\prime }$-extended
supergravity sub-theory into the scalar manifold of $\mathcal{N}=8$ theory,
as discussed in Sec. 4 of \cite{ADF-1}. Furthermore, conditions $ii)$ and $v)
$ (on the graviphotonic field strengths $T$), which are the (necessary but
noot necessary) conditions which we discuss in the present investigation,
are needed in order for $T_{ab}^{-}$ to consistently parametrize the
(dressed) graviphotons which survive the truncation under consideration (in
the case $\mathcal{N}\prime =6$, one should consider also $T_{ij}^{-}$
non-vanishing).

\subsection{\label{N=8--->N=6}$\rightarrow \mathcal{N}=6$}

In this Subsection, we discuss, at the level of the consistency conditions
yielded by supersymmetry, the case $\mathcal{N}=8\longrightarrow 6$ (for the
\textquotedblleft \textit{twin\textquotedblright } case $\mathcal{N}=2$ $%
\left( n_{V},n_{H}\right) =\left( 15,0\right) $, and further decomposition,
see point 1 of Subsubsec. \ref{Further}):
\begin{eqnarray}
J_{3}^{\mathbb{O}_{s}} &:&\mathcal{N}=8\longrightarrow J_{3}^{\mathbb{H}}:%
\mathcal{N}=6  \notag \\
&&  \notag \\
E_{7\left( 7\right) } &\supset &SO^{\ast }\left( 12\right) \times SU\left(
2\right) \supset SU(6)\times SU(2)\times U(1):  \label{ca-1} \\
&&\left\{
\begin{array}{l}
\mathbf{56}=\left( \mathbf{32},\mathbf{1}\right) +\left( \mathbf{12},\mathbf{%
2}\right)  \\
=\left( \mathbf{1},\mathbf{1}\right) _{3}+\left( \mathbf{1},\mathbf{1}%
\right) _{-3}+\left( \mathbf{15},\mathbf{1}\right) _{-1}+\left( \overline{%
\mathbf{15}},\mathbf{1}\right) _{1}+\left( \mathbf{6},\mathbf{2}\right)
_{1}+\left( \overline{\mathbf{6}},\mathbf{2}\right) _{-1};%
\end{array}%
\right.  \\
&&  \notag \\
SU\left( 8\right)  &\supset &SU(6)\times SU(2)\times U(1):\left\{
\begin{array}{l}
\mathbf{8}=\left( \mathbf{6},\mathbf{1}\right) _{-2}+\left( \mathbf{1},%
\mathbf{2}\right) _{6}; \\
\mathbf{28}=\left( \mathbf{15},\mathbf{1}\right) _{-4}+\left( \mathbf{6},%
\mathbf{2}\right) _{4}+\left( \mathbf{1},\mathbf{1}\right) _{12};%
\end{array}%
\right.   \label{ca-2} \\
&&  \notag \\
\frac{E_{7\left( 7\right) }}{SU\left( 8\right) } &\supset &\frac{SO^{\ast
}\left( 12\right) }{SU\left( 6\right) \times U\left( 1\right) }\supset \frac{%
SL\left( 2,\mathbb{R}\right) }{U\left( 1\right) }\times \frac{SO\left(
6,2\right) }{SO\left( 6\right) \times SO\left( 2\right) }.
\end{eqnarray}%
In particular, the decomposition of the $\mathbf{28}$ of $SU\left( 8\right) $
yields the following branching of the $\mathcal{N}=8$ dressed graviphotonic
field strengths ($a=1,...,6$, $i=1,2$):%
\begin{equation}
\underset{\mathbf{28}}{T_{AB}^{-}}=\underset{\left( \mathbf{15},\mathbf{1}%
\right) }{T_{ab}^{-}},~\underset{\left( \mathbf{6},\mathbf{2}\right) }{%
T_{ai}^{-}},~\underset{\left( \mathbf{1},\mathbf{1}\right) }{T_{ij}^{-}}.
\label{grav-dec}
\end{equation}%
The $SU\left( 2\right) $ commuting factor in (\ref{ca-1}) and (\ref{ca-2})
is the $\mathcal{R}$-symmetry truncated away in the supersymmetry reduction $%
\mathcal{N}=8\rightarrow 6$ (a further truncation $\mathcal{N}=6\rightarrow
\mathcal{N}=3$ is considered in Subsec. \ref{6--->3}). The truncation
condition on the two-form Abelian field strengths' fluxes reads
\begin{equation}
\left( \mathbf{12},\mathbf{2}\right) =\left( \mathbf{6},\mathbf{2}\right)
+\left( \overline{\mathbf{6}},\mathbf{2}\right) =0\Leftrightarrow
T_{ai}^{-}=0.  \label{conss}
\end{equation}%
We anticipate that this truncation condition on fluxes is complementary to
the condition (\ref{OB-1}) considered in\ Subsec. \ref{--->4-nV=6}; indeed,
the embedding (\ref{OBB-1})-(\ref{OBB-4}) is a different non-compact, real
form of the embedding (\ref{ca-2}).

As a consequence of the decomposition (\ref{ca-2}) of the $\mathbf{8}$ of $%
SU(8)$, the supersymmetry parameters, the gravitinos and the spin $1/2$
fermions respectively branch as:%
\begin{eqnarray}
\underset{\mathbf{8}}{\epsilon _{A}} &=&\underset{\left( \mathbf{6},\mathbf{1%
}\right) }{\epsilon _{a}},~~\underset{\left( \mathbf{1},\mathbf{2}\right) }{%
\epsilon _{i}}; \\
\underset{\mathbf{8}}{\psi _{A}} &=&\underset{\left( \mathbf{6},\mathbf{1}%
\right) }{\psi _{a}},~~\underset{\left( \mathbf{1},\mathbf{2}\right) }{\psi
_{i}}; \\
\underset{\mathbf{56}}{\chi _{ABC}} &=&\underset{\mathbf{20}}{\chi _{abc}},~%
\underset{\left( \mathbf{15},\mathbf{2}\right) }{\chi _{abi}},~\underset{%
\left( \mathbf{6},\mathbf{1}\right) }{\chi _{aij}}.
\end{eqnarray}%
Correspondingly, the conditions of consistent truncation read%
\begin{equation}
\left\{
\begin{array}{l}
\epsilon _{i}=0; \\
\psi _{i}=0; \\
\chi _{abi}=0,%
\end{array}%
\right. \text{such~that~}\left\{
\begin{array}{l}
\delta \psi _{i}=0; \\
\\
\delta \chi _{abi}=0.%
\end{array}%
\right.  \label{conds-SUSY-N=6}
\end{equation}%
By evaluating the supersymmetry transformations (\ref{var-SUSY-1})-(\ref%
{var-SUSY-2}) on the truncation conditions (\ref{conds-SUSY-N=6}), one
obtains%
\begin{eqnarray}
\delta \psi _{a\mu } &=&\nabla _{\mu }\epsilon _{a}+T_{ab\mid \nu \rho
}^{-}\gamma _{\mu }^{~\nu }\gamma ^{\rho }\epsilon ^{b};  \label{var-SUSY-3}
\\
\delta \chi _{abc} &=&P_{abcd,\alpha }\partial _{\mu }\phi ^{\alpha }\gamma
^{\mu }\epsilon ^{d}+\frac{1}{3}\left( T_{ab\mid \mu \nu }^{-}\gamma ^{\mu
\nu }\epsilon _{c}+T_{ca\mid \mu \nu }^{-}\gamma ^{\mu \nu }\epsilon
_{b}+T_{bc\mid \mu \nu }^{-}\gamma ^{\mu \nu }\epsilon _{a}\right) ;
\label{var-SUSY-4} \\
\delta \chi _{aij} &=&P_{abij,\alpha }\partial _{\mu }\phi ^{\alpha }\gamma
^{\mu }\epsilon ^{b}+\frac{1}{3}T_{ij\mid \mu \nu }^{-}\gamma ^{\mu \nu
}\epsilon _{a}  \label{var-SUSY-5}
\end{eqnarray}%
in the untruncated sector, and%
\begin{eqnarray}
\delta \psi _{i\mu } &=&\omega _{i\mid \mu }^{~a}\epsilon _{a}+T_{ia\mid \nu
\rho }^{-}\gamma _{\mu }^{~\nu }\gamma ^{\rho }\epsilon ^{a};
\label{var-SUSY-6} \\
\delta \chi _{abi} &=&-P_{abci,\alpha }\partial _{\mu }\phi ^{\alpha }\gamma
^{\mu }\epsilon ^{c}+\frac{2}{3}T_{i\left[ a\right\vert \mid \mu \nu
}^{-}\gamma ^{\mu \nu }\epsilon _{\left\vert b\right] }  \label{var-SUSY-7}
\end{eqnarray}%
in the truncated sector.

By then imposing (\ref{conds-SUSY-N=6}) on (\ref{var-SUSY-6}) and (\ref%
{var-SUSY-7}), one obtains%
\begin{equation}
\omega _{i}^{~a}=0=T_{ia}^{-}=P_{abci}.  \label{conss-2}
\end{equation}%
In particular, $T_{ia}^{-}=0$ is nothing but the condition (\ref{conss}).

Thus, one can conclude that the truncation (\ref{conss-2}) is fully
consistent.\medskip

A similar analysis at the level of supersymmetry can be performed in all
cases. We observe that whenever the truncation $\mathcal{N}=8\longrightarrow
\mathcal{N}^{\prime }<8$ is consistent with supersymmetry, and thus it
actually exists, there occurs an $SU(8-\mathcal{N}^{\prime })$ factor
commuting with the the $\mathcal{R}$-symmetry $U(\mathcal{N}\prime )$ of the
truncated sub-theory inside the $\mathcal{N}=8$ $\mathcal{R}$-symmetry $SU(8)
$.

\subsection{\label{8--->5--->3}$\rightarrow \mathcal{N}=5\rightarrow
\mathcal{N}=3,2$}

Next, we consider the maximal \textit{non-symmetric} embedding:
\begin{eqnarray}
J_{3}^{\mathbb{O}_{s}} &:&\mathcal{N}=8\longrightarrow M_{1,2}\left( \mathbb{%
O}\right) :\mathcal{N}=5; \\
E_{7\left( 7\right) } &\supset &SU\left( 1,5\right) \times SU\left( 3\right)
; \\
\mathbf{56} &=&\left( \mathbf{6,3}\right) +\left( \overline{\mathbf{6}}%
\mathbf{,}\overline{\mathbf{3}}\right) +\left( \mathbf{20},\mathbf{1}\right)
; \\
\frac{E_{7\left( 7\right) }}{SU\left( 8\right) } &\supset &\frac{SU\left(
1,5\right) }{U\left( 5\right) }.
\end{eqnarray}
$M_{1,2}\left( \mathbb{O}\right) $ is the Jordan triple system (not
upliftable to $D=5$) generated by $2\times 1$ matrices over $\mathbb{O}$
\cite{GST}. The $\mathbf{20}$ is the rank-$3$ antisymmetric self-real irrep.
of $SU\left( 1,5\right) $. The commuting $SU\left( 3\right) $ factor can be
interpreted as the part of the $\mathcal{R}$-symmetry truncated away in the
supersymmetry reduction $\mathcal{N}=8\rightarrow \mathcal{N}=5$. On the
two-form Abelian field strengths' fluxes, the truncation condition reads
\begin{equation}
\left( \mathbf{6,3}\right) =0.
\end{equation}
As discussed in Sec. 8 of \cite{Gnecchi-1}, the \textit{quartic} invariant
of the $\mathbf{R}=\mathbf{20}$ of $SU\left( 1,5\right) $, after
skew-diagonalization in the scalar-dressed $\mathcal{R}$-symmetry $U\left(
5\right) $-basis and use of the Hua-Bloch-Messiah-Zumino Theorem \cite%
{Hua-BMZ-Th}, is a \textit{perfect square}. On this respect the couples $%
\left( SU\left( 1,5\right) ,\text{~}\mathbf{R}=\mathbf{20}\right) $ and $%
\left( SL\left( 2,\mathbb{R}\right) \times SO\left( 6\right) ,~\mathbf{R}%
=\left( \mathbf{2},\mathbf{6}\right) \right) $ (this latter pertaining to $%
\mathcal{N}=4$ \textit{``pure''} supergravity) stand on a particular footing
among simple and respectively semisimple groups \textit{``of type }$E_{7}$%
\textit{'' }\cite{brown}. Thus, this embedding does not concern a proper
\textit{``degeneration''} of a group of type $E_{7}$, but it is however
noteworthy.

In turn, the \textit{``pure''} $\mathcal{N}=5$ theory admits two maximal
\textit{``degenerative''} truncations, which precisely match the kinematical
decomposition of the $\mathcal{N}=5$ gravity multiplet into matter $\mathcal{%
N}=2$ multiplets.

\begin{enumerate}
\item The first reads:
\begin{eqnarray}
\mathcal{N} &=&5\longrightarrow \mathcal{N}=3,\text{~}n_{V}=1\overset{\text{%
\textit{\textquotedblleft twin\textquotedblright }}}{\Leftrightarrow }%
\mathcal{N}=2\text{~}\mathbb{CP}^{3}; \\
SU\left( 1,5\right) &\supset &SU\left( 1,3\right) \times SU\left( 2\right)
\times U\left( 1\right) ; \\
\mathbf{20} &=&\left( \mathbf{4,1}\right) _{+3}+\left( \overline{\mathbf{4}}%
\mathbf{,1}\right) _{-3}+\left( \mathbf{6},\mathbf{2}\right) _{0}; \\
\frac{SU\left( 1,5\right) }{U\left( 5\right) } &\supset &\frac{SU\left(
1,3\right) }{U\left( 3\right) },
\end{eqnarray}%
and it admits two possible interpretations, due to the fact that $\mathcal{N}%
=3$~supergravity coupled to $1$ vector multiplet and $\mathcal{N}=2$
supergravity \textit{minimally coupled} to $3$ vector multiplets share the
very same bosonic sector (namely, they are \textit{\textquotedblleft
twin\textquotedblright } theories; see the discussion in\ Sec. 9 of \cite%
{Gnecchi-1}). In the $\mathcal{N}=3$ interpretation, one gets a theory with $%
1$ vector multiplets, and the $SU(2)$ commuting factor can be interpreted as
the part of the $\mathcal{R}$-symmetry truncated away in the supersymmetry
reduction $\mathcal{N}=5\rightarrow \mathcal{N}=3$. On the other hand, in
the $\mathcal{N}=2$ interpretation, one gets a theory with $3$ \textit{%
minimally coupled} vector multiplets without hypermultiplets, and the $SU(2)$
commuting factor is the global $\mathcal{N}=2$ hyper $\mathcal{R}$-symmetry.
In both cases, on the two-form Abelian field strengths' fluxes the
truncation condition reads
\begin{equation}
\left( \mathbf{6},\mathbf{2}\right) _{0}=0.  \label{trunc-1}
\end{equation}%
One can also prove that the \textit{quartic} invariant of the $\mathbf{R}=%
\mathbf{20}$ of $SU\left( 1,5\right) $, under the truncation (\ref{trunc-1})
becomes the \textit{square} of the \textit{quadratic} invariant of the $%
\mathbf{R}=\mathbf{4}$ of $SU\left( 1,3\right) $.

\item The second maximal \textit{\textquotedblleft
degenerative\textquotedblright } truncation of the \textit{\textquotedblleft
pure\textquotedblright } $\mathcal{N}=5$ theory reads
\begin{eqnarray}
\mathcal{N} &=&5\longrightarrow \mathcal{N}=2,\text{~}n_{V}=0,~n_{H}=1;
\label{OBBB-1} \\
SU\left( 1,5\right) &\supset &SU\left( 1,2\right) \times SU\left( 3\right)
\times U\left( 1\right) ; \\
\mathbf{20} &=&\left( \mathbf{1,1}\right) _{+3}+\left( \mathbf{1,1}\right)
_{-3}+\left( \mathbf{3,}\overline{\mathbf{3}}\right) _{-1}+\left( \overline{%
\mathbf{3}}\mathbf{,3}\right) _{+1};  \label{OBBB-3} \\
\frac{SU\left( 1,5\right) }{U\left( 5\right) } &\supset &\frac{SU\left(
1,2\right) }{U\left( 2\right) }.
\end{eqnarray}%
The $\mathcal{N}=2$ theory is coupled to the \textit{universal}
hypermultiplet, in absence of vector multiplets. The $SU(3)$ commuting
factor can be interpreted as the part of the $\mathcal{R}$-symmetry
truncated away in the supersymmetry reduction $\mathcal{N}=5\rightarrow
\mathcal{N}=2$, whereas the commuting $U\left( 1\right) $ factor is the
global $\mathcal{N}=2$ vector $\mathcal{R}$-symmetry. On the two-form
Abelian field strengths' fluxes, the truncation condition reads
\begin{equation}
\left( \mathbf{3,}\overline{\mathbf{3}}\right) _{-1}=0,  \label{trunc-2}
\end{equation}%
such that only the graviphoton charges $\left( \mathbf{1,1}\right)
_{+3}+\left( \mathbf{1,1}\right) _{-3}$ survive the truncation. One can also
prove that the \textit{quartic} invariant of the $\mathbf{R}=\mathbf{20}$ of
$SU\left( 1,5\right) $, under the truncation (\ref{trunc-2}) becomes nothing
but the \textit{square} of the Reissner-N\"{o}rdstrom entropy
\begin{equation}
\frac{S_{RN}}{\pi }=\frac{1}{2}\left[ \left( p^{0}\right) ^{2}+q_{0}^{2}%
\right] .  \label{RN-entropy}
\end{equation}
\end{enumerate}

\medskip

It is here worth pointing out that a consistent truncation to an
hypermultiplet(s)-coupled $\mathcal{N}=2$ theory with no vector multiplets
should necessarily contain two real singlets (namely the electric and
magnetic charge of the graviphoton) in the branching of the original flux
representation, as it holds \textit{e.g.} for (\ref{OBBB-3}) and (\ref%
{OBBBB-3}) respectively pertaining to truncations (\ref{OBBB-1}) and (\ref%
{OBBBB-1}). However, such truncations are not interesting for our
investigation, because they yield no vectors when further reduced down to $%
\mathcal{N}=1$ models (the $\mathcal{N}=2$ graviphoton is contained in the $%
\mathcal{N}=1$ gravitino multiplet, which is truncated away).

\subsection{\label{--->4-nV=6}$\rightarrow \mathcal{N}=4$ $\mathbb{R}\oplus
\mathbf{\Gamma }_{5,5}$}

Let's consider now the embedding:
\begin{eqnarray}
J_{3}^{\mathbb{O}_{s}} &:&\mathcal{N}=8\longrightarrow \mathbb{R}\oplus
\mathbf{\Gamma }_{5,5}:\mathcal{N}=4,\text{~}n_{V}=6;  \notag  \label{OBB-1}
\\
E_{7\left( 7\right) } &\supset &SL\left( 2,\mathbb{R}\right) \times SO\left(
6,6\right) ; \\
\mathbf{56} &=&\left( \mathbf{2,12}\right) +\left( \mathbf{1},\mathbf{32}%
\right) ; \\
\frac{E_{7\left( 7\right) }}{SU\left( 8\right) } &\supset &\frac{SL\left( 2,%
\mathbb{R}\right) }{U\left( 1\right) }\times \frac{SO\left( 6,6\right) }{%
SO\left( 6\right) \times SO\left( 6\right) }.  \notag  \label{OBB-4}
\end{eqnarray}%
The $\mathcal{N}=4$ theory is coupled to $6$ vector multiplets and, on the
two-form Abelian field strengths' fluxes, the truncation condition reads
\begin{equation}
\left( \mathbf{1},\mathbf{32}\right) =0.  \label{OB-1}
\end{equation}%
It still exhibits a \textit{quartic} $U$-invariant $\mathcal{I}_{4}$, but it
can be further truncated to a theory with $U$-duality group $U\left(
3,3\right) $ with \textit{quadratic} invariant.

\subsection{\label{8--->2}$\rightarrow \mathcal{N}=2$}

\begin{table}[h]
\begin{center}
{\small
\begin{tabular}{|c||c|c|c|c|c|c|}
\hline
& $%
\begin{array}{c}
\\
G_{V} \\
~%
\end{array}
$ & $%
\begin{array}{c}
\\
G_{H} \\
~%
\end{array}
$ & $%
\begin{array}{c}
\\
H_{V} \\
~%
\end{array}
$ & $%
\begin{array}{c}
\\
H_{H} \\
~%
\end{array}
$ & $%
\begin{array}{c}
~ \\
\frac{G_{V}}{H_{V}} \\
\times \\
\frac{G_{H}}{H_{H}} \\
~%
\end{array}
$ & $%
\begin{array}{c}
\\
\left( n_{V},n_{H}\right) \\
~%
\end{array}
$ \\ \hline\hline
$J_{3}^{\mathbb{H}}$ & $%
\begin{array}{c}
\\
SO^{\ast }(12) \\
~%
\end{array}
$ & $SU(2)$ & $SU(6)\times U(1)$ & $-$ & $\frac{SO^{\ast }(12)}{SU(6)\otimes
U(1)}$ & $\left( 15,0\right) $ \\ \hline
$J_{3}^{\mathbb{C}}$ & $SU(3,3)$ & $SU(2,1)$ & $%
\begin{array}{c}
SU(3)\times SU(3) \\
\times \\
U(1)%
\end{array}
$ & $SU(2)\times U(1)$ & $%
\begin{array}{c}
~ \\
\frac{SU(3,3)}{S\left( U(3)\times U(3)\right) } \\
\times \\
\frac{SU(2,1)}{SU(2)\times U(1)} \\
~%
\end{array}
$ & $\left( 9,1\right) $ \\ \hline
$J_{3}^{\mathbb{R}}$ & $%
\begin{array}{c}
\\
Sp\left( 6,\mathbb{R}\right) \\
~%
\end{array}
$ & $G_{2(2)}$ & $SU(3)\times U(1)$ & $SU(2)\times SU(2)$ & $%
\begin{array}{c}
~ \\
\frac{Sp(6,\mathbb{R})}{SU(3)\times U(1)} \\
\times \\
\frac{G_{2\left( 2\right) }}{SO(4)} \\
~%
\end{array}
$ & $\left( 6,2\right) $ \\ \hline
$STU$ & $%
\begin{array}{c}
SU(1,1) \\
\times \\
SO(2,2)%
\end{array}
$ & $SO(4,4)$ & $%
\begin{array}{c}
U(1) \\
\times \\
SO(2)\times SO(2)%
\end{array}
$ & $SO(4)\times SO(4)$ & $%
\begin{array}{c}
~ \\
\frac{SU(1,1)}{U(1)}\times \frac{SO(2,2)}{SO(2)\times SO(2)} \\
\times \\
\frac{SO(4,4)}{SO(4)\otimes SO(4)} \\
~%
\end{array}
$ & $\left( 3,4\right) $ \\ \hline
$J_{3,M}^{\mathbb{R}}$ & $SU(1,1)$ & $F_{4(4)}$ & $U(1)$ & $USp(6)\times
SU(2)$ & $%
\begin{array}{c}
~ \\
\frac{SU(1,1)}{U(1)} \\
\times \\
\frac{F_{4(4)}}{USp(6)\otimes SU(2)} \\
~%
\end{array}
$ & $\left( 1,7\right) $ \\ \hline
$J_{3,M}^{\mathbb{C}}$ & $%
\begin{array}{c}
\\
U(1) \\
~%
\end{array}
$ & $E_{6(2)}$ & $-$ & $SU(6)\times SU(2)$ & $\frac{E_{6(2)}}{SU(6)\times
SU(2)}$ & $\left( 0,10\right) $ \\ \hline
\end{tabular}%
}
\end{center}
\caption{$\mathcal{N}=2$\textbf{\ supergravities obtained as consistent
\textit{maximal} truncation of }$\mathcal{N}=8$\textbf{\ supergravity }}
\end{table}
We now consider the reduction of $\mathcal{N}=8$ supergravity to an $N=2$
theory with $n_{V}$ vector and $n_{H}$ hypermultiplets:
\begin{equation}
\left( n_{V},n_{H}\right) \equiv \left( \text{dim}_{\mathbb{C}}\left( \frac{%
G_{V}}{H_{V}}\right) ,~\text{dim}_{\mathbb{H}}\left( \frac{G_{H}}{H_{H}}%
\right) \right) ,~n_{V}\leqslant 15,~n_{H}\leqslant 20,  \label{nV,nH}
\end{equation}%
where $\frac{G_{V}}{H_{V}}$ and $\frac{G_{H}}{H_{H}}$ respectively stand for
the special K\"{a}hler and quaternionic K\"{a}hler scalar manifolds, where $%
H_{V}=mcs\left( G_{V}\right) $ and $H_{H}=mcs\left( G_{H}\right) $. $H_{V}$
always contains a factorized commuting $U(1)$ subgroup, which is promoted to
global symmetry when $n_{V}=0$; on the other hand, $H_{H}$ always contains a
factorized commuting $SU(2)$ subgroup, which is promoted to global symmetry
when $n_{H}=0$ \cite{Ferrara-Scherk-Zumino}.

We consider only $\mathcal{N}=2$ \textit{maximal} supergravities, \textit{%
i.e.} $\mathcal{N}=2$ theories (obtained by consistent truncations of $%
\mathcal{N}=8$ supergravity) which cannot be obtained by a further reduction
from some other $\mathcal{N}=2$ theory, which are also \textit{magic}. They
are called \textit{magic}, since their symmetry groups are the groups of the
famous \textit{Magic Square} of Freudenthal, Rozenfeld and Tits associated
with some remarkable geometries \cite{Freudenthal2,magic}. From the analysis
performed in \cite{ADF-1}, only six $\mathcal{N}=2$, $d=4$ \textit{maximal}
\textit{magic} supergravities\footnote{%
By $E_{7(p)}$ we denote a non-compact form of $E_{7}$, where $p\equiv \left(
\#\text{ non-compact}-\#\text{ compact}\right) $ generators of the group
\cite{Helgason,Gilmore}. In such a notation, the compact form of $E_{7}$ is $%
E_{7(-133)}$ ($dim_{\mathbb{R}}E_{7}=133$).} exist which can be obtained by
consistently truncating $\mathcal{N}=8$, $d=4$ supergravity; they are given
by Table 5. After \cite{Ferrara-Marrani-1}, we also include the case of $STU$
model \cite{Duff-stu,BKRSW,K3} with $n_{H}=4$ hypermultiplets; see below.

The models have been denoted by referring to their special geometry. $J_{3}^{%
\mathbb{H}}$, $J_{3}^{\mathbb{C}}$ and $J_{3}^{\mathbb{R}}$ stand for three
of the four $\mathcal{N}=2$, $d=4$ magic supergravities which, as their $5$%
-dim. versions, are respectively defined by the three simple Jordan algebras
$J_{3}^{\mathbb{H}}$, $J_{3}^{\mathbb{C}}$ and $J_{3}^{\mathbb{R}}$ of
degree 3 with irreducible norm forms, namely by the Jordan algebras of
Hermitian $3\times 3$ matrices over the division algebras of quaternions $%
\mathbb{H}$, complex numbers $\mathbb{C}$ and real numbers $\mathbb{R}$ \cite%
{GST,Jordan,Jacobson,Guna1,GPR}.

In Table 5, the subscript \textquotedblleft $M$\textquotedblright\ \ denotes
the model obtained by performing a $D=4$ \textit{mirror map} (\textit{i.e.}
the composition of two $c$-maps \cite{CFG} in $D=4$) from the original
manifold; such an operation maps a model with content $\left(
n_{V},n_{H}\right) $ to a model with content $\left( n_{H}-1,n_{V}+1\right) $%
, and thus the mirror $J_{3,M}^{\mathbb{H}}$ of $J_{3}^{\mathbb{H}}$, with $%
\left( n_{V},n_{H}\right) =\left( -1,16\right) $ and quaternionic manifold $%
\frac{E_{7\left( -5\right) }}{SO(12)\otimes SU(2)}$ does not exist, \textit{%
at least} in $D=4$. The $STU$ model is \textit{self-mirror}: $STU=STU_{M}$.

\subsubsection{\label{Further}Further Truncation to \textit{Minimal Coupling}%
}

Then, we consider further truncations to $\mathcal{N}=2$ theories exhibiting
scalar-vector \textit{minimal coupling}; since hyperscalars are always
minimally coupled, we study only truncations of the vector multiplets'
scalar sector.

Out of the cases reported in Table 5, some deserve immediate comments:

\begin{itemize}
\item The case pertaining to the \textit{self-mirror} $STU_{M}$ model is
included in the treatment of Sec. \ref{4--->2+hypers} starting from $%
\mathcal{N}=4$ theory coupled to $n=6$ vector multiplets (which in turn is
\textit{maximally} embedded into $\mathcal{N}=8$ theory), and considering
the splitting $(n_{1},n-n_{1})=\left( 2,4\right) $.

\item The case pertaining to the \textit{mirror} model $J_{3,M}^{\mathbb{R}}
$ is not interesting in our investigation: indeed, in the vector multiplets
sector, $J_{3,M}^{\mathbb{R}}$ is nothing but the so-called $\mathcal{N}=2$ $%
T^{3}$ model, in which the complex scalar field $T$ is \textit{not}
minimally coupled to vectors, and no further truncation to minimally coupled
$\mathcal{N}=2$ or $\mathcal{N}=1$ models is possible.
\end{itemize}

Let's now list the various relevant possibilities from the models reported
in Table 5:

\begin{enumerate}
\item Also by recalling the treatment of Subsec. \ref{N=8--->N=6} (and in
particular Eqs. (\ref{ca-1}) and (\ref{ca-2})), one can consider:%
\begin{eqnarray}
J_{3}^{\mathbb{O}_{s}} &:&\mathcal{N}=8\longrightarrow J_{3}^{\mathbb{H}%
}:\left\{
\begin{array}{c}
\mathcal{N}=2\text{ }\left( n_{V},n_{H}\right) =\left( 15,0\right) \\
\Updownarrow \text{\textquotedblleft \textit{twin\textquotedblright }} \\
\mathcal{N}=6%
\end{array}%
\right.  \notag \\
&\longrightarrow &\mathbb{R}\oplus \mathbf{\Gamma }_{1,5}\text{ }:\left\{
\begin{array}{c}
\mathcal{N}=2\text{ }\left( n_{V},n_{H}\right) =\left( 7,0\right) \\
\Updownarrow \text{\textquotedblleft \textit{twin\textquotedblright }} \\
\mathcal{N}=4~n_{V}=2%
\end{array}%
\right.  \label{OB-2} \\
&&  \notag \\
E_{7\left( 7\right) } &\supset &SO^{\ast }\left( 12\right) \times SU\left(
2\right)  \notag \\
&\supset &SO^{\ast }\left( 8\right) \times SO^{\ast }\left( 4\right) \times
SU\left( 2\right) \sim SO\left( 6,2\right) \times SL\left( 2,\mathbb{R}%
\right) \times SU(2)\times SU\left( 2\right) ;  \notag \\
&&  \label{OB-3} \\
\mathbf{56} &=&\left( \mathbf{32},\mathbf{1}\right) +\left( \mathbf{12},%
\mathbf{2}\right)  \notag \\
&=&\left( \mathbf{8}_{s},\mathbf{2},\mathbf{1},\mathbf{1}\right) +\left(
\mathbf{8}_{c},\mathbf{1},\mathbf{2},\mathbf{1}\right) +\left( \mathbf{1,2,2}%
,\mathbf{2}\right) +\left( \mathbf{8}_{v}\mathbf{,1,1},\mathbf{2}\right) ;
\label{OB-4} \\
&&  \notag \\
\frac{E_{7\left( 7\right) }}{SU\left( 8\right) } &\supset &\frac{SO^{\ast
}\left( 12\right) }{SU\left( 6\right) \times U\left( 1\right) }\supset \frac{%
SL\left( 2,\mathbb{R}\right) }{U\left( 1\right) }\times \frac{SO\left(
6,2\right) }{SO\left( 6\right) \times SO\left( 2\right) }.  \label{OB-5}
\end{eqnarray}%
The $J_{3}^{\mathbb{H}}$-based theory can either be interpreted as $\mathcal{%
N}=2$ or as its \textquotedblleft twin\textquotedblright\ $\mathcal{N}=6$
\cite{ADF-fixed,FGimK,Gnecchi-1,Samtleben}; in the former case, the $%
SU\left( 2\right) $ commuting factor is the \textit{global} hyper $\mathcal{R%
}$-symmetry, whereas in the latter case we already mentioned that the $%
\mathcal{R}$-symmetry truncated away in the supersymmetry reduction $%
\mathcal{N}=8\rightarrow 6$ (a further truncation $\mathcal{N}=6\rightarrow
\mathcal{N}=3$ is considered in Subsec. \ref{6--->3}). In both cases, the
truncation condition on the two-form Abelian field strengths' fluxes is
given by (\ref{conss}). Thence, one can proceed by truncating to the $\left(
\mathbb{R}\oplus \mathbf{\Gamma }_{1,5}\right) $-based theory still enjoys a
\textit{\textquotedblleft twin\textquotedblright } interpretation \cite%
{Gnecchi-1,Samtleben}, either $\mathcal{N}=2$ or $\mathcal{N}=4$
supergravity; in the former case, the second $SU\left( 2\right) $ commuting
factor also be interpreted as the \textit{global} hyper $\mathcal{R}$%
-symmetry, whereas in the latter case it is the $\mathcal{R}$-symmetry
truncated away in the supersymmetry reduction $\mathcal{N}=6\rightarrow
\mathcal{N}=4$. In both cases, the truncation condition is
\begin{equation}
\left( \mathbf{8}_{c},\mathbf{1},\mathbf{2},\mathbf{1}\right) =0~\text{%
\textit{or}}~\left( \mathbf{8}_{s},\mathbf{2},\mathbf{1},\mathbf{1}\right)
=0.
\end{equation}%
The resulting theory still exhibits a \textit{quartic} $U$-invariant $%
\mathcal{I}_{4}$, but it can be further truncated to a theory with $U$%
-duality group $U\left( 1,3\right) $ with \textit{quadratic} invariant. It
is here worth remarking that such a theory still admits a \textit{%
\textquotedblleft twin\textquotedblright } interpretation \cite{Gnecchi-1},
namely either as $\mathcal{N}=3$ with $n_{V}=1$ vector multiplet or as $%
\mathcal{N}=2$ \textit{minimally coupled} to $n_{V}=3$ vector multiplets
(and no hypermultiplets).

\item
\begin{eqnarray}
J_{3}^{\mathbb{O}_{s}} &:&\mathcal{N}=8\longrightarrow J_{3}^{\mathbb{C}}:%
\mathcal{N}=2\text{ }\left( n_{V},n_{H}\right) =\left( 9,1\right)
\longrightarrow \mathcal{N}=2~\mathbb{CP}^{3}\text{~}\left(
n_{V},n_{H}\right) =\left( 4,1\right) ; \\
&&  \notag \\
E_{7\left( 7\right) } &\supset &SU\left( 3,3\right) \times SU\left(
2,1\right) \supset SU\left( 1,3\right) \times SU\left( 2\right) \times
SU\left( 2,1\right) \times U\left( 1\right) ;  \notag \\
&& \\
\mathbf{56} &=&\left( \mathbf{6},\mathbf{3}\right) +\left( \overline{\mathbf{%
6}},\overline{\mathbf{3}}\right) +\left( \mathbf{20},\mathbf{1}\right)
\notag \\
&=&\left( \mathbf{1},\mathbf{2},\mathbf{3}\right) _{2}+\left( \mathbf{4},%
\mathbf{1},\mathbf{3}\right) _{-1}+\left( \mathbf{1},\mathbf{2},\overline{%
\mathbf{3}}\right) _{-2}+\left( \overline{\mathbf{4}},\mathbf{1},\overline{%
\mathbf{3}}\right) _{1}+\left( \mathbf{4},\mathbf{1,0}\right) _{+3}+\left(
\overline{\mathbf{4}},\mathbf{1,0}\right) _{-3}+\left( \mathbf{6},\mathbf{2}%
\right) _{0};  \notag \\
&& \\
\frac{E_{7\left( 7\right) }}{SU\left( 8\right) } &\supset &\frac{SU\left(
3,3\right) }{S\left( U\left( 3\right) \times U\left( 3\right) \right) }%
\times \frac{SU\left( 2,1\right) }{U\left( 2\right) }\supset \frac{SU\left(
1,3\right) }{U\left( 3\right) }\times \frac{SU\left( 2,1\right) }{U\left(
2\right) }.
\end{eqnarray}%
The $J_{3}^{\mathbb{C}}$-based theory is \textit{magic} $\mathcal{N}=2$ with
$9$ vector multiplets and $1$ universal hypermultiplet. The truncation
condition reads
\begin{equation}
\left( \mathbf{6},\mathbf{3}\right) =0.  \label{trunc-first}
\end{equation}%
A different realization of this truncation has been studied in Subsec. \ref%
{8--->5--->3}. Thence, one can proceed by truncating to $\mathcal{N}=2%
\mathbb{\ }$\textit{minimally coupled} to $3$ vector multiplets (hyper
sector untouched); the further truncation condition is
\begin{equation}
\left( \mathbf{6},\mathbf{2}\right) _{0}=0.  \label{trunc-second}
\end{equation}%
Through this chain of truncation, one can also prove that the \textit{quartic%
} invariant of the $\mathbf{R}=\mathbf{20}$ of $SU\left( 3,3\right) $
becomes the \textit{square} of the \textit{quadratic} invariant of the $%
\mathbf{R}=\mathbf{4}$ of $SU\left( 1,3\right) $.

\item From $\mathcal{N}=2$ $J_{3}^{\mathbb{C}}$ theory another truncation is
possible, namely:
\begin{eqnarray}
J_{3}^{\mathbb{O}_{s}} &:&\mathcal{N}=8\longrightarrow J_{3}^{\mathbb{C}}:%
\mathcal{N}=2\text{ }\left( n_{V},n_{H}\right) =\left( 9,1\right)
\longrightarrow \mathcal{N}=2~\mathbb{R}\oplus \mathbf{\Gamma }_{1,3}\text{~}%
\left( n_{V},n_{H}\right) =\left( 5,1\right) ;  \notag \\
&& \\
E_{7\left( 7\right) } &\supset &SU\left( 3,3\right) \times SU\left(
2,1\right)  \notag \\
&\supset &SU\left( 1,1\right) \times SU\left( 2,2\right) \times SU\left(
2,1\right) \times U\left( 1\right) \sim SL\left( 2,\mathbb{R}\right) \times
SO\left( 2,4\right) \times SU\left( 2,1\right) \times U\left( 1\right) ;
\notag \\
&& \\
\mathbf{56} &=&\left( \mathbf{6},\mathbf{3}\right) +\left( \overline{\mathbf{%
6}},\overline{\mathbf{3}}\right) +\left( \mathbf{20},\mathbf{1}\right)
\notag \\
&=&\left( \mathbf{2},\mathbf{1},\mathbf{3}\right) _{2}+\left( \mathbf{1},%
\mathbf{4},\mathbf{3}\right) _{-1}+\left( \mathbf{2},\mathbf{1},\overline{%
\mathbf{3}}\right) _{-2}+\left( \mathbf{1},\overline{\mathbf{4}},\overline{%
\mathbf{3}}\right) _{1}+\left( \mathbf{1},\mathbf{4},\mathbf{1}\right)
_{3}+\left( \mathbf{1},\overline{\mathbf{4}},\mathbf{1}\right) _{-3}+\left(
\mathbf{2},\mathbf{6}\right) _{0};  \notag \\
&& \\
\frac{E_{7\left( 7\right) }}{SU\left( 8\right) } &\supset &\frac{SU\left(
3,3\right) }{S\left( U\left( 3\right) \times U\left( 3\right) \right) }%
\times \frac{SU\left( 2,1\right) }{U\left( 2\right) }\supset \frac{SL\left(
2,\mathbb{R}\right) }{U\left( 1\right) }\times \frac{SO\left( 2,4\right) }{%
SO\left( 2\right) \times SO\left( 4\right) }\times \frac{SU\left( 2,1\right)
}{U\left( 2\right) }.
\end{eqnarray}
As for the point $2$ above, the first truncation condition is given by (\ref%
{trunc-first}), but the second one is the very opposite of (\ref%
{trunc-second}): only $\left( \mathbf{2},\mathbf{6}\right) _{0}$ does not
vanish, or equivalently:
\begin{equation}
\left( \mathbf{1},\mathbf{4},\mathbf{1}\right) _{3}=0.
\end{equation}
The resulting theory still exhibits a \textit{quartic} $U$-invariant $%
\mathcal{I}_{4}$, but it can be \textit{non-maximally} further truncated to
an $\mathcal{N}=2$ $\mathbb{CP}^{2}$ model with \textit{quadratic} invariant
through the procedure considered in Sec. \ref{2--->CPn}, to which we address
the reader for further elucidation.

\item
\begin{eqnarray}
J_{3}^{\mathbb{O}_{s}} &:&\mathcal{N}=8\longrightarrow J_{3}^{\mathbb{R}}:%
\mathcal{N}=2\text{ }\left( n_{V},n_{H}\right) =\left( 6,2\right)
\longrightarrow \mathcal{N}=2~\mathbb{R}\oplus \mathbf{\Gamma }_{1,2}\text{~}%
\left( n_{V},n_{H}\right) =\left( 4,2\right) ;  \notag \\
&& \\
E_{7\left( 7\right) } &\supset &Sp\left( 6,\mathbb{R}\right) \times
G_{2\left( 2\right) }\supset Sp\left( 2,\mathbb{R}\right) \times Sp\left( 4,%
\mathbb{R}\right) \times G_{2\left( 2\right) }\sim SL\left( 2,\mathbb{R}%
\right) \times SO\left( 2,3\right) \times G_{2\left( 2\right) };  \notag \\
&& \\
\mathbf{56} &=&\left( \mathbf{14}^{\prime },\mathbf{1}\right) +\left(
\mathbf{6},\mathbf{7}\right) =\left( \mathbf{1},\mathbf{4},\mathbf{1}\right)
+\left( \mathbf{2},\mathbf{5},\mathbf{1}\right) +\left( \mathbf{2},\mathbf{1}%
,\mathbf{7}\right) +\left( \mathbf{1},\mathbf{4},\mathbf{7}\right) ;  \notag
\\
&& \\
\frac{E_{7\left( 7\right) }}{SU\left( 8\right) } &\supset &\frac{Sp\left( 6,%
\mathbb{R}\right) }{U\left( 3\right) }\times \frac{G_{2\left( 2\right) }}{%
SO\left( 4\right) }\supset \frac{SL\left( 2,\mathbb{R}\right) }{U\left(
1\right) }\times \frac{SO\left( 2,3\right) }{SO\left( 2\right) \times
SO\left( 3\right) }\times \frac{G_{2\left( 2\right) }}{SO\left( 4\right) }.
\end{eqnarray}
The $J_{3}^{\mathbb{R}}$-based theory is \textit{magic} $\mathcal{N}=2$ with
$6$ vector multiplets and $2$ hypermultiplets. The truncation condition
reads
\begin{equation}
\left( \mathbf{6},\mathbf{7}\right) =0.
\end{equation}
Thence, one can proceed by truncating to $\left( \mathbb{R}\oplus \mathbf{%
\Gamma }_{1,2}\right) $-based $\mathcal{N}=2$ theory (hyper sector
untouched); the further truncation condition is
\begin{equation}
\left( \mathbf{1},\mathbf{4},\mathbf{1}\right) =0.
\end{equation}
The resulting theory still exhibits a \textit{quartic} $U$-invariant $%
\mathcal{I}_{4}$, but it can be \textit{non-maximally} further truncated to
an $\mathcal{N}=2$ $\mathbb{CP}^{1}$ model with \textit{quadratic} invariant
through the procedure considered in Sec. \ref{2--->CPn} (see also comment in
Subsec. \ref{Comment-CP1}), to which we address the reader for further
elucidation.

\item
\begin{eqnarray}
J_{3}^{\mathbb{O}_{s}} &:&\mathcal{N}=8\longrightarrow J_{3,M}^{\mathbb{C}}:%
\mathcal{N}=2\text{ }\left( n_{V},n_{H}\right) =\left( 0,10\right) ;  \notag
\label{OBBBB-1} \\
&& \\
E_{7\left( 7\right) } &\supset &E_{6\left( 2\right) }\times U\left( 1\right)
\supset U\left( 1\right) ;  \notag \\
&& \\
\mathbf{56} &=&\mathbf{27}_{+1}+\mathbf{27}_{-1}^{\prime }+\mathbf{1}_{+3}+%
\mathbf{1}_{-3}^{\prime };  \notag  \label{OBBBB-3} \\
&& \\
\frac{E_{7\left( 7\right) }}{SU\left( 8\right) } &\supset &\frac{E_{6\left(
2\right) }}{SU\left( 6\right) \times SU\left( 2\right) }.
\end{eqnarray}%
The resulting $\mathcal{N}=2$ theory is coupled to $10$ hypermultiplets, in
absence of vector multiplets. The commuting $U\left( 1\right) $ factor is
the global $\mathcal{N}=2$ vector $\mathcal{R}$-symmetry. On the two-form
Abelian field strengths' fluxes, the truncation condition reads
\begin{equation}
\mathbf{27}_{+1}=0,  \label{trunc-2-bis}
\end{equation}%
such that only the graviphoton charges $\mathbf{1}_{+3}+\mathbf{1}%
_{-3}^{\prime }$ survive the truncation.One can also prove that the \textit{%
quartic} invariant of the $\mathbf{R}=\mathbf{56}$ of $E_{7\left( 7\right) }$%
, under the truncation (\ref{trunc-2-bis}) becomes nothing but the \textit{%
square} of the Reissner-N\"{o}rdstrom entropy (\ref{RN-entropy}).
\end{enumerate}

\section{\label{N=6}Maximal Truncations from $\mathcal{N}=6$ ($J_{3}^{%
\mathbb{H}}$)}

\subsection{\label{6--->3}$\rightarrow \mathcal{N}=3$}

From $\mathcal{N}=6$ \textit{\textquotedblleft pure\textquotedblright }
theory, one can consider the following maximal \textit{\textquotedblleft
degenerative\textquotedblright } truncation:
\begin{eqnarray}
J_{3}^{\mathbb{H}} &:&\mathcal{N}=6\longrightarrow \mathcal{N}=3,\text{~}%
n_{V}=3; \\
SO^{\ast }\left( 12\right) &\supset &SU\left( 3,3\right) \times U\left(
1\right) ;  \label{CERN-1} \\
\mathbf{32} &=&\mathbf{6}_{-2}+\overline{\mathbf{6}}_{+2}+\mathbf{20}_{0};
\label{CCERN-1} \\
\frac{SO^{\ast }\left( 12\right) }{SU\left( 6\right) \times U\left( 1\right)
} &\supset &\frac{SU\left( 3,3\right) }{SU\left( 3\right) \times SU\left(
3\right) \times U\left( 1\right) }.
\end{eqnarray}%
The $\mathcal{N}=3$ theory is coupled to $3$ vector multiplets and, on the
two-form Abelian field strengths' fluxes, the truncation condition reads
\begin{equation}
\mathbf{20}_{0}=0.  \label{trunc-4}
\end{equation}%
One can prove that the \textit{quartic} invariant of the $\mathbf{R}=\mathbf{%
32}$ of $SO^{\ast }\left( 12\right) $, under the truncation (\ref{trunc-4})
becomes the \textit{square} of the \textit{quadratic} invariant of the $%
\mathbf{R}=\mathbf{6}$ of $SU\left( 3,3\right) $.

\subsection{\label{6--->5}$\mathcal{N}=5$}

Note that, one might consider another truncation by setting
\begin{equation}
\mathbf{6}_{-2}=0
\end{equation}
in (\ref{CCERN-1}); this corresponds to a truncation $\mathcal{N}%
=6\longrightarrow \mathcal{N}=2$~based on $J_{3}^{\mathbb{C}}$ or,
equivalently (due to the fact that $\mathcal{N}=6$ and $\mathcal{N}=2$~based
on $J_{3}^{\mathbb{H}}$ are \textit{``twin''}, \textit{i.e.} they share the
very same bosonic sector \cite{ADF-fixed,FGimK,Gnecchi-1,Samtleben}) to $%
\mathcal{N}=2~J_{3}^{\mathbb{H}}\longrightarrow \mathcal{N}=2$~$J_{3}^{%
\mathbb{C}}$. However, the resulting $\mathcal{N}=2$ ``magic'' complex
theory exhibits a generally \textit{``non-degenerate''} \textit{quartic} $U$%
-invariant $\mathcal{I}_{4}$.

On the other hand, if in (\ref{CERN-1}) $SU\left( 3,3\right) $ is changed
into $SU\left( 1,5\right) $, another, complementary, realization of the
above truncation reads
\begin{eqnarray}
\mathcal{N} &=&6\longrightarrow \mathcal{N}=5; \\
SO^{\ast }\left( 12\right) &\supset &SU\left( 1,5\right) \times U\left(
1\right) ; \\
\mathbf{32} &=&\mathbf{6}_{-2}+\overline{\mathbf{6}}_{+2}+\mathbf{20}_{0}; \\
\frac{SO^{\ast }\left( 12\right) }{SU\left( 6\right) \times U\left( 1\right)
} &\supset &\frac{SU\left( 1,5\right) }{U\left( 5\right) }.
\end{eqnarray}
The $\mathcal{N}=5$ theory is \textit{``pure''} and the commuting $U\left(
1\right) $ factor corresponds to the part of the $\mathcal{R}$-symmetry
truncated away in the supersymmetry reduction $\mathcal{N}=6\rightarrow
\mathcal{N}=5$. On the two-form Abelian field strengths' fluxes, the
truncation condition is
\begin{equation}
\mathbf{6}_{-2}=0.
\end{equation}
In turn, the \textit{``pure''} $\mathcal{N}=5$ theory admits two maximal
\textit{``degenerative''} truncations, treated in Sec. \ref{8--->5--->3},
which precisely match the kinematical decomposition of the $\mathcal{N}=5$
gravity multiplet into matter $\mathcal{N}=2$ multiplets.

\section{\label{4--->3}$\mathcal{N}=4$ $\mathbb{R}\oplus \mathbf{\Gamma }%
_{5,2n-1}\longrightarrow \mathcal{N}=3$}

We start with $\mathcal{N}=4$ supergravity coupled to $n_{V}=2n$ matter
(vector) multiplets , which is based on the rank-$3$ Jordan algebra $\mathbb{%
R}\oplus \mathbf{\Gamma }_{1,2n-1}$, with data
\begin{eqnarray}
\frac{G_{4}}{H_{4}} &=&\frac{SL_{v}\left( 2,\mathbb{R}\right) }{U\left(
1\right) }\times \frac{SO\left( 6,2n\right) }{SO\left( 6\right) \times
SO\left( n\right) }; \\
\mathbf{R} &=&\left( \mathbf{2},\mathbf{6+n}\right) .
\end{eqnarray}%
The relevant products of electric and magnetic charges read
\begin{equation}
\begin{array}{l}
p^{2}\equiv p^{\Lambda }p^{\Sigma }\eta _{\Lambda \Sigma
}=\sum_{a=1}^{6}\left( p^{a}\right) ^{2}-\sum_{I=1}^{2n}\left( p^{I}\right)
^{2}; \\
\\
q^{2}\equiv q_{\Lambda }q_{\Sigma }\eta ^{\Lambda \Sigma
}=\sum_{a=1}^{6}q_{a}^{2}-\sum_{I=1}^{2n}q_{I}^{2}; \\
\\
p\cdot q\equiv p^{\Lambda }q_{\Lambda },%
\end{array}
\label{1-bis}
\end{equation}%
where $\eta $ is the symmetric invariant structure of the vector ($\mathbf{%
Fund}$) irrep. $\mathbf{6+2n}$ of $SO\left( 6,2n\right) $, with $\Lambda
=1,...,2n+6$, where the indices $1,...,6$ pertain to the $6$ graviphotons.

We consider a complexification of the electric and magnetic charge vectors $%
p^{\Lambda }$ and $q_{\Lambda }$ as follows:
\begin{equation}
\left\{
\begin{array}{l}
P^{1}\equiv p^{1}+ip^{2}; \\
P^{2}\equiv p^{3}+ip^{4}; \\
P^{3}\equiv p^{5}+ip^{6}; \\
P^{4}\equiv p^{7}+ip^{8}; \\
.... \\
P^{n+3}\equiv p^{2n+5}+ip^{2n+6},%
\end{array}%
\right.
\end{equation}%
and analogously for the electric charges. Thus (\ref{1-bis}) can be
rewritten as
\begin{eqnarray}
p^{2} &=&\sum_{\mathcal{A}=1}^{3}\left\vert P^{\mathcal{A}}\right\vert
^{2}-\sum_{A=4}^{n+3}\left\vert P^{A}\right\vert ^{2}=P^{i}\overline{P}^{%
\overline{j}}\eta _{i\overline{j}}; \\
q^{2} &=&\sum_{\mathcal{A}=1}^{3}\left\vert Q_{\mathcal{A}}\right\vert
^{2}-\sum_{A=4}^{n+3}\left\vert Q_{A}\right\vert ^{2}=\eta ^{i\overline{j}%
}Q_{i}\overline{Q}_{\overline{j}}; \\
p\cdot q &=&\sum_{i=1}^{n+3}\text{Re}\left( P^{i}\overline{Q}_{\overline{i}%
}\right) ,
\end{eqnarray}%
with $\eta $ here denoting the invariant rank-$2$ structure in the product $%
\left( \mathbf{3+n}\right) \times \left( \overline{\mathbf{3+n}}\right) $ of
$U\left( 3,n\right) $, with $i=1,...,n+3$ (in Sec. \ref{Identities}, the
complex charge vector $\left( P^{i},Q_{i}\right) $ has been indicated by $%
\mathbf{Q}$). Therefore:
\begin{eqnarray}
\frac{1}{4}\mathcal{I}_{4,\mathbb{R}\oplus \mathbf{\Gamma }_{5,2n-1}}
&=&p^{2}q^{2}-\left( p\cdot q\right) ^{2} \\
&=&\eta _{i\overline{j}}\eta ^{k\overline{l}}P^{i}\overline{P}^{\overline{j}%
}Q_{k}\overline{Q}_{\overline{l}}-\left( \sum_{i=1}^{n+3}\text{Re}\left(
P^{i}\overline{Q}_{\overline{i}}\right) \right) ^{2} \\
&=&\frac{1}{4}\left( S_{1}^{2}-\left\vert S_{2}\right\vert ^{2}\right) ,
\end{eqnarray}%
where the following quantities have been introduced \cite{ADF-fixed,CFMZ1}:
\begin{eqnarray}
S_{1} &\equiv &p^{2}+q^{2}=\left( P^{i}\overline{P}^{\overline{j}}+Q^{i}%
\overline{Q}^{\overline{j}}\right) \eta _{i\overline{j}};  \label{9} \\
S_{2} &\equiv &\left( p^{2}-q^{2}\right) +2ip\cdot q=\left( P^{i}\overline{P}%
^{\overline{j}}-Q^{i}\overline{Q}^{\overline{j}}\right) \eta _{i\overline{j}%
}+2i\sum_{i=1}^{n+3}\text{Re}\left( P^{i}\overline{Q}_{\overline{i}}\right) .
\label{10}
\end{eqnarray}

The \textit{\textquotedblleft degeneration\textquotedblright } condition we
exploit reads as follows:
\begin{equation}
S_{2}=0\Leftrightarrow \left\{
\begin{array}{l}
\text{Re}S_{2}=0\Leftrightarrow \left( P^{i}\overline{P}^{\overline{j}}-Q^{i}%
\overline{Q}^{\overline{j}}\right) \eta _{i\overline{j}}=0; \\
\\
\text{Im}S_{2}=0\Leftrightarrow \sum_{i=1}^{n+3}\text{Re}\left( P^{i}%
\overline{Q}_{\overline{i}}\right) =0,%
\end{array}%
\right.  \label{11}
\end{equation}%
whose a solution is
\begin{equation}
Q_{j}=\pm iP^{j}~\forall j,  \label{12}
\end{equation}%
with $j$-dependent \textquotedblleft $\pm $\textquotedblright\ branches. One
thus obtains:
\begin{equation}
\left. \mathcal{I}_{4,\mathbb{R}\oplus \mathbf{\Gamma }_{5,2n-1}}\right\vert
_{S_{2}=0}=\left( S^{1}\right) ^{2}=4\left( P^{i}\overline{P}^{\overline{j}%
}\eta _{i\overline{j}}\right) ^{2}=\left( \mathcal{I}_{2,\mathcal{N}%
=3}\right) ^{2}.  \label{13-bis}
\end{equation}%
Namely, the \textit{quartic} invariant $\mathcal{I}_{4,\mathbb{R}\oplus
\mathbf{\Gamma }_{5,2n-1}}$of the real irrep. $\mathbf{R}=\left( \mathbf{2},%
\mathbf{6+2n}\right) $ of the semisimple group of type $E_{7}$ $%
G_{4}=SL_{v}\left( 2,\mathbb{R}\right) \times SO\left( 6,2n\right)
=Conf\left( \mathbb{R}\oplus \mathbf{\Gamma }_{5,2n-1}\right) $ \textit{%
\textquotedblleft degenerates\textquotedblright } into the square of the
\textit{quadratic} invariant $\mathcal{I}_{2,\mathcal{N}=3}$ of the complex
irrep. $\mathbf{R}^{\prime }=\mathbf{3+n}$ of the \textquotedblleft
degenerate\textquotedblright\ group of type $E_{7}$ $G_{4}^{\prime }=U\left(
3,n\right) $. This latter is the $U$-duality group of $\mathcal{N}=3$
supergravity coupled to $n$ vector multiplets.

In a manifestly $U\left( 3,n\right) $-covariant symplectic basis, $\mathcal{I%
}_{2,\mathcal{N}=3}$ reads:
\begin{equation}
\mathcal{I}_{2,\mathcal{N}=3}=\sum_{\mathfrak{A}=1}^{3}\left[ \left(
\mathfrak{p}^{\mathfrak{A}}\right) ^{2}+\mathfrak{q}_{\mathfrak{A}}^{2}%
\right] -\sum_{\alpha =1}^{n}\left[ \left( \mathfrak{p}^{\alpha }\right)
^{2}+\mathfrak{q}_{\alpha }^{2}\right] .  \label{14-bis}
\end{equation}
In order to make (\ref{14-bis}) consistent with (\ref{13-bis}), the
following \textit{dyonic identification} of charges can be performed:
\begin{equation}
\begin{array}{l}
P^{\mathcal{A}}\equiv \frac{1}{\sqrt{2}}\left( \mathfrak{p}^{\mathfrak{A}}+i%
\mathfrak{q}_{\mathfrak{A}}\right) ; \\
P^{A}\equiv \frac{1}{\sqrt{2}}\left( \mathfrak{p}^{\alpha }+i\mathfrak{q}%
_{\alpha }\right) .%
\end{array}
\label{15-bis}
\end{equation}

In group-theoretical terms, the \textit{\textquotedblleft
degeneration\textquotedblright } procedure under consideration goes as
follows:
\begin{eqnarray}
SL_{v}\left( 2,\mathbb{R}\right) \times SO\left( 6,2n\right) &\supset
&SL_{v}\left( 2,\mathbb{R}\right) \times U\left( 3,n\right) \supset U\left(
3,n\right) ;  \notag \\
\left( \mathbf{2},\mathbf{6+2n}\right) &=&\left( \mathbf{2},\left( \mathbf{%
3+n}\right) _{+1}\right) +\left( \mathbf{2},\left( \overline{\mathbf{3+n}}%
\right) _{-1}\right)  \notag \\
&=&2\cdot \left[ \left( \mathbf{3+n}\right) _{+1}+\left( \overline{\mathbf{%
3+n}}\right) _{-1}\right] ,
\end{eqnarray}%
with the double-counting eventually removed by the \textit{\textquotedblleft
degeneration\textquotedblright } truncating condition (\ref{11})-(\ref{12}),
which in this case sets to zero $n+3$ complex, \textit{i.e.} $2n+6$ real,
charge combinations. Notice that, also in this case, (\ref{12}) breaks $%
SL_{v}\left( 2,\mathbb{R}\right) $, and its various branches, generated by
the various possibilities in the choice of \textquotedblleft $\pm $%
\textquotedblright\ for each index $i$, are all inter-related by suitable $%
U\left( 3,n\right) $-transformations. At the level of the vector multiplets'
scalar manifolds, it holds
\begin{equation}
\underset{\mathcal{N}=4,~\mathbb{R}\oplus \mathbf{\Gamma }_{5,2n-1}}{\frac{%
SL_{v}\left( 2,\mathbb{R}\right) }{U\left( 1\right) }\times \frac{SO\left(
6,2n\right) }{SO\left( 2\right) \times SO\left( 2n\right) }}\supset \underset%
{\mathcal{N}=3}{\frac{SU\left( 3,n\right) }{U\left( 3\right) \times SU\left(
n\right) }}.
\end{equation}

\section{\label{4--->2+hypers}$\mathcal{N}=4$ $\mathbb{R}\oplus \mathbf{%
\Gamma }_{5,n-1}\longrightarrow \mathcal{N}=2$ $\mathbb{R}\oplus \mathbf{%
\Gamma }_{1,n-1}$ $+$ Hypermultiplets}

$\mathcal{N}=2$ hypermultiplets can be added to the \textit{``degenerative''}
truncation procedures (starting from the $\mathcal{N}=2$ factorized
sequence) treated above, by considering the following truncation:
\begin{eqnarray}
\mathcal{N} &=&4~\mathbb{R}\oplus \mathbf{\Gamma }_{5,n-1}\longrightarrow
\mathcal{N}=2~\mathbb{R}\oplus \mathbf{\Gamma }_{1,n_{1}-1}+\left(
n-n_{1}\right) ~\text{\textit{hypermults.}}  \label{1-2} \\
SL_{v}\left( 2,\mathbb{R}\right) \times SO\left( 6,n\right) &\supset
&SL_{v}\left( 2,R\right) \times SO\left( 2,n_{1}\right) \times SO\left(
4,n-n_{1}\right) ;  \label{2-2} \\
\left( \mathbf{2},\mathbf{6+n}\right) &=&\left( \mathbf{2},\mathbf{2+n}_{1},%
\mathbf{1}\right) +\left( \mathbf{2},\mathbf{1},\mathbf{4+n-n}_{1}\right) ,
\label{3-2}
\end{eqnarray}
where the hyperscalars fit into the quaternionic K\"{a}hler symmetric space
\begin{equation}
\frac{SO\left( 4,n-n_{1}\right) }{SO\left( 4\right) \times SO\left(
n-n_{1}\right) }.  \label{4-2}
\end{equation}
Thus, the $\mathcal{N}=2$ theory is obtained by setting
\begin{equation}
\left( \mathbf{2},\mathbf{1},\mathbf{4+n-n}_{1}\right) =0.  \label{5-2}
\end{equation}

At the level of the scalar manifolds, the truncation (\ref{1-2})-(\ref{5-2})
corresponds to
\begin{equation}
\frac{SL_{v}\left( 2,\mathbb{R}\right) }{U\left( 1\right) }\times \frac{%
SO\left( 6,n\right) }{SO\left( 6\right) \times SO\left( n\right) }\supset
\frac{SL_{v}\left( 2,\mathbb{R}\right) }{U\left( 1\right) }\times \frac{%
SO\left( 2,n_{1}\right) }{SO\left( 2\right) \times SO\left( n_{1}\right) }%
\times \frac{SO\left( 4,n-n_{1}\right) }{SO\left( 4\right) \times SO\left(
n-n_{1}\right) }.  \label{7-2}
\end{equation}
It is worth recalling that the case $n=0$ of the truncation (\ref{7-2}) has
been considered in Sec. 5 of \cite{More-on-N=8} (see also the considerations
in Subsec. \ref{Comment-CP1}).

Starting from the $\mathcal{N}=2$ theory with $n_{1}$ vector multiplets and $%
n-n_{1}$ hypermultiplets, with scalar manifolds given by the direct product
on the righthand side of\ (\ref{7-2}), \textit{iff} $n_{1}$ is \textit{even}
(\textit{i.e. iff} $n_{1}=2m$) one can then consider the further \textit{%
``degenerative''} truncation down to $\mathcal{N}=2$ \textit{minimally
coupled} supergravity with $m$ vector multiplets and $n-n_{1}=n-2m$
hypermultiplets : in practice, the procedure outlined in Subsec. \ref%
{2--->CPn}, with $n\rightarrow m$, and the hypermultiplets which are
insensitive of the truncation:
\begin{equation}
\frac{SL_{v}\left( 2,\mathbb{R}\right) }{U\left( 1\right) }\times \frac{%
SO\left( 2,n_{1}\right) }{SO\left( 2\right) \times SO\left( n_{1}\right) }%
\times \frac{SO\left( 4,n-n_{1}\right) }{SO\left( 4\right) \times SO\left(
n-n_{1}\right) }\overset{\text{\textit{iff~}}n_{1}=2m}{\supset }\frac{%
SU\left( 1,m\right) }{U\left( m\right) }\times \frac{SO\left(
4,n-n_{1}\right) }{SO\left( 4\right) \times SO\left( n-n_{1}\right) }.
\end{equation}

Let's finally mention that the quaternionic manifolds (\ref{4-2}) are
maximal in the framework under consideration, but, \textit{iff} $n-n_{1}$ is
\textit{even} (\textit{i.e. iff} $n-n_{1}=2k$) the further following
truncation in the hyper sector can be considered:
\begin{equation}
\frac{SO\left( 4,n-n_{1}\right) }{SO\left( 4\right) \times SO\left(
n-n_{1}\right) }\overset{\text{\textit{iff~}}n-n_{1}=2k}{\supset }\frac{%
SU\left( 2,k\right) }{SU\left( 2\right) \times SU\left( k\right) \times
U\left( 1\right) }.  \label{8-2}
\end{equation}

Thus, by combining the two above observations, \textit{iff}
\begin{equation}
\left.
\begin{array}{l}
n_{1}=2m; \\
\\
n-n_{1}=2k;%
\end{array}
\right\} \Rightarrow n=2\left( m+k\right) \text{\textit{even}},  \label{9-2}
\end{equation}
one can consider, along the very same lines of Subsec. \ref{2--->CPn}, the
following further \textit{non-maximal} \textit{``degenerative''} truncation
down to $\mathcal{N}=2$ \textit{minimally coupled} supergravity with $m$
vector multiplets and $k$ hypermultiplets:
\begin{equation}
\frac{SL_{v}\left( 2,\mathbb{R}\right) }{U\left( 1\right) }\times \frac{%
SO\left( 2,n_{1}\right) }{SO\left( 2\right) \times SO\left( n_{1}\right) }%
\times \frac{SO\left( 4,n-n_{1}\right) }{SO\left( 4\right) \times SO\left(
n-n_{1}\right) }\overset{\text{\textit{iff~}}n=2\left( m+k\right) }{\supset }%
\frac{SU\left( 1,m\right) }{U\left( m\right) }\times \frac{SU\left(
2,k\right) }{SU\left( 2\right) \times SU\left( k\right) \times U\left(
1\right) }.  \label{10-2}
\end{equation}

\section{\label{3--->2CPn+hypers}$\mathcal{N}=3$ $\longrightarrow \mathcal{N}%
=2$ $\mathbb{CP}^{n}$ $+$ Hypermultiplets}

Finally, let us consider the following truncation:
\begin{eqnarray}
\mathcal{N} &=&3\text{~}p\text{~\textit{vector mults.}}\longrightarrow
\mathcal{N}=2~\mathbb{CP}^{s_{1}}+\left( p-s_{1}\right) ~\text{\textit{%
hypermults.}} \\
U\left( 3,p\right) &\supset &U\left( 1,s_{1}\right) \times SU\left(
2,p-s_{1}\right) \times U\left( 1\right) ;  \label{3-2-1} \\
\left( \mathbf{3}+\mathbf{n}\right) &=&\left( \mathbf{1}+\mathbf{s}%
_{1}\right) _{+1}+\left( \mathbf{2+p-s}_{1}\right) _{-\frac{\left(
1+s_{1}\right) }{2+p-s_{1}}},  \label{3-3}
\end{eqnarray}
which, at the level of scalar manifolds corresponds to the following maximal
embedding:
\begin{equation}
\frac{SU\left( 3,p\right) }{SU\left( 3\right) \times SU\left( p\right)
\times U\left( 1\right) }\supset \frac{SU\left( 1,s_{1}\right) }{U\left(
s_{1}\right) }\times \frac{SU\left( 2,p-s_{1}\right) }{SU\left( 2\right)
\times SU\left( p-s_{1}\right) \times U\left( 1\right) }.
\end{equation}
Thus, the $\mathcal{N}=2$ \textit{minimally coupled} theory is obtained by
setting
\begin{equation}
\left( \mathbf{2+p-s}_{1}\right) =0.
\end{equation}

Notice that the starting $\mathcal{N}=3$ theory can be seen to be obtained
from $\mathcal{N}=4$ theory coupled to $2p$ matter (vector) multiplets
through the \textit{\textquotedblleft degenerative\textquotedblright }
truncation procedure outlined in Subsec. \ref{4--->3}, with $n\rightarrow p$.

\section{\label{From-N=2}Maximal Truncations within $\mathcal{N}=2$}

\subsection{\label{N=2-Exc-FHSV}$J_{3}^{\mathbb{O}}\rightarrow \mathbb{R}%
\oplus \mathbf{\Gamma }_{1,9}$ (FHSV)}

\begin{eqnarray}
J_{3}^{\mathbb{O}} &:&\mathcal{N}=2,~n_{V}=27\longrightarrow \mathcal{N}=2~%
\mathbb{R}\oplus \mathbf{\Gamma }_{1,9}\text{~}n_{V}=11; \\
E_{7\left( -25\right) } &\supset &SL\left( 2,\mathbb{R}\right) \times
SO\left( 2,10\right) ; \\
\mathbf{56} &=&\left( \mathbf{2},\mathbf{12}\right) +\left( \mathbf{1},%
\mathbf{32}\right) ; \\
\frac{E_{7\left( -25\right) }}{E_{6}\times U\left( 1\right) } &\supset &%
\frac{SL\left( 2,\mathbb{R}\right) }{U\left( 1\right) }\times \frac{SO\left(
2,10\right) }{SO\left( 2\right) \times SO\left( 10\right) }.
\end{eqnarray}
The truncation condition reads
\begin{equation}
\left( \mathbf{1},\mathbf{32}\right) =0.
\end{equation}
The resulting theory, the so-called $\mathcal{N}=2$ FHSV model \cite{FHSV},
still exhibits a \textit{quartic} $U$-invariant $\mathcal{I}_{4}$, but it
can be \textit{non-maximally} further truncated to an $\mathcal{N}=2$ $%
\mathbb{CP}^{5}$ model with \textit{quadratic} invariant through the
procedure considered in Sec. \ref{2--->CPn}, to which we address the reader
for further elucidation. Note that this case, as well as the cases treated
at points $1$, $3$ and $4$ of Sec. \ref{8--->2}, is based on the maximal
(symmetric) Jordan algebraic embedding (see \textit{e.g.} \cite{Guna-Lects}%
):
\begin{eqnarray}
J_{3}^{\mathbb{A}} &\supset &J_{2}^{\mathbb{A}}\oplus \mathbb{R}\text{,~}%
\mathbb{A}=\mathbb{O},\mathbb{H},\mathbb{C},\mathbb{R}; \\
J_{2}^{\mathbb{A}} &\sim &\mathbf{\Gamma }_{1,q+1}\text{,~}q\equiv \text{dim}%
_{\mathbb{R}}\mathbb{A}=8,4,2,1.
\end{eqnarray}

\subsection{\label{N=2-Exc-CP6}$J_{3}^{\mathbb{O}}\longrightarrow \mathbb{CP}%
^{6}$}

Interestingly, the \textit{exceptional} \textit{magic} theory admits another
relevant truncation:
\begin{eqnarray}
J_{3}^{\mathbb{O}} &:&\mathcal{N}=2,~n_{V}=27\longrightarrow \mathcal{N}=2%
\text{~}\mathbb{CP}^{6}; \\
E_{7\left( -25\right) } &\supset &SU\left( 6,2\right) \supset SU\left(
6,1\right) \times U\left( 1\right) ; \\
\mathbf{56} &=&\mathbf{28}+\overline{\mathbf{28}}=\mathbf{21}_{+1}+\mathbf{7}%
_{-3}+\overline{\mathbf{21}}_{-1}+\overline{\mathbf{7}}_{+3}; \\
\frac{E_{7\left( -25\right) }}{E_{6}\times U\left( 1\right) } &\supset &%
\frac{SU\left( 1,6\right) }{U\left( 6\right) }.
\end{eqnarray}%
The $\mathcal{N}=2$ theory is \textit{minimally coupled} to $6$ vector
multiplets and, on the two-form Abelian field strengths' fluxes, the
truncation condition reads
\begin{equation}
\mathbf{21}_{+1}=0.  \label{trunc-3-3}
\end{equation}%
It can also be proved that the \textit{quartic} invariant $\mathcal{I}_{4}$
of the $\mathbf{R}=\mathbf{56}$ of $E_{7(-25)}$, under the truncation (\ref%
{trunc-3-3}), becomes the \textit{square} of the \textit{quadratic}
invariant of the $\mathbf{R}=\mathbf{7}$ of $SU\left( 1,6\right) $.

\subsection{\label{2--->CPn}$\mathbb{R}\oplus \mathbf{\Gamma }%
_{1,2n-1}\longrightarrow \mathbb{CP}^{n}$}

A procedure very similar to the one of Sec. \ref{4--->3} can be considered
in this case.

We consider $\mathcal{N}=2$ supergravity based on the rank-$3$ Jordan
algebra $\mathbb{R}\oplus \mathbf{\Gamma }_{1,2n-1}$, with $n_{V}=2n+1$
vector multiplets, with data
\begin{eqnarray}
\frac{G_{4}}{H_{4}} &=&\frac{SL_{v}\left( 2,\mathbb{R}\right) }{U\left(
1\right) }\times \frac{SO\left( 2,2n\right) }{SO\left( 2\right) \times
SO\left( n\right) }; \\
\mathbf{R} &=&\left( \mathbf{2},\mathbf{2+n}\right) .
\end{eqnarray}%
The relevant products of electric and magnetic charges read
\begin{equation}
\begin{array}{l}
p^{2}\equiv p^{\Lambda }p^{\Sigma }\eta _{\Lambda \Sigma }=\left(
p^{0}\right) ^{2}+\left( p^{1}\right) ^{2}-\sum_{a=2}^{2n+1}\left(
p^{a}\right) ^{2}; \\
\\
q^{2}\equiv q_{\Lambda }q_{\Sigma }\eta ^{\Lambda \Sigma
}=q_{0}^{2}+q_{1}^{2}-\sum_{a=2}^{2n+1}q_{a}^{2}; \\
\\
p\cdot q\equiv p^{\Lambda }q_{\Lambda },%
\end{array}
\label{1}
\end{equation}%
where $\eta $ is the symmetric invariant structure of the vector ($\mathbf{%
Fund}$) irrep. $\mathbf{2+2n}$ of $SO\left( 2,2n\right) $, with $\Lambda
=0,1,...,2n+1$, where the indices \textquotedblleft $0$\textquotedblright\
and \textquotedblleft $1$\textquotedblright\ respectively pertain to the
graviphoton and to the axio-dilatonic Maxwell field.

We consider a complexification of the electric and magnetic charge vectors $%
p^{\Lambda }$ and $q_{\Lambda }$ as follows:
\begin{equation}
\left\{
\begin{array}{l}
P^{1}\equiv p^{0}+ip^{1}; \\
P^{2}\equiv p^{2}+ip^{3}; \\
.... \\
P^{n+1}\equiv p^{2n}+ip^{2n+1},%
\end{array}%
\right.  \label{2}
\end{equation}%
and analogously for the electric charges. Thus (\ref{1}) can be rewritten as
\begin{eqnarray}
p^{2} &=&\left\vert P^{1}\right\vert ^{2}-\sum_{A=2}^{n+1}\left\vert
P^{A}\right\vert ^{2}=P^{i}\overline{P}^{\overline{j}}\eta _{i\overline{j}};
\label{3} \\
q^{2} &=&\left\vert Q_{1}\right\vert ^{2}-\sum_{A=2}^{n+1}\left\vert
Q_{A}\right\vert ^{2}=\eta ^{i\overline{j}}Q_{i}\overline{Q}_{\overline{j}};
\label{4} \\
p\cdot q &=&\sum_{i=1}^{n+1}\text{Re}\left( P^{i}\overline{Q}_{\overline{i}%
}\right) ,  \label{5}
\end{eqnarray}%
with $\eta $ here denoting the invariant rank-$2$ structure in the product $%
\left( \mathbf{1+n}\right) \times \left( \overline{\mathbf{1+n}}\right) $ of
$U\left( 1,n\right) $, with $i=1,...,n+1$. Therefore:
\begin{eqnarray}
\frac{1}{4}\mathcal{I}_{4,\mathbb{R}\oplus \mathbf{\Gamma }_{1,2n-1}}
&=&p^{2}q^{2}-\left( p\cdot q\right) ^{2}  \label{6} \\
&=&\eta _{i\overline{j}}\eta ^{k\overline{l}}P^{i}\overline{P}^{\overline{j}%
}Q_{k}\overline{Q}_{\overline{l}}-\left( \sum_{i=1}^{n+1}\text{Re}\left(
P^{i}\overline{Q}_{\overline{i}}\right) \right) ^{2}  \label{7} \\
&=&\frac{1}{4}\left( S_{1}^{2}-\left\vert S_{2}\right\vert ^{2}\right) ,
\label{8}
\end{eqnarray}%
where, \textit{mutatis mutandis}, $S_{1}^{2}$ and $S_{2}$ are given in (\ref%
{9})-(\ref{10}) \cite{ADF-fixed,CFMZ1}.

By imposing the very same \textit{\textquotedblleft
degeneration\textquotedblright } truncating condition (\ref{11})-(\ref{12}),
and evaluating (\ref{6})-(\ref{8}) on (\ref{11})-(\ref{12}), one obtains (in
Sec. \ref{Identities}, the complex charge vector $\left( P^{i},Q_{i}\right) $
has been indicated by $\mathbf{Q}$):
\begin{equation}
\left. \mathcal{I}_{4,\mathbb{R}\oplus \mathbf{\Gamma }_{1,2n-1}}\right\vert
_{S_{2}=0}=\left( S^{1}\right) ^{2}=4\left( P^{i}\overline{P}^{\overline{j}%
}\eta _{i\overline{j}}\right) ^{2}=\left( \mathcal{I}_{2,\mathbb{CP}%
^{n}}\right) ^{2}.  \label{13}
\end{equation}%
Namely, the \textit{quartic} invariant $\mathcal{I}_{4,\mathbb{R}\oplus
\mathbf{\Gamma }_{1,2n-1}}$of the real irrep. $\mathbf{R}=\left( \mathbf{2},%
\mathbf{2+2n}\right) $ of the semisimple group of type $E_{7}$ $%
G_{4}=SL_{v}\left( 2,\mathbb{R}\right) \times SO\left( 2,2n\right)
=Conf\left( \mathbb{R}\oplus \mathbf{\Gamma }_{1,2n-1}\right) $ \textit{%
\textquotedblleft degenerates\textquotedblright } into the square of the
\textit{quadratic} invariant $\mathcal{I}_{2,\mathbb{CP}^{n}}$ of the
complex irrep. $\mathbf{R}^{\prime }=\mathbf{1+n}$ of the \textquotedblleft
degenerate\textquotedblright\ group of type $E_{7}$ $G_{4}^{\prime }=U\left(
1,n\right) $. This latter is the $U$-duality group of $\mathcal{N}=2$
supergravity \textit{minimally coupled} to $n$ vector multiplets \cite%
{Luciani}.

In a manifestly $U\left( 1,n\right) $-covariant symplectic basis, $\mathcal{I%
}_{2,\mathbb{CP}^{n}}$ reads:
\begin{equation}
\mathcal{I}_{2,\mathbb{CP}^{n}}=\left( \mathfrak{p}^{0}\right) ^{2}+%
\mathfrak{q}_{0}^{2}-\sum_{\alpha =1}^{n}\left[ \left( \mathfrak{p}^{\alpha
}\right) ^{2}+\mathfrak{q}_{\alpha }^{2}\right] .  \label{14}
\end{equation}%
In order to make (\ref{14}) consistent with (\ref{13}), the following
\textit{dyonic identification} of charges can be performed:
\begin{equation}
\begin{array}{l}
P^{1}\equiv \frac{1}{\sqrt{2}}\left( \mathfrak{p}^{0}+i\mathfrak{q}%
_{0}\right) ; \\
P^{A}\equiv \frac{1}{\sqrt{2}}\left( \mathfrak{p}^{\alpha }+i\mathfrak{q}%
_{\alpha }\right) .%
\end{array}
\label{15}
\end{equation}%
Note that in this case (\ref{11}) manifestly breaks $SL_{v}\left( 2,\mathbb{R%
}\right) $, whereas its solution (\ref{12}) further breaks $SO\left(
2,2n\right) $ down to $U\left( 1,n\right) $.

The \textit{\textquotedblleft degeneration\textquotedblright\ }of $\mathcal{I%
}_{4,\mathbb{R}\oplus \mathbf{\Gamma }_{1,2n-1}}$ can also be considered in
the scalar-dressed formalism, in which \cite{ADF-fixed,CFMZ1,CFM2}
\begin{eqnarray}
S_{1} &=&\left\vert Z\right\vert ^{2}+\left\vert Z_{s}\right\vert ^{2}-Z_{I}%
\overline{Z}^{I}; \\
S_{2} &=&2iZ\overline{Z_{s}}-Z_{I}Z^{I},
\end{eqnarray}%
where $Z$, $Z_{s}$ and $Z_{I}$ respectively are the central charge,
axio-dilatonic matter charge and non-axio-dilatonic matter charges ($%
I=1,...,2n$ denotes \textquotedblleft flatted\textquotedblright\ local
indices, also the index $s$ does). Recall that $Z_{s}\equiv \mathcal{D}_{s}Z$%
, $Z_{I}\equiv \mathcal{D}_{I}Z$, $\overline{Z}^{I}=\overline{Z_{I}}$, $%
Z^{I}\equiv Z_{I}$, where $\mathcal{D}$ is the K\"{a}hler-covariant
differential operator in \textquotedblleft flatted\textquotedblright\ local
indices. By splitting the index $I$ as $I=\left\{ \widetilde{I},\widehat{I}%
\right\} $ with $\widetilde{I}=1,...,n$ and $\widehat{I}=1,...,n$, the
\textit{\textquotedblleft degeneration\textquotedblright } condition (\ref%
{11})
\begin{equation}
S_{2}=0\Leftrightarrow 2iZ\overline{Z_{s}}=Z_{I}Z^{I}  \label{11-dressed}
\end{equation}%
can be solved by setting
\begin{equation}
Z_{s}=0,~~Z_{\widetilde{I}}=iZ_{\widehat{I}},  \label{alt-sol-to-deg}
\end{equation}%
thus implying (recall (\ref{8}))
\begin{equation}
\mathcal{I}_{4,\mathbb{R}\oplus \mathbf{\Gamma }_{1,2n-1}}=S_{1}^{2}=\left(
\left\vert Z\right\vert ^{2}-\left\vert Z_{\widetilde{I}}\right\vert
^{2}-\left\vert Z_{\widehat{I}}\right\vert ^{2}\right) ^{2}=\left(
\left\vert Z\right\vert ^{2}-2\left\vert Z_{\widetilde{I}}\right\vert
^{2}\right) ^{2}=\left( \mathcal{I}_{2,\mathbb{CP}^{n}}\right) ^{2},
\end{equation}%
where the re-writing of the invariant $\mathcal{I}_{2,\mathbb{CP}^{n}}$ in
the scalar-dressed formalism reads (see \textit{e.g.} \cite%
{ADF-fixed,CFMZ1,CFM2})
\begin{equation}
\mathcal{I}_{2,\mathbb{CP}^{n}}=\left\vert Z\right\vert ^{2}-\left\vert
Z_{\alpha }\right\vert ^{2},  \label{CPP}
\end{equation}%
thus yielding the following \textit{identification} of scalar-dressed
charges with $\alpha $-dependent \textquotedblleft $\pm $\textquotedblright\
branches:
\begin{equation}
Z_{\widetilde{I}}\equiv \pm \frac{i}{\sqrt{2}}Z_{\alpha }.
\end{equation}%
It should be stressed that (\ref{12}) and (\ref{alt-sol-to-deg}) are
different solutions, in two different (respectively \textquotedblleft
bare\textquotedblright\ and \textquotedblleft
scalar-dressed\textquotedblright ) formalisms, to the \textit{%
\textquotedblleft degeneration\textquotedblright } condition (\ref{11}) (or,
equivalently, (\ref{11-dressed})). Note that the solution (\ref%
{alt-sol-to-deg}) to the manifestly $SL_{v}\left( 2,\mathbb{R}\right) $%
-breaking \textit{\textquotedblleft degeneration\textquotedblright }
condition (\ref{11}) (or, equivalently, (\ref{11-dressed})) consistently
breaks $SO\left( 2,2n\right) $ down to $U\left( 1,n\right) $.

\textit{Mutatis mutandis}, the \textit{\textquotedblleft
degeneration\textquotedblright\ }in the scalar-dressed formalism considered
above can also be performed for of $\mathcal{I}_{4,\mathbb{R}\oplus \mathbf{%
\Gamma }_{5,2n-1}}$ of Sec. \ref{4--->3}; essentially, one has to identify
\begin{equation}
Z\equiv Z_{1},~~i\overline{Z_{s}}\equiv Z_{2},
\end{equation}%
where $Z_{1}$ and $Z_{2}$ are the skew-eigenvalues of the $\mathcal{N}=4$
central charge matrix Z$_{AB}$ ($A,B=1,...,4$) (see \textit{e.g.} \cite%
{ADF-fixed,CFMZ1,More-on-N=8,CFM2}).

In group-theoretical terms, the \textit{\textquotedblleft
degeneration\textquotedblright } truncating procedure under consideration
goes as follows:
\begin{eqnarray}
SL_{v}\left( 2,\mathbb{R}\right) \times SO\left( 2,2n\right) &\supset
&U\left( 1,n\right) ;  \notag \\
\left( \mathbf{2},\mathbf{2+2n}\right) &=&2\cdot \left[ \left( \mathbf{1+n}%
\right) _{+1}+\left( \overline{\mathbf{1+n}}\right) _{-1}\right] ,
\label{16}
\end{eqnarray}%
with the double-counting eventually removed by the \textit{\textquotedblleft
degeneration\textquotedblright } truncating condition (\ref{11})-(\ref{12}),
which sets to zero $n+1$ complex, \textit{i.e.} $2n+2$ real, charge
combinations. As mentioned, (\ref{11}) manifestly breaks $SL_{v}\left( 2,%
\mathbb{R}\right) $-invariance, and its various branches, generated by the
various possibilities in the choice of \textquotedblleft $\pm $%
\textquotedblright\ for each index $i$, are all inter-related by suitable $%
U\left( 1,n\right) $-transformations. At the level of the vector multiplets'
scalar manifolds, it holds%
\begin{equation}
\underset{\mathcal{N}=2,~\mathbb{R}\oplus \mathbf{\Gamma }_{1,2n-1}}{\frac{%
SL_{v}\left( 2,\mathbb{R}\right) }{U\left( 1\right) }\times \frac{SO\left(
2,2n\right) }{SO\left( 2\right) \times SO\left( 2n\right) }}\supset \underset%
{\mathcal{N}=2,~\mathbb{CP}^{n}}{\frac{SU\left( 1,n\right) }{U\left(
n\right) }}.  \label{17}
\end{equation}

\subsubsection{\label{Comment-CP1}A Remark On $\mathbb{CP}^{1}$}

It is worth pointing out that the $n=1$ case of the \textit{``degeneration''}
procedure (\ref{16})-(\ref{17}) is \textit{different} from the ``usual''
\textit{truncation} of the $\mathbb{R}\oplus \mathbf{\Gamma }_{1,n-1}$
sequence down to the \textit{axio-dilatonic} \textit{minimally coupled} $1$%
-modulus $\mathbb{CP}^{1}$ model, achieved by setting $n=0$:
\begin{equation}
\begin{array}{l}
SL_{v}\left( 2,\mathbb{R}\right) \times SO\left( 2,n\right) \overset{n=0}{%
\longrightarrow }SL_{v}\left( 2,\mathbb{R}\right) \times SO\left( 2\right)
\sim U\left( 1,1\right) ; \\
\\
\left( \mathbf{2},\mathbf{2+2n}\right) \overset{n=0}{\longrightarrow }\left(
\mathbf{2},\mathbf{2}\right) \sim \mathbf{2}_{+1}+\overline{\mathbf{2}}_{-1};
\\
\\
\underset{\mathcal{N}=2,~\mathbb{R}\oplus \mathbf{\Gamma }_{1,n-1}}{\frac{%
SL_{v}\left( 2,\mathbb{R}\right) }{U\left( 1\right) }\times \frac{SO\left(
2,n\right) }{SO\left( 2\right) \times SO\left( n\right) }}\overset{n=0}{%
\longrightarrow }\underset{\mathcal{N}=2,~\mathbb{CP}^{1}~\text{\textit{%
axion-dilaton}}}{\frac{SL_{v}\left( 2,\mathbb{R}\right) }{U\left( 1\right) }%
\times \frac{SO\left( 2\right) }{SO\left( 2\right) }\sim \frac{SU\left(
1,1\right) }{U\left( 1\right) }\times \frac{U\left( 1\right) }{U\left(
1\right) }}.%
\end{array}
\label{19}
\end{equation}
Thus, the $U$-duality group of the $1$-modulus \textit{minimally coupled} $%
\mathcal{N}=2$ theory is the unbroken \textit{axio-dilatonic} $SL_{v}\left(
2,\mathbb{R}\right) $ group times the factor $SO\left( 2,n=0\right)
=SO\left( 2\right) $. On the other hand, the $n=1$ case of the \textit{%
``degeneration''} procedure described in Sec. \ref{2--->CPn} manifestly
breaks $SL_{v}\left( 2,\mathbb{R}\right) $, and it determines the $U$%
-duality group of the $1$-modulus \textit{minimally coupled} $\mathcal{N}=2$
theory as the $n=1$ case of the breaking $SO\left( 2,2n\right) \rightarrow
U\left( 1,n\right) $ of the symmetry pertaining to the \textit{%
non-axio-dilatonic} matter sector.

At the level of invariant polynomials of the symplectic irrep. of the $U$%
-duality group, the truncation (\ref{19}) works as (recall (\ref{2}) and (%
\ref{15})):
\begin{eqnarray}
\left. \mathcal{I}_{4,\mathbb{R}\oplus \mathbf{\Gamma }_{1,n-1}}\right|
_{n=0} &=&4\left\{ \left[ \left( p^{1}\right) ^{2}+\left( p^{2}\right) ^{2}%
\right] \left( q_{1}^{2}+q_{2}^{2}\right) -\left(
p^{1}q_{1}+p^{2}q_{2}\right) ^{2}\right\}  \notag \\
&=&4\left( p^{1}q_{2}-p^{2}q_{1}\right) ^{2}=\left( \mathcal{I}_{2,\mathbb{CP%
}^{1}}\right) ^{2}.  \label{20}
\end{eqnarray}

The $\mathcal{N}=2$ symplectic basis obtained in this truncation is the one
in which the holomorphic prepotential reads $F=-iX^{1}X^{2}$, and it thus
differs from the one pertaining to (\ref{14}) with $n=1$, in which $F=-i%
\left[ \left( X^{0}\right) ^{2}-\left( X^{1}\right) ^{2}\right] $. Indeed,
while (\ref{20}) does not vanish \textit{iff} \textit{both} the graviphoton
(index $1$) and the matter Maxwell field (index $2$) have \textit{at least}
one non-vanishing field strength's flux (namely, \textit{iff at least} $%
p^{1},q_{2}\neq 0$ or $p^{2},q_{1}\neq 0$), (\ref{14}) can be non-vanishing
also when the graviphoton (index $0$) \textit{or} the matter Maxwell field
(index $1$) has \textit{both} electric and magnetic zero charges. The $%
Sp\left( 4,\mathbb{R}\right) /U\left( 1,1\right) $ finite transformation $%
\mathcal{S}$ relating the two symplectic bases under consideration reads
\begin{eqnarray}
\left(
\begin{array}{c}
X^{1} \\
X^{2} \\
F_{1} \\
F_{2}%
\end{array}%
\right) _{F=-iX^{1}X^{2}} &=&\mathcal{S}\left(
\begin{array}{c}
X^{0} \\
X^{1} \\
F_{0} \\
F_{1}%
\end{array}%
\right) _{F=-i\left[ \left( X^{0}\right) ^{2}-\left( X^{1}\right) ^{2}\right]
};  \label{21} \\
\mathcal{S} &\equiv &\frac{1}{2}\left(
\begin{array}{cccc}
2 & 2 & 0 & 0 \\
2 & -2 & 0 & 0 \\
0 & 0 & 1 & 1 \\
0 & 0 & 1 & -1%
\end{array}%
\right) \in Sp\left( 4,\mathbb{R}\right) /U\left( 1,1\right) .  \label{22}
\end{eqnarray}

\subsection{\label{Generalized}\textit{\textquotedblleft Generalized"}
Groups of Type $E_{7}$ and Special Geometry}

As introduced in Sec. 4.3 of \cite{FMY-T-CV}, special K\"{a}hler geometry
can be reformulated in order to capture both non-degenerate and degenerate
groups of type $E_{7}$ in a coordinate-independent (\textit{i.e.}
diffeomorphism-invariant) way. This is achieved by introducing \textit{%
\textquotedblleft generalized"} groups of type $E_{7}$, based on a quartic
\textit{\textquotedblleft entropy functional"}, expressed in terms of the
scalar-dressed basis of $\mathcal{N}=2$ central charge $Z$ (graviphoton) and
$\mathcal{N}=2$ matter charges $Z_{i}\equiv D_{i}Z$ (vector multiplets) \ as
follows \label{CFMZ1}:%
\begin{eqnarray}
\mathcal{I}_{4} &=&\left( i_{1}-i_{2}\right) ^{2}+4i_{4}-i_{5}; \\
i_{1} &\equiv &\left\vert Z\right\vert ^{2};  \label{i1} \\
i_{2} &\equiv &Z_{i}\overline{Z}^{i}; \\
i_{3} &\equiv &\frac{i}{6}\left[ Z\overline{C}_{\overline{ijk}}Z^{\overline{i%
}}Z^{\overline{j}}Z^{\overline{k}}+\overline{Z}C_{ijk}\overline{Z}^{i}%
\overline{Z}^{j}\overline{Z}^{k}\right] \\
i_{4} &\equiv &\frac{i}{6}\left[ Z\overline{C}_{\overline{ijk}}Z^{\overline{i%
}}Z^{\overline{j}}Z^{\overline{k}}-\overline{Z}C_{ijk}\overline{Z}^{i}%
\overline{Z}^{j}\overline{Z}^{k}\right] ; \\
i_{5} &\equiv &g^{i\overline{i}}C_{ijk}\overline{C}_{\overline{ilm}}%
\overline{Z}^{j}\overline{Z}^{k}Z^{\overline{l}}Z^{\overline{m}}.  \label{i5}
\end{eqnarray}%
Note that $\mathcal{I}_{4}=\left( i_{1}-i_{2}\right) ^{2}$ if $C_{ijk}=0$;
this corresponds to symmetric $\mathcal{N}=2$ $\mathbb{CP}^{n}$ models,
which upon reduction to $\mathcal{N}=1$ yield \textit{minimal coupling}.
Another way to obtain $\mathcal{N}=2$ $\mathbb{CP}^{n}$ models by truncating
an $\mathcal{N}=2$ theory with $C_{ijk}\neq 0$ is discussed in Subsec. \ref%
{2--->CPn}.

One can make a model-independent analysis holding for any special K\"{a}hler
geometry, by relating the invariants $i_{1}$, $i_{2}$, $i_{3}$, $i_{4}$ and $%
i_{5}$ defined in (\ref{i1})-(\ref{i5}) to the three roots $\lambda _{1}$, $%
\lambda _{2}$, $\lambda _{3}$ of the universal cubic equation (\textit{cfr.}
Eqs. (5.11)-(5.18) of \cite{CDFY-2})%
\begin{equation}
\lambda ^{3}-i_{2}\lambda ^{2}+\frac{i_{5}}{4}\lambda -\frac{\left(
i_{3}^{2}+i_{4}^{2}\right) }{4i_{1}}=0.  \label{Univ-Eq}
\end{equation}%
Within this formalism, the \textit{\textquotedblleft degeneration"}
corresponds to truncating the $\mathcal{N}=2$ vector multiplets such that%
\begin{gather}
i_{3}=i_{4}=i_{5}=0  \label{condd} \\
\Downarrow  \notag \\
\mathcal{I}_{4}=\left( i_{1}-i_{2}\right) ^{2}.
\end{gather}%
The condition (\ref{condd}) implies that a unique non-vanishing independent
root of (\ref{Univ-Eq}) exists, namely $\lambda =i_{2}$.

All reductions treated in Sec. \ref{From-N=2} satisfy the condition (\ref%
{condd}), which can be regarded as a necessary, but not necessarily
sufficient, condition for truncating \textit{any} $\mathcal{N}=2$ model down
to an $\mathcal{N}=2$ $\mathbb{CP}^{n}$ model, and thus to $\mathcal{N}=1$
supergravity models with \textit{minimal coupling}.

\section{\label{N=2--->N=1}$\mathcal{N}=2\rightarrow \mathcal{N}=1$
Truncation and \textit{Minimal Coupling}}

Truncation of $\mathcal{N}=2$ theories to $\mathcal{N}=1$ theories was
studied in \cite{ADF-1,ADF-2}. From Eq. (\ref{1.4}) it is clear that, after
projecting out the graviphoton, the anti-holomorphic vector kinetic matrix
becomes%
\begin{equation}
\mathcal{N}_{\alpha \beta }=\overline{\mathcal{F}_{\alpha \beta }}=\overline{%
\partial }_{\overline{\alpha }}\overline{\partial }_{\overline{\beta }}%
\overline{F}\left( \overline{X}\right) ,
\end{equation}%
where the projective symplectic sections $t^{a}\equiv X^{a}/X^{0}$ have been
split as%
\begin{equation}
t^{a}\equiv \left( t^{\alpha },t^{i}\right) ,
\end{equation}%
with index $\alpha $ referring to the scalar directions of the would-be $%
\mathcal{N}=1$ vector multiplets, whereas index $i$ refers to the would-be $%
\mathcal{N}=1$ chiral multiplets. As pointed out above, \textit{minimal
coupling }of vectors requires $F\left( X\right) $ to be \textit{quadratic}
in the $\mathcal{N}=2$ symplectic sections corresponding to $\mathcal{N}=1$
vector multiplets, such that when truncating down to $\mathcal{N}=1$, the
kinetic vector matrix $\mathcal{N}_{\alpha \beta }$ is a \textit{%
scalar-independent} symmetric rank-$2$ tensor. Note that we here use a
symplectic frame of special K\"{a}hler geometry in which an holomorphic
prepotential exists%
\begin{equation}
F\left( X\right) =\left( X^{0}\right) ^{2}F\left( \frac{X}{X^{0}}\right)
\equiv \left( X^{0}\right) ^{2}f\left( t\right) ,
\end{equation}%
so that the $C$-tensor of special geometry reads%
\begin{equation}
C_{abc}=e^{K}\partial _{a}\partial _{b}\partial _{c}f\left( t\right) .
\end{equation}%
In particular, in this basis, $d$-geometries (which include all symmetric
special geometries but the $\mathbb{CP}^{n}$ models) correspond to%
\begin{equation}
\partial _{a}\partial _{b}\partial _{c}f=d_{abc}\text{ constant,}
\end{equation}%
whereas $\mathbb{CP}^{n}$ models correspond to%
\begin{equation}
f\left( t\right) =-\frac{i}{2}\left[ 1-\sum_{a}\left( t^{a}\right) ^{2}%
\right] .
\end{equation}

It is worth remarking that minimal coupling requires, in addition to%
\begin{equation}
C_{\alpha \beta \gamma }=0=C_{\alpha ij},
\end{equation}%
also \cite{ADF-1,ADF-2}%
\begin{equation}
C_{\alpha \beta i}=0,
\end{equation}%
and thus the only non-vanishing components of the $C$-tensor can lie along
the directions $C_{ijk}$ corresponding to the would-be $\mathcal{N}=1$
chiral multiplets.

For symmetric cosets, this is only possible for $\mathbb{CP}^{n}$ scalar
manifolds, with $n=n_{c}+n_{V}$ \ (with $n_{c}$ and $n_{V}$ here denoting
the number of $\mathcal{N}=1$ chiral and vector multiplets, respectively).
The only other possibility would consist in taking the models based on the
\textit{semi-simple} $U$-duality group $SL\left( 2,\mathbb{R}\right) \times
SO\left( 2,n\right) $, and considering only one vector multiplet, but this
is nothing but the $\mathbb{CP}^{1}$ model itself (see the comment in
Subsubsec. \ref{Comment-CP1}).

For non-symmetric special geometry, other solutions exist. In Calabi-Yau
compactifications, the effective $\mathcal{N}=2$ prepotential for particular
orbifold realizations can have a cubic dependence on the \textit{untwisted}
moduli $X_{U}$ and a quadratic dependence on the \textit{twisted} moduli $%
X_{T}$ (see \textit{e.g.} \cite{DKL}, and Refs. therein):%
\begin{equation}
F\left( X_{U},X_{T}\right) =C_{ijk}X_{U}^{i}X_{U}^{j}X_{U}^{k}+C_{\alpha
\beta }X_{T}^{\alpha }X_{T}^{\beta }.  \label{F-DKL}
\end{equation}%
If one performs a truncation in which the $\mathcal{N}=1$ chiral multiplets
correspond to \textit{untwisted} moduli and $\mathcal{N}=1$ vector
multiplets correspond to \textit{twisted} ones (as suggested by the index
splitting in (\ref{F-DKL}), one obtains a scalar-independent kinetic vector
matrix : $f_{\alpha \beta }=C_{\alpha \beta }$ (\textit{minimal} $\mathcal{N}%
=1$ vector coupling).

Theories which exhibit \textit{minimal coupling} under truncation can for
instance be given by suitable projections of an original $\mathcal{N}=3$
theory down to $\mathcal{N}=1$. Indeed, if some vector multiplets survive
the truncation down to $\mathcal{N}=1$, they necessarily exhibit a minimal
coupling, because the matrix $f_{\alpha \beta }$ is independent of the
remaining $\mathcal{N}=1$ chiral multiplets' complex scalar fields. This can
be understood by considering the intermediate truncation $\mathcal{N}%
=3\rightarrow \mathcal{N}=2$, corresponding to the following branching of
the $U$-duality group (see Sec. \ref{3--->2CPn+hypers}):%
\begin{equation}
U\left( 3,n\right) \supset U\left( 1,n_{V}\right) \times SU\left(
2,n_{H}\right) \times U\left( 1\right) ,~n=n_{V}+n_{H}.
\end{equation}%
The kinetic matrix of the $\mathcal{N}=2$ $n_{V}$ vector multiplets is
independent of the $n_{H}$ $\mathcal{N}=2$ hyperscalars, and after
projecting out the $\mathcal{N}=2$ graviphoton and thus reducing to $%
\mathcal{N}=1$, it also becomes independent of the scalars corresponding to
the $\mathcal{N}=2$ vector multiplets, thus becoming constant and giving
rise to an $\mathcal{N}=1$ \textit{minimal} vector coupling.

Other non-symmetric special geometries are obtained in $\mathcal{N}=1$
Calabi-Yau orientifold compactifications \cite{29-FK,FK-creation}. The
kinetic vector matrix generally depends on the moduli, and in the simplest
case reads as
\begin{equation}
\overline{\mathcal{N}}_{\alpha \beta }=d_{\alpha \beta i}z^{i},  \label{d-or}
\end{equation}%
where as above $\alpha $, $\beta $ run over $\mathcal{N}=1$ vector
multiplets, and $i$ runs over $\mathcal{N}=1$ chiral multiplets. (\ref{d-or}%
) corresponds to orientifold projections of $\mathcal{N}=2$ special $d$%
-geometries \cite{dWVVP}, as they naturally occur in Calabi-Yau
compactifications (where the $d$-tensor is related to the triple
intersection numbers).

\section{\label{Comment-FD}On \textit{Freudenthal Duality} and its \textit{%
``Degeneration''}}

All the cases in which $\mathcal{I}_{4}$ degenerates to $\left( \mathcal{I}%
_{2}\right) ^{2}$ provide instances of the so-called \textit{Freudenthal
duality} \cite{Duff-FD-1,FMY-FD-1}, whose manifest invariance (by
construction, and apart from possible ``hidden'' symmetries) is given by the
$U$-duality group of the theory obtained \textit{after} truncation.

In the \textit{\textquotedblleft degenerative\textquotedblright }
truncations under consideration, the corresponding \textit{\textquotedblleft
degeneration\textquotedblright } of the (\textit{on-shell}, non-polynomial)
\textit{Freudenthal duality} is given by the (\textit{on-shell}, linear)
formula:
\begin{equation}
\widetilde{\mathcal{Q}}^{M}\equiv \mathbb{C}^{MN}\frac{\partial \mathcal{I}%
_{2}}{\partial \mathcal{Q}^{N}},  \label{deg-FD}
\end{equation}%
where $\mathcal{Q}$ is the dyonic charge vector, and
\begin{equation}
\mathbb{C}^{MN}\equiv \left(
\begin{array}{cc}
0^{\Lambda \Sigma } & -\delta _{\Sigma }^{\Lambda } \\
\delta _{\Lambda }^{\Sigma } & 0_{\Lambda \Sigma }%
\end{array}%
\right)
\end{equation}%
is the symplectic metric. Due to the very structure of $\mathcal{I}_{2}$, it
holds that
\begin{equation}
\widetilde{\mathcal{I}_{2}}\left( \mathcal{Q}\right) \equiv \mathcal{I}%
_{2}\left( \widetilde{\mathcal{Q}}\right) =\mathcal{I}_{2}\left( \mathcal{Q}%
\right) .  \label{FD-inv-I2}
\end{equation}

In the manifestly $U\left( 1,n\right) $-covariant $\mathcal{N}=2$ symplectic
basis specified by (\ref{14}), the \textit{\textquotedblleft
degenerate\textquotedblright } Freudenthal duality (\ref{deg-FD}) can be
made explicit as follows:
\begin{eqnarray}
\widetilde{\mathcal{Q}}^{M} &\equiv &\mathbb{C}^{MN}\mathcal{A}_{NP}\mathcal{%
Q}^{P};  \label{deg-FD-1} \\
\mathcal{A}_{MN} &\equiv &\left(
\begin{array}{cc}
\eta _{\Lambda \Sigma } & 0_{\Lambda }^{\Sigma } \\
0_{\Sigma }^{\Lambda } & -\eta ^{\Lambda \Sigma }%
\end{array}
\right) ,  \label{deg-FD-2}
\end{eqnarray}
namely, in components ($\mathcal{Q}=\left( \mathfrak{p}^{\Lambda },\mathfrak{%
q}_{\Lambda }\right) ^{T}$, consistent with (\ref{14})):
\begin{equation}
\left(
\begin{array}{c}
\widetilde{\mathfrak{p}}^{\Lambda } \\
\\
\widetilde{\mathfrak{q}}_{\Lambda }%
\end{array}
\right) =\left(
\begin{array}{c}
-\eta ^{\Lambda \Sigma }\mathfrak{q}_{\Sigma } \\
\\
\eta _{\Lambda \Sigma }\mathfrak{p}^{\Sigma }%
\end{array}
\right) ,  \label{deg-FD-3}
\end{equation}
where $\eta $ is the metric of (the fundamental irrep. of) $SO\left(
1,n\right) $. Note that this explicit treatment can be generalized to $%
\mathcal{N}=3$ supergravity in the manifestly $U\left( 3,n\right) $%
-covariant symplectic basis specified by (\ref{14-bis}) by simply
considering $\eta $ as the metric of (the fundamental irrep. of) $SO\left(
3,n\right) $.

It can be easily checked that the \textit{``degenerate''} Freudenthal
duality transformation $\mathbb{C}\mathcal{A}$ (\ref{deg-FD-1})-(\ref%
{deg-FD-3}) is nothing but a particular \textit{anti-involutive} symplectic
transformation of the relevant $U$-duality group $G_{4}$. Thus, the
invariance (\ref{FD-inv-I2}) is trivial, and in the simple, \textit{%
degenerate} groups of type $E_{7}$ relevant to $D=4$ supergravity (namely, $%
U\left( 1,n\right) $ or $U\left( 3,n\right) $) the corresponding Freudenthal
duality is an \textit{anti-involutive} $U$-duality transformation.

\section{\label{Fermions}\textit{Non-Minimal} Coupling and Fermions}

Certain aspects of \textit{non-minimal} vector coupling reflect on fermions
and their interactions. In particular, one finds that in case that the
holomorphic function $f_{\alpha \beta }(z)$ depends on $z$ the mass of
gaugino's may have a non-vanishing tree level contribution of the form (in
the notation of \cite{Kallosh:2000ve})
\begin{equation}
{\frac{1}{4}}f_{\alpha \beta i}\,g^{-1i}{}_{j}e^{K/2}D^{j}W\bar{\lambda}%
_{R}^{\alpha }\lambda _{R}^{\beta }+(R\Leftrightarrow L).  \label{mass}
\end{equation}%
Such a mass term for $D^{j}W\neq 0$ may play an important role in particle
physics. In the \textit{minimal coupling} case, $f_{\alpha \beta i}\equiv {%
\frac{\partial f_{\alpha \beta }}{\partial z^{i}}}=0$, and the mass of
gaugino's may only come from soft breaking terms and from quantum effects.

Another case of \textit{non-minimal} coupling in the fermion sector involves
a Pauli coupling of a vector to a fermion of the chiral multiplet and a
gaugino (see also App. \ref{Pauli} further below)
\begin{equation}
{\frac{1}{4}}f_{\alpha \beta }{}^{i}\bar{\chi}_{i}\gamma ^{\mu \nu }F_{\mu
\nu }^{-\alpha }\lambda _{L}^{\beta }+h.c.  \label{dec}
\end{equation}%
This process is interesting in the context of creation of matter in the
Universe, after inflation. The bosonic cubic vertices $\phi F^{2}$ or $aF%
\tilde{F}$ provide a possibility of creation of vectors fields from the
inflaton (scalar $\phi $, or the axion $a$). A Pauli coupling above will
allow the fermionic partner of the inflaton, $\chi $ to decay and create a
vector and a gaugino, standard model particles. Thus the dependence of the
vector coupling on scalars due to supersymmetry is present also in the
fermionic sector of the theory and may also be useful. Clearly, both terms
in (\ref{mass}) and in (\ref{dec}) are absent in models of $\mathcal{N}=1$
supergravity with minimal coupling, but necessarily present in models
originating from higher supersymmetries.

\section{\label{Conclusion}Conclusion}

The minimal vector coupling in $\mathcal{N}=1$ supergravity corresponds to
the choice of the constant vector kinetic term as shown in eq. (\ref{min}),
when instead of a holomorphic function of scalars, $f_{\alpha\beta}(z)$, as
in eq. (\ref{nonminimal}), one has $f_{\alpha\beta}= \delta_{\alpha\beta}$.
Meanwhile, there is an interesting possibility to use the couplings like $%
\phi F^2$, and $aF\tilde F$ and the ones with fermions, for cosmological
applications, see for example \cite{Kallosh:2011qk}.

It is therefore interesting to study the origin of such couplings,
attractive for cosmology and for particle physics, from well motivated
superstring theory and their compactification, and related to these
four-dimensional supergravities with higher suppersymmetries.

As resulting from the present paper, generalizing and refining the
investigation carried out in \cite{FK-creation}, the answer to this question
follows from duality symmetry and has a group theoretical origin. The
question is why the vector kinetic matrix $\mathcal{N}_{\Lambda \Sigma
}(\varphi )$ in $\mathrm{Im}\mathcal{N}_{\Lambda \Sigma }F_{\mu \nu
}^{\Lambda }F^{\mu \nu \Sigma }+i\mathrm{Re}\mathcal{N}_{\Lambda \Sigma
}F_{\mu \nu }^{\Lambda }\tilde{F}^{\Sigma \mu \nu }$ (\ref{0}) in $\mathcal{N%
}\geqslant 2$ depends, generically, or does not depend, in degenerate cases,
on scalars, when the theory is reduced to $\mathcal{N}=1$ case. In $\mathcal{%
N}\geqslant 2$ there is a duality symmetry group $G$, embedded into an $%
Sp(2n_{v},\mathbb{R})$, such that the $n_{v}$ vector $2$-form field
strengths and their duals fit into a symplectic representation
\begin{equation}
\mathbf{R}^{\prime }={\mathcal{S}}\mathbf{R}\ ,\qquad {\mathcal{S}}=%
\begin{pmatrix}
A & B \\
C & D%
\end{pmatrix}%
\;\qquad \mathcal{S}^{t}\Omega \;\mathcal{S}=\Omega \ ,\qquad \Omega =%
\begin{pmatrix}
0 & -\mathbb{I} \\
\mathbb{I} & 0%
\end{pmatrix}%
\ .
\end{equation}%
The gauge kinetic term $\mathcal{N}$ generically depends on scalars since it
transforms via fractional transformations
\begin{equation}
\mathcal{N}^{\prime }=(C+D\mathcal{N})(A+B\mathcal{N})^{-1}\,.
\label{sctransf}
\end{equation}

The symplectic symmetric tensor (see \textit{e.g.} \cite{CDF-rev}, and Refs.
therein)%
\begin{eqnarray}
\mathcal{M}_{MN}\left( \mathcal{N}\right) &\equiv &\left(
\begin{array}{cc}
\mathcal{A} & \mathcal{B} \\
\mathcal{C} & \mathcal{D}%
\end{array}%
\right) ;  \label{M-calll} \\
\mathcal{A} &\equiv &\text{Im}\mathcal{N}+\text{Re}\mathcal{N}\left( \text{Im%
}\mathcal{N}\right) ^{-1}\text{Re}\mathcal{N};~~\mathcal{B}\equiv -\text{Re}%
\mathcal{N}\left( \text{Im}\mathcal{N}\right) ^{-1};  \notag
\label{M-calll-2} \\
\mathcal{C} &\equiv &-\left( \text{Im}\mathcal{N}\right) ^{-1}\text{Re}%
\mathcal{N};~~\mathcal{D}\equiv \left( \text{Im}\mathcal{N}\right) ^{-1}
\notag  \label{M-calll-3}
\end{eqnarray}%
is never constant (\textit{i.e.} scalar-independent) in $\mathcal{N}%
\geqslant 2$ supergravity, because, as shown in \cite{FK-creation}, this
would imply the existence of an invariant quadratic form with Euclidean
signature (due to the negative definiteness of $\mathcal{M}$ (\ref{M-calll}%
)-(\ref{M-calll-3}) itself). However, in the present investigation we
exploited a systematic investigation of the cases in which \textit{degenerate%
} groups of type $E_{7}$, when reduced to $\mathcal{N}=1$, may provide a
\textit{scalar-independent} kinetic vector matrix $\mathcal{N}$, and thus a
\textit{scalar-independent} $\mathcal{M}$. For $\mathcal{N}=2$ theories,
this can only occur when the matrix $\mathcal{F}_{\Lambda \Sigma }\equiv
\partial _{\Lambda }\partial _{\Sigma }F$ projected onto the directions
pertaining to the would-be $\mathcal{N}=1$ vector multiplets, is constant,
namely when the holomorphic prepotential $F$ is \textit{quadratic} in the
scalar degrees of freedom corresponding to the would-be $\mathcal{N}=1$
vector multiplets. In \textit{symmetric} special K\"{a}hler geometry, this
implies that $\mathcal{M}\left( \mathcal{F}\right) $ (defined as (\ref%
{M-calll})-(\ref{M-calll-3}) with $\mathcal{N}_{\Lambda \Sigma }\rightarrow
\mathcal{F}_{\Lambda \Sigma }$) is a scalar-independent matrix with
Lorentzian signature, and the corresponding quadratic form $\mathcal{Q}%
\mathcal{M}\left( \mathcal{F}\right) \mathcal{Q}^{T}$ defines the quadratic
symmetric invariant structure of \textit{degenerate} groups of type $E_{7}$
(recall (\ref{CPP}) and Eqs. (34) and (35) of \cite{FK})%
\begin{equation}
\mathcal{I}_{2,\mathbb{CP}^{n}}=i_{1}-i_{2}=-\frac{1}{2}\mathcal{Q}\mathcal{M%
}\left( \mathcal{F}\right) \mathcal{Q}^{T}.  \label{CPPP-2}
\end{equation}%
For \textit{non-degenerate} groups of type $E_{7}$, $\mathcal{M}\left(
\mathcal{F}\right) $ is never scalar-independent, and thus \textit{minimal
coupling} is not allowed.

In the present paper, we carried out a detailed classification and analysis
of all cases of degeneration of groups of type $E_{7}$ responsible for the
duality symmetry of extended supergravity: in this way, our investigation
provides an explanation for the fact that the \textit{minimal coupling} case
is \textit{non-generic} in $\mathcal{N}=1$ supergravity originating from
higher supersymmetries, thus supporting the proposal to use a \textit{%
non-minimal} vector coupling for applications in particle physics and
cosmology.

\section*{Acknowledgments}

S. F. would like to thank Laura Andrianopoli, Riccardo D'Auria and Mario
Trigiante for enlightening discussions. R. K. is grateful to A. Linde and K.
Olive for the discussion of the non-minimal vector couplings in applications
to cosmology and particle physics. The work of S.F. has been supported by
the ERC Advanced Grant no. 226455, Supersymmetry, Quantum Gravity and Gauge
Fields (SUPERFIELDS). The work of R.K. was supported by the Stanford
Institute of Theoretical Physics and NSF grant 0756174.

\appendix

\section{\label{Pauli}Pauli Terms}

\subsection{General Structure}

In a $D=4$ $\mathcal{N}$-extended supergravity theory, the general structure
of Pauli terms read (we use the notation and conventions of \cite%
{DF-gradient-flow}, to which the reader is addressed for further
elucidation):
\begin{equation}
\left[ (\sqrt{-g})^{-1}\mathrm{\mathcal{L}}\right] _{\text{Pauli}}=\mathcal{F%
}_{\mu \nu }^{-\Lambda }\text{Im}\mathcal{N}_{\Lambda \Sigma }\left(
L_{AB}^{\Sigma }\overline{\psi }^{\mu A}\psi ^{\nu B}+L_{IA}^{\Sigma }%
\overline{\psi }^{\mu A}\gamma ^{\nu }\lambda ^{I}+L_{IJ}^{\Sigma }\overline{%
\lambda }^{I}\gamma ^{\mu \nu }\lambda ^{J}\right) +h.c.,
\label{Pauli-general-N}
\end{equation}
\noindent where $\lambda _{I}$ and $\psi _{A\mu }$ respectively denote the
spin-$\frac{1}{2}$ fermions and the gravitino fields, and${\mathcal{F}}_{\mu
\nu }^{(\mp )\Lambda }$ are the self-dual/anti-self-dual combinations of the
vector field strengths:
\begin{eqnarray}
{\mathcal{F}}_{\mu \nu }^{(\mp )\Lambda }\, &\equiv &\,\frac{1}{2}\left( {%
\mathcal{F}}_{\mu \nu }^{\Lambda }\,\mp \,i\star \mathrm{{\mathcal{F}}}_{\mu
\nu }^{\Lambda }\right) ;  \notag \\
\star {\mathcal{F}^{\Lambda }}_{\mu \nu } &\equiv &\frac{1}{2}\epsilon _{\mu
\nu \rho \sigma }{\mathcal{F}}^{\rho \sigma \Lambda },  \notag \\
\star {\mathcal{F}^{\Lambda (\pm )}}_{\mu \nu }\, &=&\,\mp i\mathrm{{%
\mathcal{F}}}_{\mu \nu }^{\Lambda (\pm )}.  \label{defs-1}
\end{eqnarray}
$A,B,\dots $ indices range in the fundamental representation of the $%
\mathcal{R}$-symmetry $SU(\mathcal{N})$ $\times U(1)$ (the $U\left( 1\right)
$ term is missing in the maximal case $\mathcal{N}=8$), their lower (upper)
position denoting left (right) chirality. Besides enumerating the fields,
the indices $I$ actually are a short-hand notation, which encompasses
various possibilities: if the fermions belong to vector multiplets $%
I\rightarrow IA$, since they also transform under $\mathcal{R}$-symmetry; if
they refer to fermions of the gravitational multiplet they are a set of
three $SU(\mathcal{N})$ antisymmetric indices: $I\rightarrow \lbrack ABC]$.
(In the particular case of $\mathcal{N}=2$ $n_{H}$ hypermultiplets : $%
I\rightarrow \alpha $, where $\alpha $ is in the fundamental of $USp(2n_{H})$
).

The matrices entering the Lagrangian are in general all dependent on the
scalar fields $q^{i}$. $\mathcal{N}_{\Lambda \Sigma }$ is the kinetic vector
matrix, generally depending on (a subset $q^{i}$ of) the scalar fields $%
q^{u} $. According to \cite{GZreview}, the indices $\Lambda ,\Sigma $ sit in
the relevant symplectic representation of the $U$-duality group $G$. The
structures $L_{AB}^{\Sigma },\,L_{IA}^{\Sigma },\,L_{IJ}^{\Sigma }$ are
coset representatives of the $\sigma $-model $G/H$ for $\mathcal{N}>2$,
while they are objects of special K\"{a}hler geometry for $\mathcal{N}=2$.
For $\mathcal{N}=1$, they are related to the kinetic matrix of the vectors
(with $L_{AB}^{\Sigma }=0$, because there are no vectors in the $\mathcal{N}%
=1$ gravity multiplet).

In the following, we will specify (\ref{Pauli-general-N}) to $\mathcal{N}=8$%
, to $\mathcal{N}=2$ (in particular, when $G$ is a ``degenerate'' group
\textit{``of type }$E_{7}$\textit{''}) and to $\mathcal{N}=1$ theories (also
in presence of \textit{minimal coupling}).

\subsection{$\mathcal{N}=8$}

In this case, $A=1,...,8$ range in the $\mathbf{8}$ of the $\mathcal{R}$%
-symmetry $SU\left( 8\right) $. Only gravitational multiplet is present; the
gauginos $\lambda _{\left[ ABC\right] }$ are in the rank-$3$ antisymmetric
irrep. $\mathbf{56}$ of $SU\left( 8\right) $, whereas the scalars $q^{\left[
ABCD\right] }$ sit into the rank-$4$ antisymmetric \textit{self-real} irrep.
$\mathbf{70}$ of $SU\left( 8\right) $. (\ref{Pauli-general-N}) thus
specifies to:
\begin{eqnarray}
\mathcal{N} &=&8:\left[ (\sqrt{-g})^{-1}\mathrm{\mathcal{L}}\right] _{\text{%
Pauli}}=\mathcal{F}_{\mu \nu }^{-\Lambda }\text{Im}\mathcal{N}_{\Lambda
\Sigma }L_{AB}^{\Sigma }\overline{\psi }^{\mu A}\psi ^{\nu B}  \notag \\
&&+\mathcal{F}_{\mu \nu }^{-\Lambda }\text{Im}\mathcal{N}_{\Lambda \Sigma
}L_{AB}^{\Sigma }\overline{\psi }_{~C}^{\mu }\gamma ^{\nu }\lambda ^{ABC}
\notag \\
&&+\mathcal{F}_{\mu \nu }^{-\Lambda }\text{Im}\mathcal{N}_{\Lambda \Sigma
}\epsilon _{ABCDEFGH}\overline{\lambda }^{ABC}\gamma ^{\mu \nu }\lambda
^{DEF}\overline{L}^{\Sigma \mid GH}+h.c.~.  \label{Pauli-N=8}
\end{eqnarray}
Thus, by introducing
\begin{eqnarray}
T_{\mu \nu ,~AB}^{-} &\equiv &\mathcal{F}_{\mu \nu }^{-\Lambda }\text{Im}%
\mathcal{N}_{\Lambda \Sigma }L_{AB}^{\Sigma }; \\
T_{\mu \nu }^{-\mid AB} &\equiv &\mathcal{F}_{\mu \nu }^{-\Lambda }\text{Im}%
\mathcal{N}_{\Lambda \Sigma }\overline{L}^{\Sigma \mid AB},
\end{eqnarray}
(\ref{Pauli-N=8}) can be rewritten as
\begin{eqnarray}
\mathcal{N} &=&8:\left[ (\sqrt{-g})^{-1}\mathrm{\mathcal{L}}\right] _{\text{%
Pauli}}  \notag \\
&=&T_{\mu \nu ,~AB}^{-}\overline{\psi }^{\mu A}\psi ^{\nu B}+T_{\mu \nu
,~AB}^{-}\overline{\psi }_{~C}^{\mu }\gamma ^{\nu }\lambda ^{ABC}  \notag \\
&&+\epsilon _{ABCDEFGH}\overline{\lambda }^{ABC}\gamma ^{\mu \nu }\lambda
^{DEF}T_{\mu \nu }^{-\mid GH}+h.c.~.
\end{eqnarray}

\subsection{$\mathcal{N}=2\noindent $}

$\mathcal{N}=2$ supergravity the scalar manifold is a product manifold \cite%
{dlv,dff,CFG},
\begin{equation}
\mathcal{M}_{scalar}={}\mathcal{M}_{vec}\times {}\mathcal{M}_{hyper}
\label{mani}
\end{equation}
since there are two kinds of matter multiplets, the vector multiplets and
the hypermultiplets. The geometry of ${}\mathcal{M}_{vec}$ is described by
the \textit{special\thinspace \thinspace K\"{a}hler\thinspace \thinspace
geometry} \cite{dlv,SKG}, while the geometry of ${}\mathcal{M}_{hyper}$ is
described by \textit{quaternionic\thinspace \thinspace geometry} \cite%
{dlv,dff,HG}; for a thorough geometric treatment, see \textit{e.g.} \cite{a1}%
.\newline
With respect to the general case (\ref{Pauli-general-N})
\begin{equation}
\Lambda =0,1,\dots ,n_{V};\,\,A,B=1,2;\,\,i=1,\dots
,4n_{H}+2n_{V};\,\,I=1,\dots n_{H}+n_{V},  \label{conv}
\end{equation}
where the index $0$ pertains to the \textit{graviphoton}.

As it will be the case in $\mathcal{N}=1$ supergravity, we denote the
complex scalars parameterizing ${}^{(vec)}$ by {${z^{i},\bar{z}^{\bar{\imath}%
}}$}, while the scalars parameterizing $\mathcal{M}_{hyper}$ will be denoted
by {$q^{u}$}. When the index $I$ runs over the vector multiplets it must be
substituted by $IA$ in all the formulae relevant to the vector multiplet,
since the fermions $\lambda ^{IA}$ are in the fundamental of the $\mathcal{R}
$-symmetry group $U(2)$.

$L^{\Lambda }(z,\,\bar{z})$ and its "magnetic" counterpart $M_{\Lambda }(z,\,%
\bar{z})={}\mathcal{N}_{\Lambda \,\Sigma }\,L^{\Sigma }$ actually form a $%
2n_{V}$ dimensional covariantly holomorphic section $V=(L^{\Lambda
},\,M_{\Lambda })$ of a flat symplectic bundle.

When the index $I$ runs over the hypermultiplets, we rename them as follows:
$(I,J) \rightarrow (\alpha,\, \beta) $ and since there are no vectors in the
hypermultiplets we have $f^{\Lambda A} _{\alpha} =0$

The \textit{Vielbein} of the quaternionic manifold ${}\mathcal{M}_{hyper}$
are usually denoted by ${}\mathcal{U}^{\alpha \,A}\equiv \mathcal{U}%
{}_{u}^{\alpha \,A}dq^{u}$, where $\alpha =1,\dots ,2n_{H}$ is an index
labelling the fundamental representation of $USp(2n_{H})$. The inverse
matrix \textit{Vielbein} is ${}_{\alpha \,A}^{u}$. We raise and lower the
indices $\alpha ,\beta ,\dots $ and $A,B,\dots $ with the symplectic
matrices $\mathbb{C}^{\alpha \beta }$ and $\epsilon _{A\,B}$.

Thus, (\ref{Pauli-general-N}) specifies to:%
\begin{eqnarray}
\mathcal{N} &=&2:\left[ (\sqrt{-g})^{-1}\mathrm{\mathcal{L}}\right] _{\text{%
Pauli}}  \notag \\
&=&\mathcal{F}_{\mu \nu }^{-\Lambda }\text{Im}\mathcal{N}_{\Lambda \Sigma }%
\left[
\begin{array}{l}
4L^{\Sigma }\overline{\psi }^{A\mu }\psi ^{B\nu }\epsilon _{AB}-4i\overline{D%
}_{\overline{i}}\overline{L}^{\Sigma }\overline{\lambda }_{A}^{\overline{i}%
}\gamma ^{\nu }\psi _{B}^{\mu }\epsilon ^{AB} \\
\\
+\frac{i}{2}C_{ijk}g^{k\overline{k}}\overline{D}_{\overline{k}}\overline{L}%
^{\Sigma }\overline{\lambda }^{iA}\gamma ^{\mu \nu }\lambda ^{jB}\epsilon
_{AB}-L^{\Sigma }\overline{\zeta }_{\alpha }\gamma ^{\mu \nu }\zeta _{\beta }%
\mathbb{C}^{\alpha \beta }%
\end{array}%
\right] +h.c.,  \label{Pauli-N=2}
\end{eqnarray}%
where $\zeta _{\alpha }$, $\overline{\zeta }_{\alpha }$ denote the spin-$%
\frac{1}{2}$ fermions of the hypermultiplets (\textit{hyperinos}). The
kinetic vector matrix $\mathcal{N}_{\Lambda \Sigma }$ can be constructed in
terms of $L^{\Lambda }$ through the procedure \textit{e.g.} given in \cite%
{a1}.

By introducing the gravity- and matter- \textit{vector projectors}
\begin{eqnarray}
T_{\mu \nu }^{-} &\equiv &2i\text{Im}\mathcal{N}_{\Lambda \Sigma }L^{\Sigma }%
\mathcal{F}_{\mu \nu }^{-\Lambda }; \\
T_{\mu \nu }^{-i} &\equiv &-\text{Im}\mathcal{N}_{\Lambda \Sigma }\mathcal{F}%
_{\mu \nu }^{-\Lambda }g^{i\overline{j}}\overline{D}_{\overline{j}}\overline{%
L}^{\Sigma },
\end{eqnarray}%
(\ref{Pauli-N=2}) can be rewritten as
\begin{eqnarray}
\mathcal{N} &=&2:\left[ (\sqrt{-g})^{-1}\mathrm{\mathcal{L}}\right] _{\text{%
Pauli}}  \notag \\
&=&-\frac{i}{2}T_{\mu \nu }^{-}\left[ 4\overline{\psi }^{A\mu }\psi ^{B\nu
}\epsilon _{AB}-\overline{\zeta }_{\alpha }\gamma ^{\mu \nu }\zeta _{\beta }%
\mathbb{C}^{\alpha \beta }\right]  \label{Pauli-N=2-gravity} \\
&&+\frac{i}{2}T_{\mu \nu }^{-k}\left[ 8g_{k\overline{i}}\overline{\lambda }%
_{A}^{\overline{i}}\gamma ^{\nu }\psi _{B}^{\mu }\epsilon ^{AB}-C_{ijk}%
\overline{\lambda }^{iA}\gamma ^{\mu \nu }\lambda ^{jB}\epsilon _{AB}\right]
+h.c..  \label{Pauli-N=2-matter}
\end{eqnarray}%
Note that for $\mathcal{N}=2$ \textit{minimally coupled} theories, whose $U$%
-duality group is a \textit{degenerate} group of type $E_{7}$ : $%
G_{4}=U\left( 1,n_{V}\right) $, it holds that $C_{ijk}=0$, and thus the
second Pauli term in the \textquotedblleft matter sector\textquotedblright\ (%
\ref{Pauli-N=2-matter}) is absent.

\subsection{$\mathcal{N}=1$}

In order to specify the general formula (\ref{Pauli-general-N}) to $\mathcal{%
N}=1$, we recall that the scalar manifold is in this case a K\"{a}hler-Hodge
manifold and that the $\mathcal{R}$-symmetry reduces simply to $U(1)$.; for
a general treatment, see \textit{e.g.} \cite{Cremmer:1978hn,CFGVP-1}.~It is
convenient in this case to use as \textit{``Vielbeins''} the differentials
of the complex coordinates $dz^{i},d\bar{z}^{\overline{i}}$, where $z^{i}(x)$
are the complex scalar fields parameterizing the K\"{a}hler-Hodge manifold
of (complex) dimension $n_{C}$; thus, in this case we set $q^{u}\rightarrow
(z^{i},\bar{z}^{\overline{i}})$. The spin $\frac{1}{2}$ fermions are either
in chiral or in vector multiplets; so, the index $I$ runs over the number $%
n_{V}+n_{C}$ of vector and chiral multiplets: $I=1,\dots ,n_{V}+n_{C}$.
Furthermore, it is convenient to assign the index $\Lambda $, the same as
for the vectors, to the fermions of the vector multiplets : we will denote
them as $\lambda ^{\Lambda }$,\thinspace\ $\Lambda =1,\dots ,n_{V}$; the
fermions of the chiral multiplets will instead be denoted by $\chi ^{i},\chi
^{\overline{i}}$ in the case of left-handed or right-handed spinors,
respectively. Since the gravitino and the gaugino fermions have no $SU(%
\mathcal{N})$ indices, their chirality will be denoted by a lower or an
upper dot for left-handed or right handed fermions respectively, namely {($%
\psi _{\bullet }$, $\psi ^{\bullet }$)} and {($\lambda _{\bullet }^{\Lambda
} $, $\lambda ^{\bullet \Lambda }$)}. Thus, (\ref{Pauli-general-N})
specifies to:
\begin{equation}
\mathcal{N}=1:\left[ (\sqrt{-g})^{-1}\mathrm{\mathcal{L}}\right] _{\text{%
Pauli}}=\text{Im}\mathcal{N}_{\Lambda \Sigma }\mathcal{F}_{\mu \nu
}^{-\Lambda }\bar{\lambda}^{\bullet \Sigma }\gamma ^{\mu }\psi _{\bullet
}^{\nu }-\frac{i}{8}\partial _{i}{\overline{\mathcal{N}}}_{\Lambda \Sigma }%
\mathcal{F}_{\mu \nu }^{-\Lambda }\bar{\chi}^{i}\gamma ^{\mu \nu }\lambda
_{\bullet }^{\Sigma }+h.c.,  \label{Pauli-N=1}
\end{equation}
where ${\mathcal{F}}_{\mu \nu }^{(\mp )\Lambda }$ are defined in (\ref%
{defs-1}). Within the adopted conventions, $\mathcal{N}_{\Lambda \Sigma }$
is \textit{anti-holomorphic} in the chiral multiplets' complex scalars:
\begin{equation}
\partial _{i}\mathcal{N}_{\Lambda \Sigma }=0.  \label{anti-holomorphic}
\end{equation}

It is instructive to compare (\ref{Pauli-N=1}) with its $\mathcal{N}=2$
counterpart (\ref{Pauli-N=2-gravity})-(\ref{Pauli-N=2-matter}). When
performing the supersymmetry reduction $\mathcal{N}=2\rightarrow \mathcal{N}%
=1$, the ``gravity sector'' (\ref{Pauli-N=2-gravity}) of the $\mathcal{N}=2$
Pauli terms is projected out because, as mentioned, the $\mathcal{N}=1$
gravity multiplet des \textit{not} contain any \textit{graviphoton}. On the
other hand, the ``matter sector'' (\ref{Pauli-N=2-matter}) of the $\mathcal{N%
}=2$ Pauli terms (simpler in the $\mathcal{N}=2$ \textit{minimally coupled}
theory due to $C_{ijk}=0$) becomes (\ref{Pauli-N=1}) itself.

Furthermore, it should be noted that when the $\mathcal{N}=1$ scalars are
\textit{minimally coupled} to the vectors ($\partial _{i}{\overline{\mathcal{%
N}}}_{\Lambda \Sigma }=0$; thus, from (\ref{anti-holomorphic})), the second
term in (\ref{Pauli-N=1}) vanishes, and the Pauli term (\ref{Pauli-N=1})
acquires its \textit{minimally coupled} form
\begin{equation}
\mathcal{N}=1~\text{\textit{minimal~coupling}}:\left[ (\sqrt{-g})^{-1}%
\mathrm{\mathcal{L}}\right] _{\text{Pauli}}=\text{Im}\mathcal{N}_{\Lambda
\Sigma }\mathcal{F}_{\mu \nu }^{-\Lambda }\bar{\lambda}^{\bullet \Sigma
}\gamma ^{\mu }\psi _{\bullet }^{\nu }+h.c..  \label{Pauli-N=1-mc}
\end{equation}

\end{document}